\documentclass[11pt, oneside, hidelinks]{article}

\usepackage{booktabs}
\usepackage[margin=1in]{geometry} 
\usepackage{amsmath,amsthm,amssymb, graphicx, multicol, array,hyperref, bbm, comment, subcaption,  float, booktabs, xpatch, color}
\usepackage{setspace} 
\usepackage{longtable}
\usepackage{enumerate}
\usepackage{algorithm}
\usepackage{algorithmicx}
\usepackage{algpseudocode}

\def\*#1{\mathbf{#1}}
\def\+#1{\mathbb{#1}}
\def\wt#1{\widetilde{#1}}

\newcommand{\tr}{\mathrm{trace}}

\newcommand{\RNum}[1]{\uppercase\expandafter{\romannumeral #1\relax}}

\usepackage[T1]{fontenc}
\usepackage[utf8]{inputenc}
\usepackage{lmodern}
\usepackage[english]{babel}
\usepackage[autostyle]{csquotes}
\usepackage{amsmath}
\usepackage{natbib}
\setcitestyle{open={(},close={)}}

\usepackage{hyperref} 
\usepackage{xr}
\makeatletter
\newcommand*{\addFileDependency}[1]{
	\typeout{(#1)}
	\@addtofilelist{#1}
	\IfFileExists{#1}{}{\typeout{No file #1.}}
}
\makeatother

\usepackage{enumitem}
\setlist{itemsep=.01em}
\setlist{topsep=.5em}

\newtheorem{innercustomgeneric}{\customgenericname}
\providecommand{\customgenericname}{}

\newtheorem{theorem}{Theorem}
\newtheorem{lemma}{Lemma}
\newtheorem{proposition}{Proposition}
\newtheorem{remark}{Remark}

\newtheorem{assumption}{{Assumption}}

\def\beq{\begin{equation}}
\def\eeq{\end{equation}}
\def\beqr{\begin{eqnarray}}
\def\eeqr{\end{eqnarray}}
\def\beqrs{\begin{eqnarray*}}
\def\eeqrs{\end{eqnarray*}}
\def\bet{\begin{theorem}}
\def\eet{\end{theorem}}
\def\bel{\begin{lemma}}
\def\eel{\end{lemma}}
\def\bep{\begin{proposition}}
\def\eep{\end{proposition}}
\def\bg{\begin{figure}[tbph]\begin{center}}
\def\eg{\end{center}\end{figure}}

\def\bc{\begin{center}}
\def\ec{\end{center}}

\def\wt{\widetilde}
\def\wh{\widehat}

\newcommand{\bB}{{\mathbf B}}
\newcommand{\bF}{{\mathbf F}}

\newcommand{\bG}{{\mathbf G}}
\newcommand{\bH}{{\mathbf H}}
\newcommand{\bI}{{\mathbf I}}

\newcommand{\bR}{{\mathbf R}}

\newcommand{\bU}{{\mathbf U}}
\newcommand{\bV}{{\mathbf V}}

\newcommand{\bZ}{{\mathbf Z}}

\newcommand{\bb}{{\mathbf b}}

\newcommand{\be}{{\mathbf e}}
\newcommand{\bff}{{\mathbf f}}
 \newcommand{\bgg}{{\mathbf g}}

\newcommand{\bu}{{\mathbf u}}

\newcommand{\bx}{{\mathbf x}}
\newcommand{\by}{{\mathbf y}}

\newcommand{\bbeta}  {\boldsymbol{\beta}}

\newcommand{\bdelta} {\boldsymbol{\delta}}

\newcommand{\bSigma}{\boldsymbol{\Sigma}}
\newcommand{\bDelta}{\boldsymbol{\Delta}}
\newcommand{\bgamma}{\boldsymbol{\gamma}}

\newcommand{\bve}{\mbox{\boldmath$\varepsilon$}}

\newcommand{\bxi} {\boldsymbol{\xi}}

\newcommand{\ve}{{\varepsilon}}
\renewcommand{\epsilon}{{\ve}}

\def\wt{\widetilde}

\newcolumntype{L}[1]{>{\raggedright\arraybackslash}p{#1}}

\usepackage[margin=10pt,labelfont=bf]{caption}

\begin{document}
	
	\title{Supervised Dynamic PCA: Linear Dynamic Forecasting with Many Predictors}
\date{\today}
		\author{
		Zhaoxing Gao\thanks{\scriptsize  Center for Data Science, Zhejiang University,  Email: \url{zhaoxing_gao@zju.edu.cn}.}
		\and
		Ruey S. Tsay\thanks{ \scriptsize  Booth School of Business, University of Chicago, Email: \url{ruey.tsay@chicagobooth.edu}.}
	}

	\begin{onehalfspacing}
		\begin{titlepage}
		\maketitle
                \thispagestyle{empty}
			\vspace{0cm}
			
			\begin{abstract}
				This paper proposes a novel dynamic forecasting method using a new supervised Principal Component Analysis (PCA) when a large number of predictors are available. The new supervised PCA provides an effective way to bridge the gap between predictors and the target variable of interest by scaling and combining the predictors and their lagged values, resulting in an 
    effective dynamic forecasting. Unlike the traditional diffusion-index approach, which does not learn the relationships between the predictors and the target variable before conducting PCA, we first re-scale each predictor according to their significance in forecasting the targeted variable in a dynamic fashion, and a PCA is then applied to a re-scaled and additive panel, which establishes a connection between the predictability of the PCA factors and the target variable. Furthermore, we also propose to use penalized methods such as the LASSO approach to select the significant factors that have superior predictive power over the others.  Theoretically, we show that our estimators are consistent and outperform the traditional methods in prediction under some mild conditions. We conduct extensive simulations to verify that the proposed method produces satisfactory 
			forecasting results and outperforms most of the existing methods using the traditional PCA. A real example of predicting U.S. macroeconomic variables using a large number of predictors showcases that our method fares better than most of the existing ones in applications. The proposed method thus provides a comprehensive and effective approach for dynamic forecasting in high-dimensional data analysis.
				
				\vspace{0.5cm}
				
				\noindent\textbf{Keywords:}  Dynamic Forecasting, Factor Analysis, Supervised Principal Components,  Large-Dimension, LASSO

				\noindent\textbf{JEL classification:} C22, C23, C38, C53
			\end{abstract}

		\end{titlepage}

\section{Introduction}

Dynamic forecasting is the process of analyzing time series data using statistical and dynamic modeling techniques to make predictions and to inform strategic decision-making. Recent advances in information technologies make it possible to collect large amounts of data over time, which naturally form high-dimensional time series and characterize many contemporary application problems in business, economics, finance, environmental sciences, and other scientific fields. Extracting useful information from such high-dimensional dependent data to make accurate predictions, dimension reduction becomes a necessity. In the past decades, there have been many  dimension-reduction methods developed in the literature, and one of the most successful ones is the Principal Component Analysis (PCA). The PCA method is a versatile tool for dimension reduction and is capable of extracting useful features by transforming the observed variables into a few new uncorrelated factors while retaining much information in the data. See \cite{anderson1958introduction} and \cite{anderson1963asymptotic} for its basic concept 
and properties. The PCA has many applications in various scientific areas, including business, economics, finance, and management. An incomplete list of recent works includes the asset pricing in \cite{gu2020empirical}, \cite{giglio2021asset}, and \cite{he2022shrinking}, the econometric modeling and forecasting in \cite{mullainathan2017machine}, \cite{gaotsay2022}, and \cite{huang2022scaled}, the business and management in \cite{kim2005customer}, among others. Despite its importance, existing works in these fields mainly focus on employing a static decomposition, where the extracted principal components represent the cross-sectional structure of a contemporaneous panel and ignore the information embedded in the past lagged variables of the series.

Arguably the most widely used forecasting method that relies on PCA is 
the ``diffusion-index” approach, which is also known as factor-augmented model developed by \cite{Stock2002b,Stock2002a} and has received much attention among  researchers and practitioners. In a data-rich environment, it has become common that the number of predictors is large and may exceed the number of data points. The factor-based diffusion-index approach is widely applicable under such situations. 
Specifically, for a high-dimensional observed predictor vector $\bx_t\in R^{N}$, 
the goal is to predict $y_{t+h}$ based on the available information $\{\bx_j, j=1,...,t\}$. To avoid the curse of dimensionality, the factor-based approach assumes that the predictors and the variable of interest admit the following structure:
\begin{equation}\label{SW-DI}
\begin{array}{c}
     \bx_t=\bB\bff_t+\bu_t,  \\
     y_{t+h}=\bbeta'\bff_t+\bve_{t+h},
\end{array}    
\end{equation}
where $\bx_t$ and $y_t$ are both taken to have mean zero, $\bff_t$ is an $r$-dimensional vector of latent factors, which may also contain the dynamic factors in \cite{Stock2002b,Stock2002a}, $\bB$ is the associated factor loading matrix, $\bu_t$ is an $N$-dimensional idiosyncratic term, which is uncorrelated with the factors $\bff_t$, and $\bbeta$ is an $r$-dimensional slope parameter linking 
the factors $\bff_t$ to $y_{t+h}$ with $h > 0$. 

Motivated by the success of the diffusion-index approach and to explore further 
available information in $\bx_t$ for predicting $y_{t+h}$, we consider 
the following four perspectives to develop the proposed approach:
\begin{enumerate}
    \item[(i).] Information perspective: Because of the serial dependence in time-series data, past lagged variables of $\bx_t$ are often helpful in predicting $y_{t+h}$. 
    For example, the lagged variables are used in a factor-augmented VAR model in \cite{Bernanke2004} for studying the effect of monetary policy on the economy.
    Yet there is no unified approach available under the existing   
    methods to incorporating lagged information. 
    Different users make use of the lagged variables in different ways. Consider the diffusion approach. One can augment lagged values of the predictors to form an 
    extended predictor vector before applying PCA to extract common factors. The extracted factors are then linear combinations of $\bx_t$ and its lagged variables. Another possibility is to use lagged values of the extracted factors obtained from the original predictors. This procedure assumes {\em a priori} that the 
    linear combinations of predictors carry over lags. Consequently, the lagged information is exploited differently and there are no guidelines available to best use the lagged variables.  In addition, there is no simple way to select the 
    number of past lagged variables needed in an application. 
    \item[(ii).] Adaptive perspective: Empirical applications are often interested in both short-term and long-term predictions so that various choices of $h$ in Model (\ref{SW-DI}) are employed. Yet the same common factors are used for all choices of $h$ because an unsupervised PCA does not consider the target variable in extracting common factors. It is possible in applications that useful predictors for $y_{t+h}$ may not be good ones for $y_{t+h'}$ under the setting in Model (\ref{SW-DI}), where $h\neq h'$. 
 
    \item[(iii).] Supervision perspective: In machine learning, applying principal component analysis to extract the latent factors from $\bx_t$ is an unsupervised procedure. It does not require any information of the target variable $y_{t+h}$. 
    This might become a drawback in prediction as the target variable is of main interest.  In addition, PCA is not scale-invariant, and the extracted factors can change easily when scales of the components of $\bx_t$ change. This may lead to 
    inferior prediction if care is not exercised in applying diffusion index models. 
    For this reason, the components of $\bx_t$ are often standardized before applying PCA. 
    \item[(iv).] Accuracy in factor extraction: The traditional PCA method may not provide accurate estimation of the common factors, especially when some common factors are strong and the remaining ones are weak; see, for example, the explanation in \cite{huang2022scaled} and \cite{bai2021approximate}. For time series data, 
    common factors that have stronger serial dependence tend to dominate those with 
    weaker serial dependence as strong serial dependence often results in higher variances, which in turn 
    leads to larger eigenvalues in the covariance matrix of $\bx_t$.
\end{enumerate}

In view of the above discussions,  we propose a new diffusion-index model in this paper for dynamic forecasting by developing a supervised dynamic PCA (sdPCA) method. First, our proposed sdPCA takes explicitly the lagged variables of predictors into account. Each predictor is allowed 
to have its own number of past lagged variables. In fact, the number of lagged variables for 
each predictor may be selected by an information criterion function such as the Akaike Information Criterion. 
Second, our new diffusion-index model can also employ lagged variables of the extracted common factors. 
Therefore, the proposed approach explores the dynamic dependence in two ways. The extracted factors by 
the proposed sdPCA contain lagged variables of the predictors and the new diffusion-index model also 
employs lagged variables of the common factors. Consequently, our new approach is different from 
the settings in \cite{Stock2002b} and in \cite{bai2006}, where, as discussed earlier, the lagged 
dependence is not fully explored. Third, our sdPCA incorporates information of the target variable 
$y_{t+h}$ in constructing the predictor vector for PCA so that the scale of each component of 
the predictor vector is properly adjusted. Consequently, the proposed new approach does not 
depend on the scales of observed 
predictor $\bx_t$. This property is similar to that of the sPCA of \cite{huang2022scaled}. 
Fourth, the sdPCA is adaptive in extracting common factors so that the common factors used 
to predict $y_{t+h}$ and $y_{t+h'}$ may be different. This can enhance the predictive power of the 
proposed new diffusion-index model.

There are two steps in the proposed sdPCA that produces useful factors to serve as predictors. First, sdPCA runs predictive linear regressions of the target variable on each observed predictor and its past lagged variables with the number of lags being selected by an information criterion. 
In this way, useful lagged variables of each predictor are properly 
addressed. The fitted values of the individual predictive linear 
regressions form a new high-dimensional predictor vector on which 
PCA is carried out to extract common factors. Since the predictive linear regression 
automatically adjusts the scale of each predictor, the scaling effect of 
observed predictors on PCA is also properly addressed. As a matter of fact, one 
can think of the scale of each predictor being adjusted based on its predictive 
power to the target variable. This adjustment may depend on $h$, the number of 
step ahead prediction of interest. Second, sdPCA allows users to add lagged variables of extracted factors to the contemporaneous ones in the linear prediction model. This enables us to explore farther the dynamic information of the data to improve the forecasts. A similar technique is discussed  in \cite{Stock2002a} under different settings. 
In short, these two steps of the proposed 
sdPCA not only capture the dynamic information of the data in an additive manner, 
but also adjust automatically the scaling effect of the data. 
More importantly, the steps enable users to explore fully the linear dependence 
of the data to improve the accuracy of prediction. We show that, under some 
general conditions, the proposed sdPCA can outperform the traditional 
diffusion-index approach using unsupervised PCA, both in theory and in simulation.

Asymptotic properties of the proposed sdPCA are established under the modern setting that the number of predictors $N$ and the sample size $T$ diverge to infinity. We also compare the proposed method with some commonly used methods, and derive the conditions under which, the proposed method can outperform the existing ones theoretically. In addition, to embrace the modern development of machine learning techniques, we also apply the well-known least absolute shrinkage and selection operator (Lasso) approach to select the most  relevant factors that have predictive power for the target variable. Theoretically, we also establish the consistency of the Lasso estimators under some identification conditions.

We  illustrate and assess the performance of  the proposed sdPCA and the new diffusion-index forecasting method with an application to macroeconomic index forecasting.  We forecast the U.S. industrial production (IP) growth, change in the unemployment rate (UNRATE), the consumer price index: all (CPI-All), the S\&P 500 index volatility change (Volatility Change), and the S\&P 500 index return using 123 macroeconomic variables from FRED-MD, as that in \cite{huang2022scaled}, \cite{mccracken2016fred}, \cite{Stock2002b,Stock2002a}, among others. 
Similarly to those in \cite{huang2022scaled}, the sdPCA loadings have re-assigned the weights to the predictors and a smaller subset of the macro variables tend to have more predictive power compared with the unsupervised PCA loadings. Furthermore, the proposed sdPCA together with the Lasso procedure produces comparable or even better forecasting results compared with some commonly used factor-based forecasting methods, such as the traditional PCA method, the sPCA in \cite{huang2022scaled}, and the diffusion-index model in \cite{Stock2002b,Stock2002a}.

The scaled PCA (sPCA) of \cite{huang2022scaled} is the closest method to ours. Both 
methods reconstruct a new high-dimensional prediction vector before extracting common factors using PCA. However, our method differs from the sPCA in three ways. First, our proposed diffusion-index forecasts may include lagged variables of common factors as predictors while sPCA 
only uses the contemporaneous one. Thus, the predictive model  in \cite{huang2022scaled} is similar to that in \cite{Stock2002b} whereas ours generalizes the traditional method to a dynamic fashion.  
As a matter of fact,  sPCA is a special case of the sdPCA  if one does not include any lagged variables of each predictor in constructing the high-dimensional 
prediction vector, nor use any lagged variables of the extracted 
factors. Second, instead of running simple linear regression of the target variable on each predictor alone, we run a time series regression by including lagged variables of the predictors with the number of lags being selected 
by an information criterion. In other words, the proposed sdPCA 
explores the dynamic dependence of the target variable on each 
observed predictor in constructing the new high-dimensional predictor vector 
for PCA. 
Third, the proposed prediction model uses the Lasso method to select the factors that have significant predictive power in prediction. This provides a data-driven approach to identifying relevant predictors.

There are other related works in the literature concerning supervised learning or building connections between predictors and the target variable in prediction; see the references in \cite{huang2022scaled}. We briefly discuss some differences
between the proposed sdPCA approach and some related methods. 
\cite{bai2008forecasting} first applied a screening method to select a subset of predictors that are tested to have relatively more predictive power to the target under either soft- or hard-thresholding rules, and the PCA is conducted on the selected subset to extract common factors for use in the diffusion-index forecasting.  Our proposed method, on the other hand, assigns different weights to each predictor and its lagged variables according to their predictive power without using any thresholding rule. An incomplete list of other related works that share similar insights to ours includes the partial least squares (PLS) regression in \cite{wold1966estimation} and \cite{kelly2015three} with financial applications in \cite{kelly2013market}, \cite{huang2015investor}, and \cite{light2017aggregation}, among others. The comparison of the aforementioned methods and the sPCA has been extensively studied in \cite{huang2022scaled}, and the sPCA is shown to have advantages in forecasting. Therefore, we only compare our proposed method with the sPCA and some commonly used factor-based linear forecasting methods in this paper.

The rest of the paper is organized as follows. Section \ref{sec2} introduces the sdPCA method and the new diffusion-index model, and presents their asymptotic properties and some comparison results. Section \ref{sec3} studies the finite-sample performance 
of the proposed approach via simulation, and \ref{sec4} illustrates the proposed procedure with an empirical application. Section \ref{sec5} concludes. All the proofs and derivations for the asymptotic results are relegated to an online Appendix.

{\bf Notation:}  We use the following notation. For a $p\times 1$ vector
$\bu=(u_1,..., u_p)'$, $\|\bu\|_1=\sum_{i=1}^p|u_i|$ is the $\ell_1$-norm and $\|\bu\|_\infty=\max_{1\leq i\leq p}|u_i|$ is the $\ell_\infty$-norm. $\bI_p$ denotes the $p\times p$ identity matrix. For a matrix $\bH$, its Frobenius norm is $\|\bH\|=[\tr(\bH'\bH)]^{1/2}$ and its operator norm is $\|\bH
\|_2=\sqrt{\lambda_{\max} (\bH' \bH ) }$, where
$\lambda_{\max} (\cdot) $ denotes the largest eigenvalue of a matrix, and $\|\bH\|_{\min}$ is the square root of the minimum non-zero eigenvalue of $\bH\bH'$. The superscript ${'}$ denotes the 
transpose of  a vector or matrix. We also use the notation $a\asymp b$ to denote $a=O(b)$ and $b=O(a)$.


\section{Methodology} \label{sec2}
	\subsection{Model Setup}\label{model_overview}
		
		Let $\bx_t=(x_{1,t},...,x_{N,t})'$ be an $N$-dimensional observable time series, for $t=1,...,T$, and $y_{t+h}$ be the target variable of 
  interest, where $h \geq 1$. The goal is to predict $y_{t+h}$ using $\bx_t$ and its past information. Based on the discussion in the {\it Introduction} section, we assume the data are centered and consider the following model:	
		\begin{equation}\label{GT-DI}
\begin{array}{c}
     \bx_t=\bB\bff_t+\bu_t,  \\
     y_{t+h}=\bbeta(L)'\bff_t+\ve_{t+h},
\end{array}    
\end{equation}
	where $\bbeta(L)=\bbeta_0+\bbeta_1L+...+\bbeta_{q-1}L^{q-1}$ with $L$ being the backshift (or lag) operator such that $L\bff_t = \bff_{t-1}$, and $q\geq 1$ is the number of lagged variables of factors used in predictions. When $q=1$, Model (\ref{GT-DI}) reduces to the diffusion-index forecasting equation in (\ref{SW-DI}). 
	
	The model in (\ref{GT-DI}) is similar to that in \cite{Stock2002b} and \cite{bai2006}, but the mechanisms to produce predictions of the models are fundamentally different. In \cite{Stock2002b}, a dynamic factor model is considered for a large panel of time series and some dynamic factors  are 
 used as predictors, which are assumed to explain most of the variability of the panel, measured by the covariance matrix of the panel. Their factors not necessarily contain information of the past lagged values of $\bx_t$. In Model (\ref{GT-DI}), the observed time series $\bx_t$  admits a factor structure with static factor processes, and these static factors and their past lagged variables are used as predictors in the prediction equation. For similar reasons, Model (\ref{GT-DI}) is also different from the setting in \cite{bai2006} and \cite{huang2022scaled}. On the other hand, there are some similar insights between Model (\ref{GT-DI}) and Model (2.3) in \cite{Stock2002a} because both models 
 attempt to explore the dynamic dependence of the data in the prediction. But the proposed model includes past lagged variables of the common factors whereas the one in \cite{Stock2002a} employs the past lagged variables of $y_t$ in the prediction.  A recursive expansion of $y_t$ in Model (2.3) of \cite{Stock2002a} also leads to using past lagged variables of common factors as predictors albeit with 
 some constraints in the coefficient parameters. For simplicity, we only consider Model (\ref{GT-DI}) in this paper and investigate its predictive ability.

	Under some identification conditions, it is natural to apply PCA to extract common factors from $\bx_t$ and use the extracted factors and their lagged 
 variables as predictors in predicting $y_{t+h}$. Alternatively, one may perform PCA on a stacked vector $(\bx_t',\bx_{t-1}',...,\bx_{t-q+1}')'$ and apply the extracted factors directly in forecasting. Since the factors so obtained already include the lagged variables of $\bx_t$ so that no lagged variables of the factors 
 are used. However, there are some drawbacks of this approach. First, the application of PCA in the first step does not learn any information from the target variable $y_{t+h}$ and, therefore, the factors extracted directly from the 
 stacked vector may not have the best predictability for $y_{t+h}$. Second, the consistency of the PCA procedure is usually shown under the assumption of strong factors in the literature; see, for example, \cite{bai2002determining} and \cite{fan2013large}. In practice, there is no guarantee that all the factors are strong because the noise effect can be prominent when adding more variables to $\bx_t$. As a result, the factors extracted from all the components of $\bx_t$ (or 
 stacked vector) may not have better  predictability than those from a 
 subset of the panel. See \cite{boivin2006more} for further information.
	
	To overcome these drawbacks, we introduce a new supervised dynamic PCA method to extract the common factors and to employ relatively informative factors in a three-stage procedure to explore the dynamic dependence in the data. Our goal is to extract the factors that have more predictive power by learning from the target variable. In other words, we need to estimate the factors $\bff_t$ using a new approach for predictions. Details are discussed next.

	\subsection{Estimation Procedure}\label{est}
	
	In this section, we propose a new PCA useful for dynamic forecasting. 
 The procedure consists of the following three steps:
	\begin{enumerate}
	    \item For $i=1,...,N$, estimate the slope parameters by regressing the target variable on the $i$-th predictor and its past lagged variables:
	    \begin{equation}\label{y:reg}
	  y_{t+h}\approx \wh\mu_i+\wh\gamma_{i,0}x_{i,t}+\wh\gamma_{i,1}x_{i,t-1}+...+\wh\gamma_{i,q_i-1}x_{i,t-q_i+1},\,\,t=q_i,...,T-h,
	    \end{equation}
     where $q_i$ is selected by an information criterion such as the Akaike 
     Information Criterion (AIC). Let $q=\max_{1\leq i \leq N}\{q_i\}$.
	  \item For $t=q,q+1,...,T$, let $\wh\bx_t=(\wh x_{1,t},...,\wh x_{N,t})'$ with $\wh x_{i,t}=\wh\gamma_{i,0}x_{i,t}+\wh\gamma_{i,1}x_{i,t-1}+...+\wh\gamma_{i,q_i-1}x_{i,t-q_i+1}$. Apply PCA to $\wh\bx_t$ and obtain the estimated factors $\wh\bgg_t$, which contains the information of the original factors $\bff_t$ as well as its lagged variables relevant to $y_{t+h}$. One can think of $\wh\bgg_t$ as an estimator for $(\bff_t',...,\bff_{t-q+1}')'$ in Model~(\ref{GT-DI}) under some proper conditions.
	  \item For $t=q,q+1,..., T-h$, apply the Ordinary Least-Squares method to 
   the linear regression of the target variable on $\wh \bgg_t$ and obtain the estimated coefficients:
	  \begin{equation}\label{y:pred}
	    y_{t+h}\approx \wh\alpha+ \wh\bbeta'\wh\bgg_t,\,\, t=q,...,T-h.
	  \end{equation}
	  	Finally, the prediction of $y_{T+h}$ is given by $\wh y_{T+h}=\wh\alpha+\wh\bbeta'\wh\bgg_T$.
	\end{enumerate}
 
Note that the intercepts $\wh\mu_i$ and $\wh\alpha$ in Step 1 and Step 3 above will be removed if the data are assumed to be centered as that in Model (\ref{GT-DI}).	Some remarks are as follows. First, the idea of learning from the target variable using linear regression is similar to that in \cite{huang2022scaled}. However, we focus on dynamic forecasting with lagged variables in the regression, while the sPCA of the aforementioned paper only uses contemporaneous information. Second, unlike the traditional method, the number of lagged variables of each observed predictor used in prediction is selected during the first step. Once the additive panel is formed according to the regression results in Step 1, the extracted factors in Step 2 will automatically contain the dynamic information and can be used as predictors in Step 3, which captures the dynamic dependence in time series forecasting. Third, we only use 
 $\wh\bgg_t$ in Step 3, but we can also employ some of its lagged variables if necessary. Furthermore, as will be seen in the next section, 
 we apply Lasso regularization 
 to obtain $\wh\bbeta$ so that adding additional lagged variables of $\wh\bgg_t$ 
 does not cause any difficulties to the proposed procedure. Fourth, 
 when the number of predictors $N$ and the sample size $T$ are large, one can 
 use a sufficiently large value $q$ in lieu of $q_i$ in Step 1 to simplify the 
 computation.

\subsection{When Diffusion-Index Forecast Meets Sparsity}\label{sec2.3}
As discussed before, even though we have constructed an $N$-dimensional 
vector of predictors in Step 2, it is possible that only a subset of the 
extracted common factors has predictive power for the target variable. Similar insights are also mentioned in \cite{huang2022scaled}, but a statistical approach to automatically select the significant factors is still not available therein.  

In this section, we propose to use penalized regression methods to select the factors that have predictive power for the target variable. For simplicity, we assume the intercepts are zero and only introduce the Lasso approach since other penalized methods can be similarly established. Let $\wh\bbeta_{lasso}$ be the Lasso solution that solves the following optimization problem:
\begin{equation}\label{lasso:s}
\wh\bbeta_{lasso}=\arg\min_{\bbeta\in R^{rq}}\left\{\frac{1}{T}\sum_{t=q}^{T-h}\|y_{t+h}-\bbeta'\wh\bgg_t\|^2+\lambda_T\|\bbeta\|_1\right\},
\end{equation}
	where $\wh\bgg_t$ is the extracted factor process from Step 2 of the 
 proposed procedure in Section \ref{est}, and $\lambda_T>0$ is a penalty parameter to be determined later. The above optimization is a convex one and can be solved by many existing algorithms. See \cite{hastie2009elements} for detailed illustrations.
	
	The basic idea of using regularization estimation, such as the  Lasso regression, is that possibly only a subset of $\wh\bgg_t$ has predictive power for $y_{t+h}$. This is particularly so if $N$ is large. In this case, 
 $\bbeta$ is a sparse vector with only a few non-zero elements. The penalty parameter $\lambda_T$ is used to control the number of significant factors to be used in the predictions.

It is natural to ask why the proposed sdPCA and the new diffusion-index forecasts can outperform the traditional PCA and the scaled PCA in \cite{huang2022scaled} in 
prediction. There are several reasons that can answer this question. First, 
consider the information aspect. Our proposed sdPCA generalizes the scaled PCA by including more relevant lagged variables of observed predictors. As such, the scaled PCA is a special case of the proposed method with $q_i=1$ at Step 1, for 
$i=1, \ldots, N$. Consequently, the proposed sdPCA captures 
 more dynamic dependence information in the data and, hence, can improve the forecasting performance. Second, the proposed Lasso regression can further screen out irrelevant predictors, which in turn can improve parameter estimation and avoid the difficulty of over-parameterization, especially when $N$ is large.

 \begin{remark}
     The Lasso procedure could be very helpful in practice because we do not know how many factors should be included as predictors in Step 3 of Section \ref{est}. Note that $\wh\bgg_t$ in Step 3 can be treated as a proxy for $(\bff_t',...,\bff_{t-q+1}')'$, which contains the dynamic information learned by regressing the target variable on lagged predictors in Step-1 
     of the proposed procedure. In fact, we may also add the lagged variables of $\wh\bgg_t$ and adopt Lasso to select the significant factors in prediction. 
     This expanded approach can collect additional dynamic information that may be missed in Step 1. We do not explore any further this approach to save space.
 \end{remark}

 \begin{remark}
 It is well-known that the Lasso estimate $\wh\bbeta_{lasso}$ is a biased estimate of the true coefficient $\bbeta$. To correct the bias, one can 
 re-run the linear regression using only those predictors selected by the 
 Lasso procedure. This is referred to as a post-selection inference 
 in the literature; see, for instance, \cite{belloni2013}.
\end{remark}

\subsection{Selection of the Lag Parameters}
We use information criteria in Step 1 of the proposed procedure to 
select the number $q_i$ of lagged variables for the predictor $x_{it}$; 
see Equation (\ref{y:reg}). In applying any information criterion, there is 
a need to select the maximum order allowed. This maximum order is unknown 
in practice, but there are statistical methods available to guide the choice. 
We mention two possibilities in this section. First, 
from a time series analysis point of view, the maximum order provides 
an approximation to the {\em true} lagged linear dependence between the 
target $y_{t+h}$ and $\{x_{it},x_{i,t-1},x_{i,t-2},\ldots\}$. When the sample size 
$T$ is large, one can improve the accuracy in approximation by increasing the maximum order allowed. Therefore, a common practice in the literature 
is to use $q_{max} \approx \log(T)$. 
In this way, the information criterion used selects $q_i \in [1,\ldots, q_{max}]$.
This approach can also be applied to the selection of $q$ in the diffusion-index 
forecasting model of Step 3.

 A second method is cross-validation, which is particularly useful when $N$ and $T$ are both large. We partition the data into two sub-samples, say $\{\bx_1,...,\bx_{T_1-h},y_{h},...,y_{T_1}\}$ and $\{\bx_{T_1-h+1},...,\bx_{T-h},y_{T_1+1},...,y_{T}\}$ for some $T_1<T$. For a small integer $q_{max} \geq 1$ and each $1\leq q\leq q_{max}$, we perform the proposed estimation of 
 Section \ref{est} to the first sub-sample with $q_i = q$, for all $i$,  and 
 predict $y_{T_1+h}$ to compute the associated forecasting error. 
  We then move the data $\{\bx_{T_1-h+1},y_{T_1+1}\}$ from the 
  second sub-sample to the first one and repeat the above procedure to calculate 
  another forecasting error. This estimation-forecasting exercise is repeated 
  until we obtain the forecast error of $y_T$. Then $\wh q$ is chosen as the one that produces the smallest out-of-sample forecasting errors, where we may adopt the mean-squared forecasting error (RMSE) defined in (\ref{msfe:d}) in Section \ref{sec27} below.

  \begin{remark}
      In the empirical study of macroeconomic forecasting, we adopt the second approach mentioned above because the number of predictors is large, and many macroeconomic indicators only depend on a small number of the most recent lagged variables. The AIC or BIC criterion can be treated as another option if one wants to further improve the forecasting accuracy of some variables of interest, though information criteria may not always outperform cross-validation in lag selection.
  \end{remark}

\subsection{Selection of the Number of Factors}\label{sec25}
The above analysis depends on a known number of factors $r$, which is unknown in practice. As discussed in \cite{bai2021approximate}, if we want to estimate the number of factors with $\nu>0$, where $\nu$ is a strength parameter defined in Assumption \ref{asm2} below, the criteria in \cite{bai2002determining} remain useful. In addition, there are other estimation methods such as the criteria of \cite{onatski2010determining} and \cite{ahn2013eigenvalue} which separate the bounded eigenvalues from diverging ones of the covariance matrix. For time series factor models with weak factors, \cite{lam2012factor} proposed a multi-step eigenvalue-ratio method to estimate the number of factors with different strengths. This method remains valid under our framework and it is especially useful when the factors have different strengths of weaknesses. Therefore, we may apply those existing methods to estimate the number of factors. Note that it might be helpful to include more factors as predictors and let the Lasso procedure select the factors that have more predictive power. Limited simulation results suggest that the Lasso method can accurately identify the number of significant factors.

We mention that there is an extreme case that the loading matrix associated with $\bgg_t$ in Step 2 may not be of full rank when $\bbeta_i=\bbeta_j$ for some $0\leq i,j\leq q-1$. Then the number of factors identified by the aforementioned methods would be fewer than the true one. But this is not an issue because we can treat $\bff_{i,t}+\bff_{j,t}$ as a new factor with components sharing a common regression coefficient in the linear forecasting step via a regression method. Simulation results in Section \ref{sec3} suggest that the proposed method still works well in out-of-sample prediction.

\begin{remark}
    It is natural that the information criterion in \cite{bai2002determining} should be refined to cover the case of weak factors. However, a valid criterion depends on an accurate estimation of the strength parameter $\nu$ in Assumption \ref{asm2} below, which is difficult to obtain in practice. 
    This difficulty can be addressed by the Lasso procedure introduced in Section \ref{sec2.3}. Specifically, we adopt a data-driven procedure by including more factors as predictors in our empirical studies. The Lasso procedure can then select the factors that play important roles in forecasting the target variable.
\end{remark}

\subsection{Assumptions}
In this section, we introduce the assumptions needed to derive theoretical results of the asymptotic forecasting performance of the proposed method. Most assumptions below are commonly used in the PCA or approximate-factor modeling literature, and the derivation of the consistency of the LASSO estimates needs some slightly stronger assumptions. We use $c$ or $C$ to denote a generic positive constant the value of which may change at different places. 
\begin{assumption}\label{asm01}
The process $\{\bff_t\}$ is $\alpha$-mixing with the mixing coefficients satisfying the condition $\alpha_N(k)<\exp(-k)$, where $\alpha_N(k)$ is defined as
\begin{equation}\label{amix}
\alpha_N(k)=\sup_{i}\sup_{A\in\mathcal{F}_{-\infty}^i,B\in \mathcal{F}_{i+k}^\infty}|P(A\cap B)-P(A)P(B)|,
\end{equation}
where $\mathcal{F}_i^j$ is the $\sigma$-field generated by $\{\bff_t:i\leq t\leq j\}$. 
\end{assumption}
\begin{assumption}\label{asm1}
    For $\bgg_t=(\bff_t',\bff_{t-1}',...,\bff_{t-q+1}')'$ with a fixed $q \geq 1$, $\sup_tE\|\bgg_t\|^4\leq C$ and $\frac{1}{T}\sum_{t=q}^{T-h}\bgg_t\bgg_t'\rightarrow_p \bSigma_g$, which is a $qr\times qr$ positive-definite matrix.
\end{assumption}


Note that the $i$-th row of loading $\bb_i$ can be either a random or a fixed constant vector. Either way, we may define $\bb_i={\bf 0}$ symbolically if $x_{i,t}$ does not depend on the common factors. Thus, $\bb_i\neq {\bf 0}$ implies that $x_{i,t}$ depends on the common factors with some positive probability. We define $\mathcal{I}_b=\{i:\bb_i\neq {\bf 0}, i=1,\ldots,N\}$, which consists of the indexes for which the corresponding predictors depend on the common factors with positive probabilities. The following assumption is related to the strength of the factor loading.
\begin{assumption}\label{asm2}
$\sup_{1\leq i\leq N}  E\|\bb_i\|^4\leq C$ holds. The cardinality of the set $\mathcal{I}_b$ satisfies $\text{Card}(\mathcal{I}_b)\asymp N^\nu$, 
for some $0 < \nu \leq 1$, and $\frac{1}{N^\nu} \sum_{i=1}^N\bb_i\bb_i'\rightarrow_p\bSigma_B$, which is an $r\times r$ positive definite matrix.
\end{assumption}
If $\nu = 1$ in Assumption 3, all factors are strong as that in \cite{bai2002determining} and \cite{fan2013large}. Similar to Assumption A2 in \cite{bai2021approximate}, we exclude the case of $\nu=0$ because the factors and the idiosyncratic terms are indistinguishable in such a situation.

\begin{assumption}\label{asm3}
    The idiosyncratic term $u_{i,t}=\sigma_i e_{i,t}$ for some $c\leq \sigma_i\leq C$, where $e_{i,t}$ is independent and identically distributed over $i$ and $t$ with the eighth moment bounded.
\end{assumption}

\begin{assumption}\label{asm4}
    $\{\bb_i,1\leq i\leq N\}$, $\{\bff_t,1\leq t\leq T\}$, and $\{\bu_{i,t},1\leq i\leq N,1\leq t\leq T\}$ are mutually independent with each other.
\end{assumption}

\begin{assumption}\label{asm5}
    $\{\bve_t\}$ is independent with the three sets of variables in Assumption~\ref{asm4} and it is a martingale-difference sequence such that $E(\bve_{t+h}|\mathcal{F}_t)=0$ for any integer $h>0$, where $\mathcal{F}_t$ is the $\sigma$-field generated by $\{\bgg_t,\bu_t,\bgg_{t-1},\bu_{t-1},...\}$. Furthermore, $\sup_tE(\bve_t^4)\leq C$.
\end{assumption}

Assumption~\ref{asm01} is standard to characterize the dynamic dependence of 
the factor processes. See, for example, \cite{gao2019banded}. It is used to control the magnitude of joint partial sums in the derivations as well as the consistency of the Lasso estimators in Section \ref{sec2.3}. Assumptions~\ref{asm1}-\ref{asm5} are similar to those in \cite{huang2022scaled}, and they also imply that the results in Assumptions A1-A3 of \cite{bai2021approximate} hold, except for the assumption of distinct eigenvalues in A2(iii) therein. In fact, they are adequate for proving the consistency of the estimators and deriving the asymptotic forecasting performance. The illustrations of all the assumptions are stated in \cite{huang2022scaled} or \cite{bai2021approximate}, and we omit the details to save space. The assumption of distinct eigenvalues in A2(iii) of \cite{bai2021approximate} is only used to show the limiting distributions of $\wt\bF'\bF/T$, where $\wt\bF$ is the PC estimator for $\bF$ therein, and we will make similar assumptions below in order to show the consistency of the Lasso estimators in Section \ref{sec2.3}. 

It is worth mentioning that the independence assumption in Assumption \ref{asm3} is only made to simplify the theoretical derivations. It can be relaxed to a weaker assumption such as those in \cite{bai2021approximate}, and those bounds and inequalities in Assumptions A1 and A3 therein can be verified if we impose some mixing condition or weak dependence assumption on the idiosyncratic vector both cross-sectionally over space and dynamically over time.

For the consistency of the Lasso estimators, we also need the following assumptions.
Let $\bB_\gamma'=(\bgamma_1\otimes\bb_1,...,\bgamma_N\otimes\bb_N)$, where $\bgamma_i=(\bI_q\otimes \bb_i')\bbeta$, for $1\leq i\leq N$.
\begin{assumption}\label{asm7}
   $\bG'\bG/T=\bI_{rq}$ and  $\bB_\gamma'\bB_\gamma$ is a diagonal matrix with distinct eigenvalues, where $\bG=(\bgg_q,...,\bgg_{T-h})'$ consists of $\bgg_t$'s defined in Assumption \ref{asm1} as its row vectors.
\end{assumption}

\begin{assumption}\label{asm8}
    For any $1\leq i\leq r$,  and $1\leq t\leq T$,  $P(|f_{i,t}|>x)\leq C_0\exp(-C_1x)$ and $P(|\ve_{t}|>x)\leq C_0\exp(-C_1x)$,  where $C_0>0$ and $C_1>0$ are constants.
\end{assumption}

Assumption \ref{asm7} is an identification condition that guarantees the uniqueness of the estimated factors. See the illustration in \cite{bai2013principal} for details. Assumption \ref{asm8} controls the tails of the factors and the random errors in the forecasting model, and is essentially a sub-exponential assumption. This assumption is stronger than the moment conditions above, but they are adequate for establishing the consistency of the Lasso estimators with the theory developed in \cite{merlevede2011bernstein}.

\subsection{Asymptotic Forecasting Performance}\label{sec27}
In this subsection, we  present some theoretical properties of the proposed estimators and compare the asymptotic forecasting performance of the proposed method with some existing ones.

 Letting $\wh\bG_{\text{sdPCA}}$ be the estimated factors using the proposed method, we have the following consistency result.
\begin{theorem}\label{thm1}
Suppose that Assumptions \ref{asm01}-\ref{asm5} hold. If $N^{1-\nu}/T^2\rightarrow 0$, there exits an invertible rotation matrix $\bH_{\text{sdPCA}}$ such that the estimated factors satisfy
\[\frac{1}{\sqrt{T}}\|\wh\bG_{\text{sdPCA}}-\bG\bH_{\text{sdPCA}}'\|=O_p(N^{-\nu/2}+T^{-1}+\frac{N^{1-\nu}}{T^2}).\]
\end{theorem}

The following proposition provides the condition under which the traditional PCA estimators are also consistent and the condition under which the traditional PCA does not produce consistent factor estimates.

\begin{proposition}\label{prop1}
Suppose that Assumptions \ref{asm01}-\ref{asm5} hold.\\
(i) If $\lim\inf N^{1-\nu}/T \geq C$, for some positive constant $C>0$, then, for any invertible matrix $\bH_{PCA}$, the following result holds,
\[\frac{1}{\sqrt{T}}\|\wh\bF_{\text{PCA}}-\bF\bH'_{PCA}\|\geq C>0,\]
implying that $\wh\bF_{\text{PCA}}$ is not a consistent estimator under such conditions.\\
(ii)  If $N^{1-\nu}/T\rightarrow 0$, there exists a rotation matrix $\bH_{PCA}$ such that, the tradition PCA estimator, denoted by $\wh\bF_{PCA}$, satisfies
\[\frac{1}{\sqrt{T}}\|\wh\bF_{\text{PCA}}-\bF\bH'_{PCA}\|=O_p(N^{-\nu/2}+\frac{N^{1-\nu}}{T}).\]
\end{proposition}
Some remarks on the results in Theorem~\ref{thm1} and Proposition~\ref{prop1} are in order. First, The conditions in Theorem~\ref{thm1} and in Proposition~\ref{prop1} indicate that the requirement for the consistency of the factor estimation using the proposed method is relatively weak. For example, if $N^{1-\nu}/T^2\rightarrow 0$ but $\lim\inf N^{1-\nu}/T>C$ for some constant $C>0$, the extracted factors using the proposed method are consistent while those by the traditional method are not. Second, even if $N^{1-\nu}/T^2\rightarrow 0$, under which both methods produce consistent factors, the factors obtained by the proposed method can still produce more accurate predictions as shown in Theorem ~\ref{thm2} below, where the smaller term in (\ref{msfe:prop}) converges to zero faster than that in (\ref{msfe:trad}). Finally, if all the factors are strong ones, that is, $\nu=1$, the convergence rates of the factors extracted by two methods are the same.

Next, we compare the asymptotic forecasting performance of several existing linear forecasting methods using factors. Define the  mean-squares-forecast error (MSFE) as 
\begin{equation}\label{msfe:d}
  \text{MSFE}={\frac{1}{T}\sum_{t=q}^{T-h}(y_{t+h}-\wh y_{t+h})^2}, 
\end{equation}
where $\wh y_{t+h}=\wh\bbeta'\wh\bgg_t$. Let $w_{1,NT}=N^{-\nu/2}+T^{-1}+\frac{N^{1-\nu}}{T^2}$, $w_{2,NT}=N^{-\nu/2}+\frac{N^{1-\nu}}{T}$, $\by=(y_{h+1},...,y_{T})'$ and $\wh\by=(\wh y_{h+1},...,\wh y_{T})'$. In the following error analysis, we assume the true forecasting model is the one in (\ref{GT-DI}), and denote the MSFE produced by the proposed method, the method of  \cite{huang2015investor}, and the one in \cite{Stock2002b} by $\text{MSFE}_{sdPCA}$, $\text{MSFE}_{sPCA}$, and $\text{MSFE}_{SW}$, respectively. The MSFE of the forecasting method that stacks all extracted factors by the traditional PCA without rescaling is denoted by $\text{MSFE}_{PCA}$. Note that both $\text{MSFE}_{sdPCA}$ and $\text{MSFE}_{PCA}$ make use of lagged factors in prediction while the other two methods only use contemporaneous factors.  We have the following theorem concerning the forecasting performance of different methods.
\begin{theorem}\label{thm2}
Suppose that Assumptions \ref{asm01}-\ref{asm5} hold.\\
(i) If $N^{1-\nu}/T^2\rightarrow 0$, 
\begin{equation}\label{msfe:prop}
    \text{MSFE}_{sdPCA}=\frac{1}{{T}}\|\by-\wh\by\|^2= \frac{1}{{T}}\|(\bI-\frac{1}{T}\bG\bG')\bve\|^2+O_p(w_{1,NT}^2).
\end{equation}
(ii) If $N^{1-\nu}/T\rightarrow 0$,
\begin{equation}\label{msfe:trad}
     \text{MSFE}_{PCA}=\frac{1}{{T}}\|\by-\wh\by\|^2= \frac{1}{{T}}\|(\bI-\frac{1}{T}\bG\bG')\bve\|^2+O_p(w_{2,NT}^2).
\end{equation}
(iii) If $\lim\inf N^{1-\nu}/T\geq c>0$ and $N^{1-\nu}/T^2=o(1)$, then
\[\text{MSFE}_{PCA}-\text{MSFE}_{sdPCA}\geq C>0,\]
implying that the proposed method outperforms the traditional one in theory.\\
(iv) If $N^{1-\nu}/T\rightarrow 0$,  we have
\[\text{MSFE}_{SW}-\text{MSFE}_{sdPCA}\geq C>0,\]
with a strictly positive probability, where, as defined earlier, 
 $\text{MSFE}_{SW}$ denotes the mean-squares of forecasting errors using the diffusion-index model in \cite{Stock2002b}. \\
(v) If $N^{1-\nu}/T^2\rightarrow 0$, then 
\[\text{MSFE}_{sPCA}-\text{MSFE}_{sdPCA}\geq C>0,\]
with a strictly positive probability. 
\end{theorem}
Some remarks on the results in Theorem~\ref{thm2} are given below. First, the results in Theorem \ref{thm2}(iii) show that the proposed forecasting method outperforms the one using the traditional PCA if $\lim\inf N^{1-\nu}/T\geq c>0$ and $N^{1-\nu}/T^2=o(1)$. Second, even if $N^{1-\nu}/T\rightarrow 0$, the smaller term in (\ref{msfe:prop}) converges to zero faster than that in (\ref{msfe:trad}), implying that the proposed forecasting method still has a certain probability to make more accurate predictions. Third, Theorem~\ref{thm2}(iv)-(v) show that our forecasting method can produce smaller errors than the ones in \cite{Stock2002b} and \cite{huang2022scaled}, where the latter two methods only include the contemporaneous  factors and overlook the relevant information in the lagged variables.

Next, we present the asymptotic behavior of the Lasso estimator in the forecasting step.
\begin{theorem}\label{thm3}
    Suppose that Assumptions \ref{asm01}-\ref{asm8} hold. If the sparsity is $0<s^*<rq$ and the penalty parameter $\lambda_T\geq\max\{M\sqrt{\frac{rq}{T}},Mw_{1,NT}\}$ for some $M>0$, then, with probability tending to one, we have
    \[\|\wh\bbeta_{lasso}-\bbeta\|\leq C\lambda_T,\]
    where $C$ may depend on $s^*$. The Lasso estimate is consistent under the conditions in Theorem \ref{thm1} if we choose $\lambda_T=C^*\max(\sqrt{\frac{rq}{T}},w_{1,NT})$ for some appropriate constant $C^*>0$.
\end{theorem}
The result in Theorem \ref{thm3} is a classical one in the Lasso literature; see, for example, \cite{buhlmann2011statistics}.
When the penalty parameter is properly chosen, Theorem~\ref{thm3} indicates that we can correctly recover the non-zero elements in the linear regression asymptotically. Simulation results in Section \ref{sec3} suggest that the Lasso procedure works sufficiently well in finite samples. In addition, it can also improve the forecasting performance as shown in the empirical example in Section \ref{sec4}.

\section{Simulation Studies}\label{sec3}

In this section, we use Monte-Carlo experiments to compare the forecasting performance of the proposed method with some existing ones. The data-generating process (DGP) used is given below. Consider a two-factor model in the experiment, that is, the number of factors $r=2$ in Model (\ref{GT-DI}). The factor process $\bff_t$ is independently generated from normal distributions with zero mean and identity covariance, that is, $\bff_t\sim N({\bf 0},{\bf \bI_2})$. The idiosyncratic terms $u_{i,t}$ are independently and normally distributed with zero mean and unit variance, i.e., $u_{i,t}\sim N(0,1)$. The elements of the loading matrix $\bB$ are drawn independently from the uniform distribution $U(-2,2)$. The target variables are generated by $y_{t+1}=\bbeta_0'\bff_t+\bbeta_1'\bff_{t-1}+\ve_{t+1}$ with independent and identical errors $\ve_t\sim N(0,1)$, that is, $q=2$ in Model (\ref{GT-DI}). We use 100 replications for each configuration $(T,N)$, where $T$ and $N$ are the sample size and the number of predictors, respectively. To make the results below replicable, the seed is set to be \texttt{1234} in the \texttt{R} programming.

\subsection{In-Sample Forecasting}\label{sec31}
We first examine the in-sample forecasting errors with different competing methods. The coefficients in the linear forecasting model are set as $\bbeta_0=(1,-0.8)'$ and $\bbeta_1=(-1,2)'$. To create weak factors, we randomly choose $n$ rows in $\bB$ as nonzero ones with $n\ll N$, and set all the remaining rows to zero. 
We consider two configurations of $(T,N)$ with $(T,N)=(200,300)$ and $(T,N)=(200,500)$, respectively. In practice, the time span of the predictors is from 1 to $T$ and that of the target variable $y$ is from $3$ to $T+1$. We compare the forecasting performance of the factors extracted by the proposed method (denoted by  sdPCA), of the factors and their lagged variables extracted by the traditional PCA approach (denoted by PCA), of the factors extracted by the method in \cite{huang2022scaled} (denoted by sPCA), and of the ones extracted by the diffusion-index model of \cite{Stock2002b} (denoted by SW).

Table~\ref{Table-a1} presents the forecasting performance of the proposed supervised dynamic PCA method (sdPCA) with the aforementioned competing methods such as the PCA, sPCA, and SW, where the factors are of different degrees of weakness. We focus on their in-sample MSFEs defined as (\ref{msfe:d}) in this experiment under different settings, that is, we first estimate the factors and the coefficients in the forecasting model, and then examine the in-sample sum-of-squared residuals and report the MSFEs. We report the mean and the median of the in-sample MSFEs for each configuration of $(T, N,n)$ and each method in Table~\ref{Table-a1}.

\begin{table}[h]
\caption{The in-sample one-step ahead root
mean squared forecast errors of $y_t$. Four methods are used: the proposed method (sdPCA), the one using the factors and their lagged ones extracted by PCA (denoted by PCA), the scaled PCA of \cite{huang2022scaled} (denoted sPCA), and the original diffusion-index one in \cite{Stock2002b} using the factors extracted by PCA without lagged ones (denoted by SW).  Two configurations of the sizes are considered, $(T,N)=(200,300)$ and $(200,500)$. The factors are weak ones in the sense that only $n=40,30,20,10$ out of $N=300$ or $500$ predictors have non-zero loading coefficients on the common factors. The data generating process is that in (\ref{GT-DI}) with $r=2$, where the factors, idiosyncratic terms, and the random errors in the linear forecasting model are independently and identically drawn from the standard normal distribution. The non-zero loading elements are drawn independently from the Uniform distribution $U(-2,2)$. The regression coefficients are $\bbeta_0=(1,-0.8)'$ and $\bbeta_1=(-1,2)'$. 100 replications are used throughout the experiments.} 
          \label{Table-a1}
{\begin{center}
\begin{tabular}{cccccccccccc}
\toprule
\multicolumn{12}{c}{$(T,N)=(200,300)$}\\
\hline
&\multicolumn{2}{c}{sdPCA}&&\multicolumn{2}{c}{PCA}&&\multicolumn{2}{c}{sPCA}&&\multicolumn{2}{c}{SW}\\
\cline{2-3}\cline{5-6}\cline{8-9}\cline{11-12}
$n$&mean&median&&mean&median&&mean&median&&mean&median\\
40&1.064&1.065&&1.080&1.084&&2.395&2.399&&2.414&2.419\\
30&1.077&1.082&&1.112&1.113&&2.394&2.389&&2.415&2.410\\
20&1.123&1.123&&1.203&1.205&&2.396&2.393&&2.425&2.422\\
10&1.258&1.248&&1.858&1.844&&2.422&2.432&&2.526&2.529\\
\midrule
\multicolumn{12}{c}{$(T,N)=(200,500)$}\\
\hline
40&1.054&1.055&&1.136&1.137&&2.406&2.392&&2.440&2.431\\
30&1.093&1.089&&1.266&1.273&&2.403&2.380&&2.454&2.436\\
20&1.129&1.123&&1.525&1.525&&2.387&2.376&&2.484&2.472\\
10&1.325&1.329&&2.571&2.567&&2.303&2.307&&2.697&2.676\\
\bottomrule
\end{tabular}
  \end{center}}
\end{table}
From Table~\ref{Table-a1}, we see that the proposed supervised dynamic PCA outperforms all the other three methods and it fits the model better in terms of the mean and median of the in-sample MSFEs. Specifically, the proposed sdPCA and the PCA methods produce the most accurate predictions in terms of the mean or median of the MSFEs, which is understandable since they both use the correct number of factors while the other two methods do not produce accurate predictions because they only use the contemporaneous factors without including any lagged information in prediction.  Our proposed sdPCA method performs slightly better than the traditional PCA because the supervised procedure can strengthen the components that have non-zero loadings and mitigate the effect of the components with no loadings on the factors. Furthermore, we note that as $n$ decreases, i.e., when there are more zero-loadings or equivalently, the strength of the factors becomes weaker, the in-sample MSFEs tend to become larger, which is in agreement with our theory in the sense that the asymptotic rates in the Theorems of Section 3 will be higher if $\nu$ becomes smaller. In addition, the MSFEs tend to become larger when the number of predictors $N$ increases and this is also in line with the asymptotic theory. Finally, we see that the MSFEs produced by sPCA are smaller than those by SW, which is also reasonable since the scaled PCA may recover the first factor process more accurately using a supervised procedure than the latter unsupervised method.

\subsection{Out-of-Sample Forecasting}\label{sec32}
Next, we examine the out-of-sample performance of the sdPCA and the other three methods  using simulated data. The settings of all the parameters are the same as those in Section 3.1. For each configuration of $(T,N,n)$ and each iteration, we split the data into two sub-samples, one with the first $T_1=[3T/5]$ observations for training and the rest with $T_2=T-T_1$  data points for out-of-sample testing. Specifically, we adopt a rolling-window framework as follows. For the proposed method and the training samples $\{\bx_1,...,\bx_{T_1}\}$ and 
 $\{y_3,...,y_{T_1}\}$, the supervised learning procedure and the estimation of the regression of the linear forecasting model are based on the training samples $\{\bx_1,...,\bx_{T_1-1}\}$ and $\{y_3,...,y_{T_1}\}$, because $h=1$. Then, the predictors are formed as $\{\wh\bx_2,...,\wh\bx_{T_1}\}$, where $\bx_{T_1}$ of the 
 forecast origin is used to obtain $\wh\bx_{T_1}$. Next, the factors are extracted as $\{\wh\bgg_2,...,\wh\bgg_{T_1}\}$. The regression coefficients in the linear forecasting model, denoted by $\wh\bbeta$, are estimated using $\{\wh\bgg_2,...,\wh\bgg_{T_1-1}\}$ and $\{y_3,...,y_{T_1}\}$, and finally, the forecast of $y_{T_1+1}$ is 
 \[\wh y_{T_1+1}=\wh\bbeta'\bgg_{T_1}.\]
We continue this estimation and forecasting procedure by moving the next available data point to the first sub-sample and repeating the above process, and the root-MSFE is defined as
\begin{equation}\label{rmsfe}
   \text{RMSFE}=\sqrt{\frac{1}{T_2}\sum_{\tau=1}^{T_2}(y_{T_1+\tau+h-1}-\wh y_{T_1+\tau+h-1})^2},
\end{equation}
where $h=1$ is considered in the experiments. Table~\ref{Table-a2} reports the mean and median of the out-of-sample RMSFEs for each method and each configuration of $(T,N,n)$, where $(T,N,n)$ are the same as those in Section 3.1.

\begin{table}[h]
\caption{The out-sample one-step ahead root
mean squared forecast errors of $y_{t+1}$. Four methods are used: the proposed method (sdPCA), the one using the factors and their lagged ones extracted by PCA (denoted by PCA), the scaled PCA by \cite{huang2022scaled} (denoted sPCA), and the original diffusion-index one in \cite{Stock2002b} using the factors extracted by PCA without lagged variables (denoted by SW).  Two configurations of the sizes are considered:  $(T,N)=(200,300)$ and $(200,500)$. The factors are weak ones in the sense that only $n=40,30,20,10$ out of $N=300$ or $500$ predictors have non-zero loadings on the common factors. The data generating process is that in (\ref{GT-DI}) with $r=2$, where the factors, idiosyncratic terms, and the random errors in the linear forecasting are independently and identically drawn from the standard normal distribution. The non-zero loading elements are drawn independently from the Uniform distribution $U(-2,2)$. The regression coefficients are $\bbeta_0=(1,-0.8)'$ and $\bbeta_1=(-1,2)'$. 100 replications are used throughout the experiments.} 
          \label{Table-a2}
{\begin{center}

\begin{tabular}{cccccccccccc}
\toprule
\multicolumn{12}{c}{$(T,N)=(200,300)$}\\
\hline
&\multicolumn{2}{c}{sdPCA}&&\multicolumn{2}{c}{PCA}&&\multicolumn{2}{c}{sPCA}&&\multicolumn{2}{c}{SW}\\
\cline{2-3}\cline{5-6}\cline{8-9}\cline{11-12}
$n$&mean&median&&mean&median&&mean&median&&mean&median\\
40&1.120&1.126&&1.112&1.117&&2.427&2.425&&2.428&2.409\\
30&1.134&1.137&&1.523&1.157&&2.428&2.440&&2.431&2.447\\
20&1.188&1.181&&1.262&1.258&&2.438&2.467&&2.449&2.480\\
10&1.136&1.350&&2.030&2.017&&2.508&2.508&&2.596&2.584\\
\midrule
\multicolumn{12}{c}{$(T,N)=(200,500)$}\\
\hline
40&1.115&1.116&&1.179&1.183&&2.471&2.468&&2.482&2.479\\
30&1.176&1.180&&1.334&1.343&&2.475&2.479&&2.490&2.505\\
20&1.241&1.241&&1.644&1.654&&2.474&2.490&&2.543&2.539\\
10&1.678&1.655&&2.639&2.612&&2.707&2.691&&2.734&2.714\\
\bottomrule
\end{tabular}
  \end{center}}
\end{table}
From Table~\ref{Table-a2}, we see that the proposed sdPCA tends to outperform all the other three methods except for the case when $(n,T,N) = (40,200,300)$, which might be due to the errors incurred in finite samples when the idiosyncratic terms and the estimation errors in the supervised procedure are not dominated by the factors. As the factors become weaker, i.e., when $n$ becomes smaller, our proposed sdPCA outperforms all the other methods. Other findings are similar to those in the in-sample case in Section \ref{sec31}, and we do not describe them to save space.

Furthermore, we conduct simulations to confirm that the singularity of $\bB_{\gamma}$ in the Assumption \ref{asm7} does not affect our estimation and forecasting. This case can happen when $\bbeta_0=\bbeta_1$ as mentioned in Section \ref{sec25}. We consider the case of $\bbeta_0=\bbeta_1=(1,1)'$ in this experiment. Table~\ref{Table-a3} presents the mean and median of the RMSFE for different methods and degrees of weakness with $(T,N)=(200,500)$. The settings  of other parameters of the DGP are the same as before. One difference is that we recover two factors in our proposed sdPCA in this scenario while there are 4 factors extracted in the examples in Sections \ref{sec31} and \ref{sec32}. From Table~\ref{Table-a3}, we can see that the performance of various methods is the same as those in Tables~\ref{Table-a1} and \ref{Table-a2}, and the sdPCA continues to work well under the extreme case that $\bB_\gamma$ is singular.

\begin{table}[h]
\caption{The out-sample one-step ahead root
mean squared forecast errors of $y_{t+1}$. Four methods are used: the proposed method (sdPCA), the one using the factors and their lagged ones extracted by PCA (denoted by PCA), the scaled PCA of \cite{huang2022scaled} (denoted sPCA), and the original diffusion-index one in \cite{Stock2002b} using the factors extracted by PCA without lagged ones (denoted by SW).  Two configurations of the sizes are considered: $(T,N)=(200,300)$ and $(200,500)$. The factors are weak ones in the sense that only $n=40,30,20,10$ out of $N=300$ or $500$ predictors have non-zero loadings on the common factors. The data generating process is that in (\ref{GT-DI}) with $r=2$, where the factors, idiosyncratic terms, and the random errors in the linear forecasting are independently and identically drawn from the standard normal distribution. The non-zero loading elements are drawn independently from the Uniform distribution $U(-2,2)$. The regression coefficients are $\bbeta_0=(1,1)'$ and $\bbeta_1=(1,1)'$. 100 replications are used throughout the experiments.} 
          \label{Table-a3}
{\begin{center}

\begin{tabular}{cccccccccccc}
\toprule
\multicolumn{12}{c}{$(T,N)=(200,500)$,$\bbeta=(1,1,1,1)'$}\\
\hline
&\multicolumn{2}{c}{sdPCA}&&\multicolumn{2}{c}{PCA}&&\multicolumn{2}{c}{sPCA}&&\multicolumn{2}{c}{SW}\\
\cline{2-3}\cline{5-6}\cline{8-9}\cline{11-12}
$n$&mean&median&&mean&median&&mean&median&&mean&median\\
40&1.052&1.061&&1.079&1.083&&1.753&1.774&&1.759&1.780\\
30&1.061&1.063&&1.106&1.110&&1.760&1.747&&1.773&1.762\\
20&1.087&1.085&&1.193&1.203&&1.777&1.786&&1.810&1.808\\
10&1.130&1.126&&1.465&1.466&&1.790&1.797&&1.903&1.913\\
\bottomrule
\end{tabular}
  \end{center}}
\end{table}


\subsection{The LASSO Estimation}
In this section, we conduct a simulation study to verify the efficacy of the proposed Lasso procedure. We set $\bbeta_0=(6,3)'$ and $\bbeta_1=(-5,0)'$ where the last coefficient in the linear forecasting model is zero. All the other settings of parameters and data-generating processes are the same as before. Due to the identification issues, there is a subtle change in the final regression step. Specifically, once we have obtained $\wh\bgamma_i$ in the first step and hence a new loading matrix $\bB_{\wh\bgamma}$, we perform a singular-value decomposition on $\bB_{\wh\bgamma}$ and obtain the right singular matrix $\bV$, then the final step is to perform a linear regression of $y_{t+1}$ on $\bV\wh\bgg_t$. We report the frequencies of the correct recoveries of the nonzero coefficients using Lasso through 100 replications in Table~\ref{Table-a4}. We see that, overall, the Lasso approach tends to recover the correct number of non-zero coefficients for moderately large $N$ and $T$ with weak factors. There are two findings from Table~\ref{Table-a4}. First, when the sample size increases with a fixed dimension $N$, the accuracy in recoveries may be improved and the frequencies may slightly decrease when the dimension $N$ increases for a fixed $T$, which is understandable as the Lasso approach can be more accurate when the sample size increases but it may create more errors when the dimension $N$ increases, which is in line with our asymptotic analysis in Theorem~\ref{thm3}.

\begin{table}
\caption{Frequencies of the correct recoveries of the nonzero coefficients using Lasso.   Four configurations of the sizes are considered, $(T,N)=(100,250)$, $(200,250)$, $(200,400)$, and $(300,500)$. The factors are weak ones in the sense that only $n=60,50,40,30$ out of $N$ predictors have non-zero loadings on the common factors. The data generating process is that in (\ref{GT-DI}) with $r=2$, where the factors, idiosyncratic terms, and the random errors in the linear forecasting are independently and identically drawn from standard normal distributions. The non-zero loading elements are drawn independently from the Uniform distribution $U(-2,2)$. The regression coefficients are $\bbeta_0=(6,3)'$ and $\bbeta_1=(-5,0)'$, where only one position is zero. 100 replications are used throughout the experiments.} 
          \label{Table-a4}
{\begin{center}

\begin{tabular}{ccccc}
\toprule 
 & \multicolumn{4}{c}{$(T,N)$} \\ \cline{2-5} 
 $n$ &$(100,250)$&$(200,250)$&$(200,400)$&$(300,500)$\\
\hline
60&0.94&0.94&0.86&0.66\\
50&0.90&0.99&0.80&0.82\\
40&0.87&0.91&0.79&0.85\\
30&0.92&0.95&0.89&0.87\\
\bottomrule
\end{tabular}
  \end{center}}
\end{table}

\section{Empirical Studies} \label{sec4}
In this section, we apply the proposed sdPCA and linear dynamic forecasting method to macroeconomic forecasting with the widely used U.S. monthly macroeconomic variables. To highlight the forecasting power of the proposed method, we compare the performance of sdPCA with some factor-based forecasting methods which are commonly used in the literature. Since the comparisons between the sPCA of \cite{huang2022scaled} and the target PCA, PLS, and regularized methods have been studied in \cite{huang2022scaled}, showing that the sPCA tends to dominate other methods in terms of the forecasting accuracy, we only compare our method with the  sPCA, the factors and their lagged ones extracted by the traditional PCA, and the diffusion-index method in \cite{Stock2002b} without including the lagged ones.

\subsection{Data and In-Sample Results}

We consider the macroeconomic variables studied by \cite{Stock2002a}, \cite{mccracken2016fred}, and \cite{huang2022scaled}, among many others. The data are obtained from the FRED-MD data base which are maintained by St. Louis Fed\footnote{\url{https://research.stlouisfed.org/econ/mccracken/fred-databases/}}. As described in \cite{mccracken2016fred}, this data set extends the widely used \cite{Stock2002a} set and covers broad economic categories including the output and income (OUT), Labor market(LM), Housing (HS), Consumption, orders, and inventories (COI), Money and credit (MC), Interest and exchange rates (IER), Prices (PR), and Stock market (SM). The groups of these variables are the same as those in \cite{mccracken2016fred} while \cite{huang2022scaled} re-grouped them into six ones. The detailed variables and transformation codes to ensure the stationarity of each macro variable are provided in the online data appendix. There are 127 variables in the online data set, but 4 of them are removed due to missing values therein. The remaining 123 macro variables are slightly different from those used in \cite{huang2022scaled} because we 
only focus on the variables contained in the data file without adding new variables or replacing old ones with new variables. We consider the 123 macro variables spanning from July 1962 to December 2019 as all the series have no missing values during this period. Therefore, we have $N=123$ and $T=690$.

We apply the proposed sdPCA to these 123 macro variables  to forecast the 1-month ahead U.S. industrial production (IP) growth, change in the unemployment rate (UNRATE), change in the consumer price index: all (CPI-All), 
growth of real manufacturing and trade industries sales (M\&T Sales), the S\&P 500 index volatility change (Volatility Change), and the S\&P 500 index return (Return), where the S\&P 500 index volatility and the S\&P 500 index return are obtained from the online data appendix of \cite{welch2008comprehensive}\footnote{\url{https://sites.google.com/view/agoyal145}}. The S\&P 500 index return are obtained from the \texttt{CRSP\_SPvw} column, and
the S\&P 500 index volatility is the squared root of the \texttt{svar} column in the online data file, which is slightly different from those calculated in \cite{ludvigson2007empirical} since the former does not subtract the risk-free rate when calculating the monthly volatility.

\begin{figure}
\begin{center}
\caption{Bar charts of in-sample $R^2$s in predicting 1-month ahead industrial production growth (Panel A), unemployment rate (Panel B), consumer price index: all (Panel C), Real Manufacturing and Trade Industries Sales (Panel D), S\&P 500 index Volatility Change (Panel E), and the S\&P 500 index return (Panel F) by each of the 123 macro variables from the FRED-MD data set of \cite{mccracken2016fred}, consisting of eight groups including the output and income (OUT), Labor market(LM), Housing (HS), Consumption, orders, and inventories (COI), Money and credit (MC), Interest and exchange rates (IER), Prices (PR), and Stock market (SM). We set $q=2$ in each regression. Macro variables are collected at a monthly frequency and the sample period is 1962:07--2019:12.}\label{fig1-a}
{\includegraphics[width=0.9\textwidth]{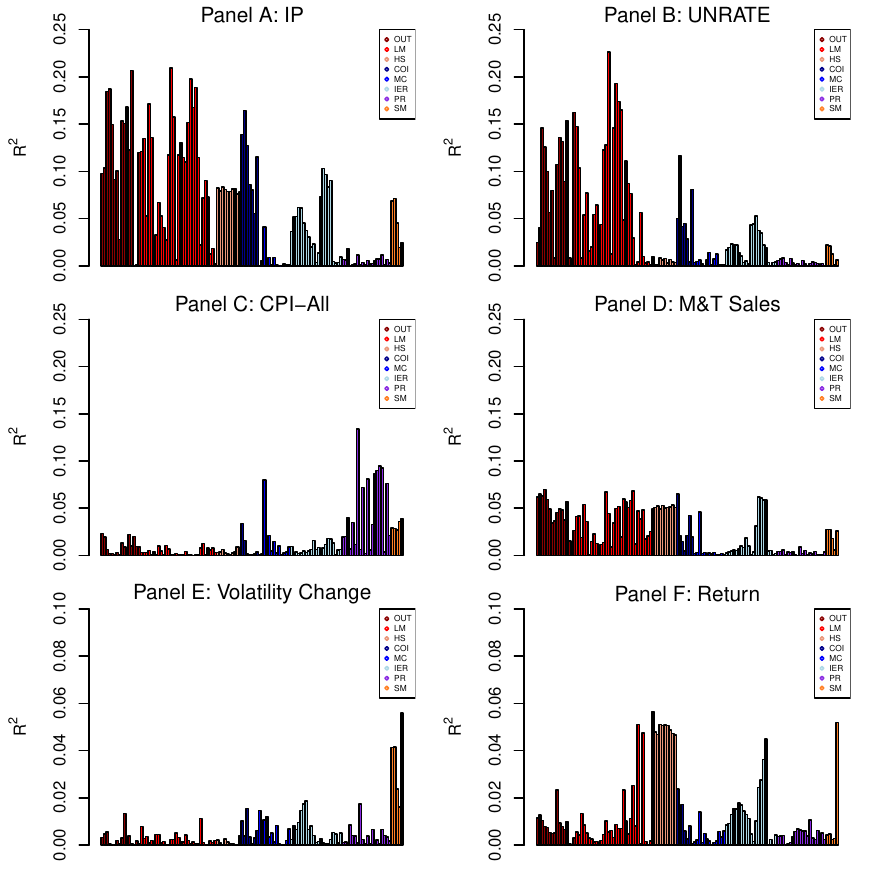}}
\end{center}
\end{figure}

\begin{figure}
\begin{center}
\caption{Bar charts of in-sample $R^2$s in predicting 1-month ahead industrial production growth (Panel A), unemployment rate (Panel B), consumer price index: all (Panel C), Real Manufacturing and Trade Industries Sales (Panel D), S\&P 500 index Volatility Change (Panel E), and the S\&P 500 index return (Panel F) by each of the 123 macro variables from the FRED-MD data set of \cite{mccracken2016fred}, consisting of eight groups including the output and income (OUT), Labor market(LM), Housing (HS), Consumption, orders, and inventories (COI), Money and credit (MC), Interest and exchange rates (IER), Prices (PR), and Stock market (SM). We set $q=3$ in each regression. Macro variables are collected at a monthly frequency and the sample period is 1962:07--2019:12.}\label{fig2-a}
{\includegraphics[width=0.9\textwidth]{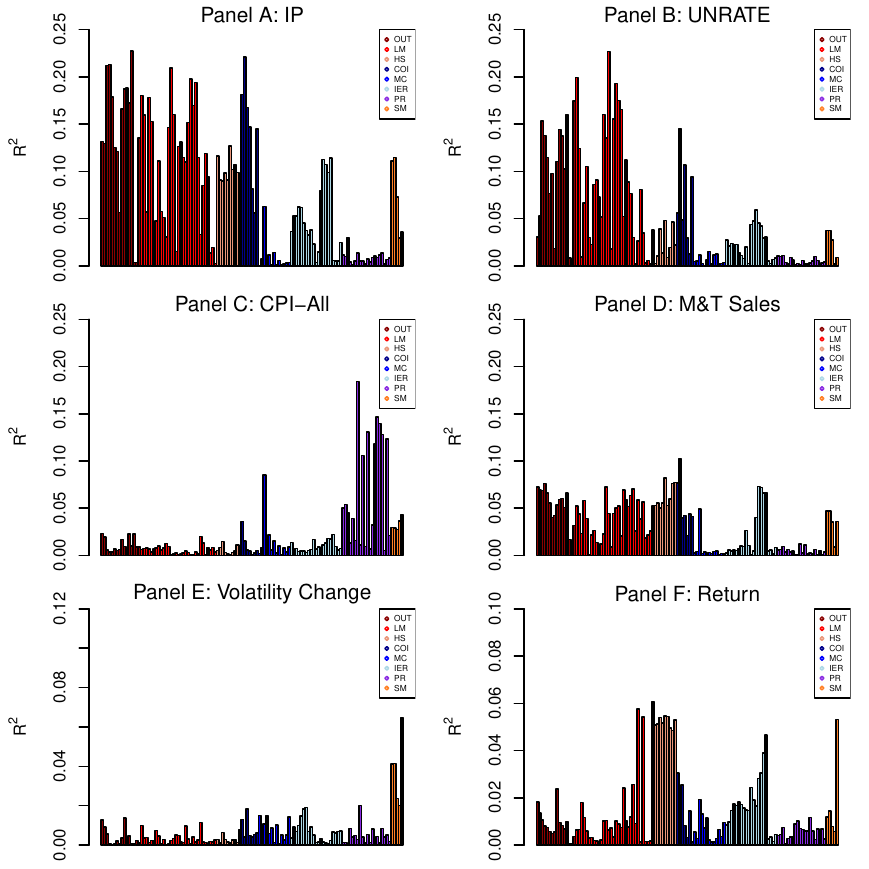}}
\end{center}
\end{figure}

To begin, we study the predictive power of each individual predictor 
and its lagged values on the variables of interest. In Figures~\ref{fig1-a} and \ref{fig2-a}, we plot the in-sample $R^2$s of predicting the 1-month ahead IP growth, change in UNRATE, change in M\&T Sales, change in S\&P 500 index volatility, and the S\&P 500 index return by each of the 123 macro variables, respectively, where $q=2$ is used in Figure~\ref{fig1-a} and $q=3$ in Figure~\ref{fig2-a}. Panels A and B indicate that the Labor Market (LM) conditions have the highest predictive power for future IP growth and unemployment rate, which is similarly found in \cite{huang2022scaled}. In addition, the Output and income, Consumption, orders, and inventories, and 
 the Interest and exchange rates also have higher predictive power than the remaining groups of variables.
From Panel C, we see that the Prices have the highest predictive power for CPI-all, which is reasonable as the prices are directly related to inflation. Panel D indicates that the Out, LM, HS, COI, and IER have comparable predictive power for M\&T sales, while the remaining variables do not have significant predictive power. For the Stock Market predictions in Panes E and F, we see that the Stock market prices  have more predictive power for the volatility change, and the HS and LM conditions also have higher predictive power for the S\&P 500 index returns. Furthermore, the plots in Figure~\ref{fig2-a} suggest that the in-sample $R^2$ may be increased when we add more lagged variables in the linear forecasting. This is a common phenomenon in the autoregression context provided that the number of lagged values used is not too large.  Overall,
Figures~\ref{fig1-a} and \ref{fig2-a} indicate that each predictor has different forecasting ability and their weights should be carefully assigned when extracting factors.


\begin{figure}
\begin{center}
\caption{Bar charts of in-sample $R^2$s in predicting 1-month ahead industrial production growth (Panel A), unemployment rate (Panel B), consumer price index: all (Panel C), Real Manufacturing and Trade Industries Sales (Panel D), S\&P 500 index Volatility Change (Panel E), and the S\&P 500 index return (Panel F) by each of the 123 macro variables from the FRED-MD data set of \cite{mccracken2016fred}, consisting of eight groups including the output and income (OUT), Labor market(LM), Housing (HS), Consumption, orders, and inventories (COI), Money and credit (MC), Interest and exchange rates (IER), Prices (PR), and Stock market (SM). We set $q=5$ in each regression and use AIC to select the model in Step 1. Macro variables are collected at a monthly frequency and the sample period is 1962:07--2019:12.}\label{fig3-a0}
{\includegraphics[width=0.9\textwidth]{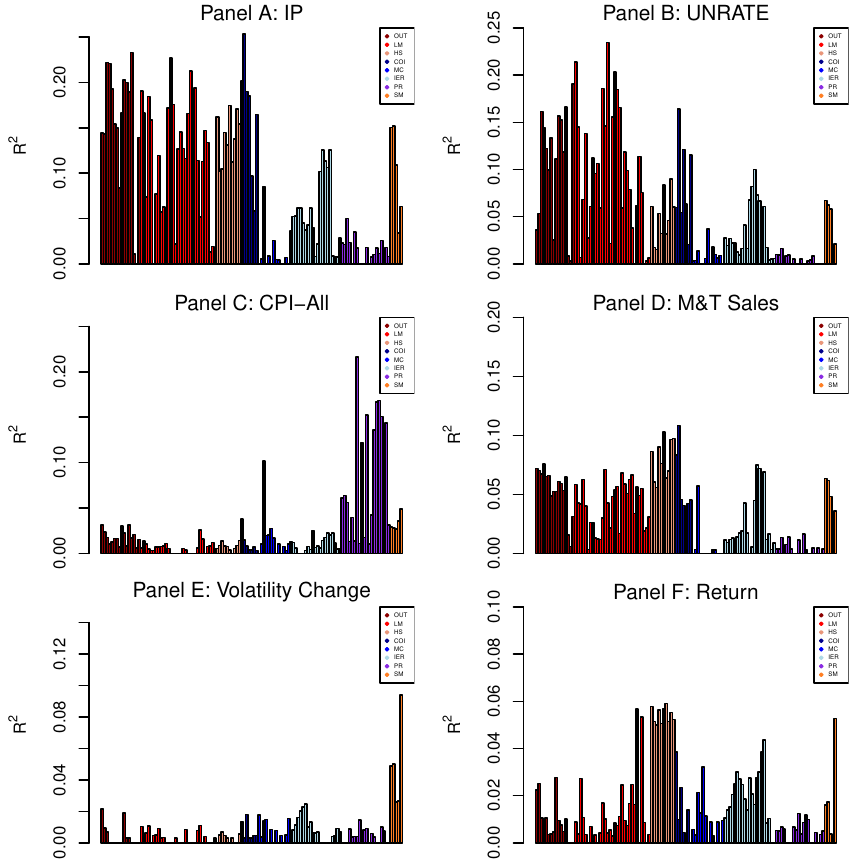}}
\end{center}
\end{figure}

Furthermore, we also use AIC to select the linear models used in Step 1 and plot the in-sample $R^2$s of predicting the 1-month ahead IP growth, change in UNRATE, change in M\&T Sales, change in S\&P 500 index volatility, and the S\&P 500 index return by each of the 123 macro variables, respectively,  in Figure~\ref{fig3-a0}, where the maximal order is set to be $q=5$. From Figure~\ref{fig3-a0}, we see that the 
overall pattern of the $R^2$'s in each plot is similar to its counterparts in Figures~\ref{fig1-a} and \ref{fig2-a}. There are some minor differences between the $R^2$ explained by the models with fixed lags and those selected by AIC. For example, the Consumption, orders, and inventories (COI) related variables have more predictive power for IP than those in Figures~\ref{fig1-a}--\ref{fig2-a}, and the Prices (PR) related variables produce slightly higher predictive power for the consumer price index: all (CPI-All). Nevertheless, the findings in Figures~\ref{fig1-a}--\ref{fig2-a} remain valid in terms of variables that are related to the target variables according to their predictive power.


Next, we consider the in-sample data analysis. First, we standardize each macroeconomic variable and calculate the eigenvalues of the resulting covariance of the 123 variables. That is, we perform eigen-value decomposition of the sample correlation matrix of the predictors. Based on the resulting eigenvalues, the first PCA factor explains about 18\% of the total variation. When we apply the proposed sdPCA method to the data, the first sdPCA factor explains 21\% to 52\% of the total variation depending on the target variable of interest, which is higher than that explained by the first PCA factor. This suggests that the supervised 
PCA  may improve the predictive power of the available predictors.

\begin{figure}
\begin{center}
\caption{Bar charts of loadings on the 123 macro variables of the first six PCA factors. The macro variables are collected at a monthly frequency from the FRED-MD data set of \cite{mccracken2016fred}, consisting of eight groups including the output and income (OUT), Labor market(LM), Housing (HS), Consumption, orders, and inventories (COI), Money and credit (MC), Interest and exchange rates (IER), Prices (PR), and Stock market (SM).  The sample period is 1962:07--2019:12.}\label{fig3-a}
{\includegraphics[width=0.9\textwidth]{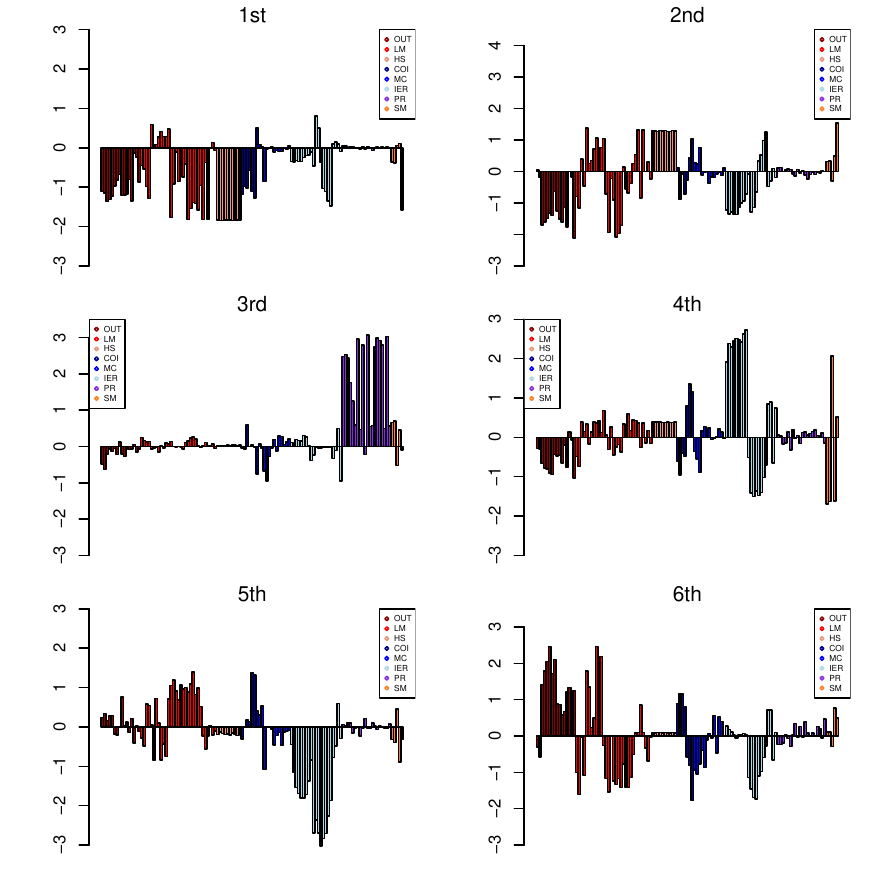}}
\end{center}
\end{figure}

 We plot the loadings of the first to the sixth factors using the traditional PCA method in Figure~\ref{fig3-a}, where, for ease of reading, each loading vector is obtained by multiplying the corresponding eigenvector by 10. From the plot, 
 we see that the first and the second PCA factors are more related to the real economic conditions and they have heavier loads on output and income, labor, and housing variables, followed by the interest and exchange rates. The third PCA factor depends mainly on price-related variables. The fourth factor has heavier loads on the interest rates and stock market conditions. The fifth factor 
has larger loadings on the interest rates while the sixth factor shows similar loading grouping as those of the first two factors. 

\begin{figure}
\begin{center}
\caption{Bar charts of loadings on the 123 macro variables of the first six factors of the proposed sdPCA in predicting the 1-month ahead industrial production growth (OP) and the unemployment rate (UNRATE). The macro variables are collected at a monthly frequency from the FRED-MD data set of \cite{mccracken2016fred}, consisting of eight groups including the output and income (OUT), Labor market(LM), Housing (HS), Consumption, orders, and inventories (COI), Money and credit (MC), Interest and exchange rates (IER), Prices (PR), and Stock market (SM).  The sample period is 1962:07--2019:12.}\label{fig4-a}
{\includegraphics[width=8cm,height=14cm]{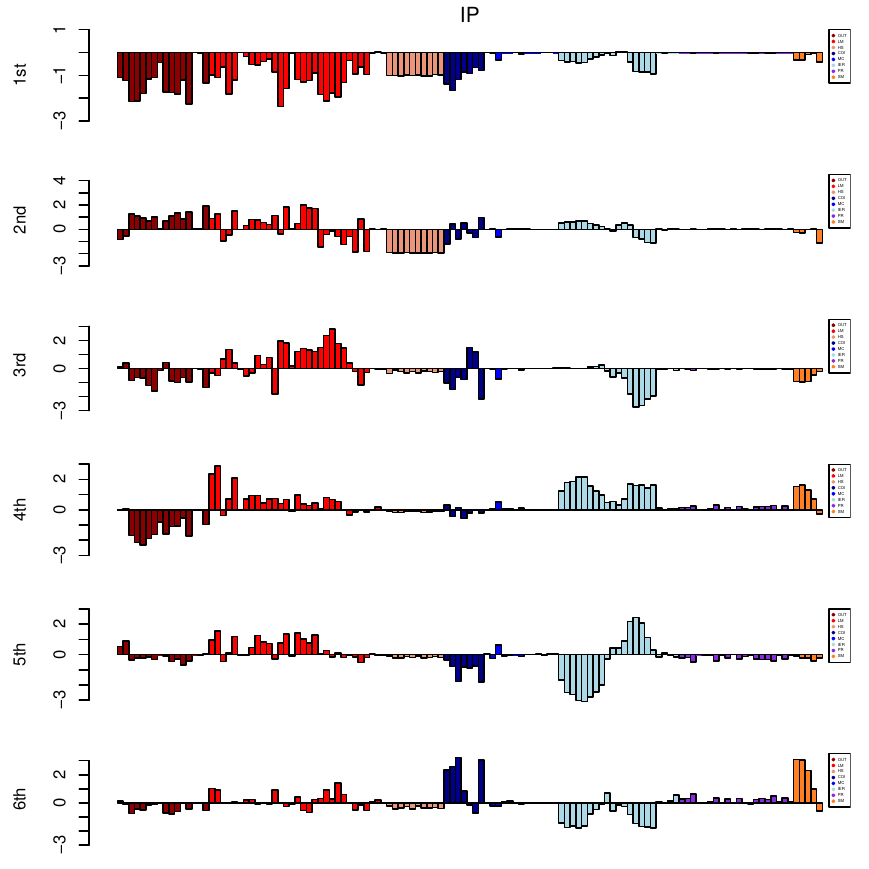}}
{\includegraphics[width=8cm,height=14cm]{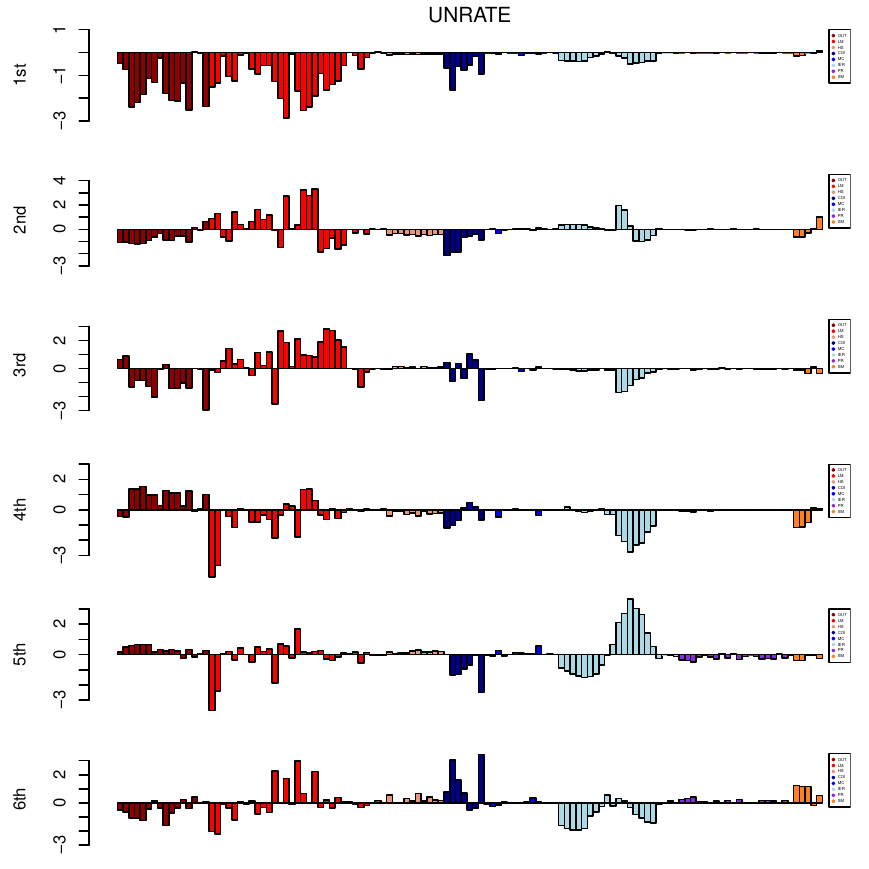}}
\end{center}
\end{figure}

\begin{figure}
\begin{center}
\caption{Bar charts of loadings on the 123 macro variables of the 
first six factors of the proposed sdPCA in predicting the 1-month ahead consumer price index: all (CPI-All) and Real Manufacturing and Trade Industries Sales (M\&T Sales). The macro variables are collected at a monthly frequency from the FRED-MD data set of \cite{mccracken2016fred}, consisting of eight groups including the output and income (OUT), Labor market(LM), Housing (HS), Consumption, orders, and inventories (COI), Money and credit (MC), Interest and exchange rates (IER), Prices (PR), and Stock market (SM).  The sample period is 1962:07--2019:12.}\label{fig5-a}
{\includegraphics[width=8cm,height=14cm]{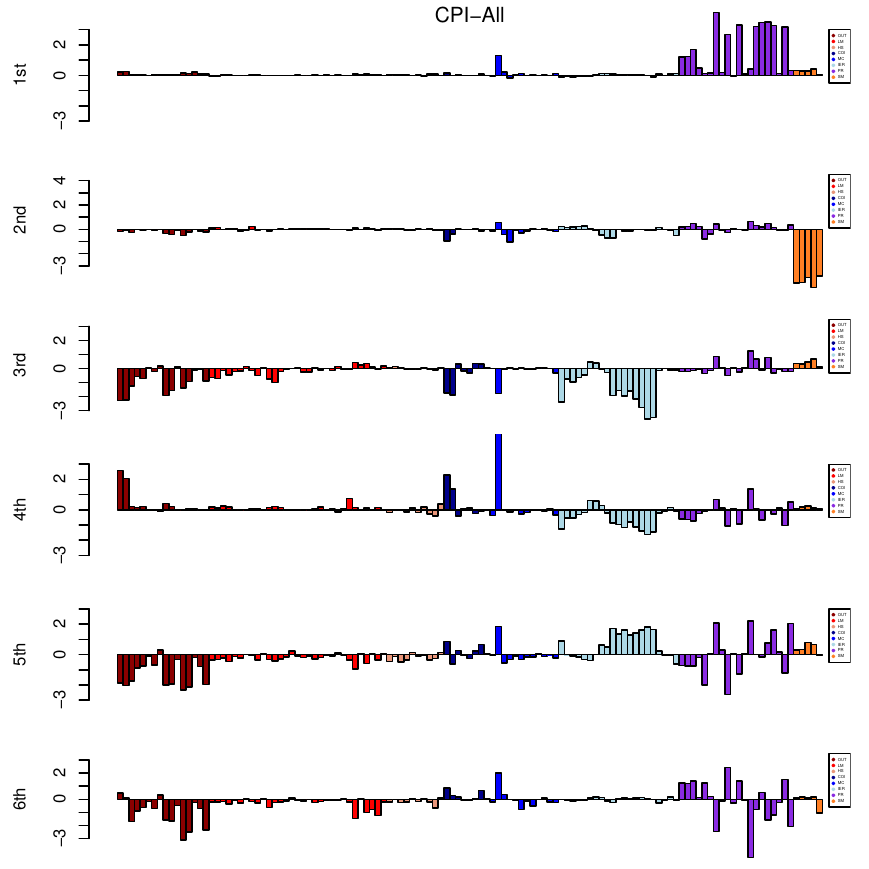}}
{\includegraphics[width=8cm,height=14cm]{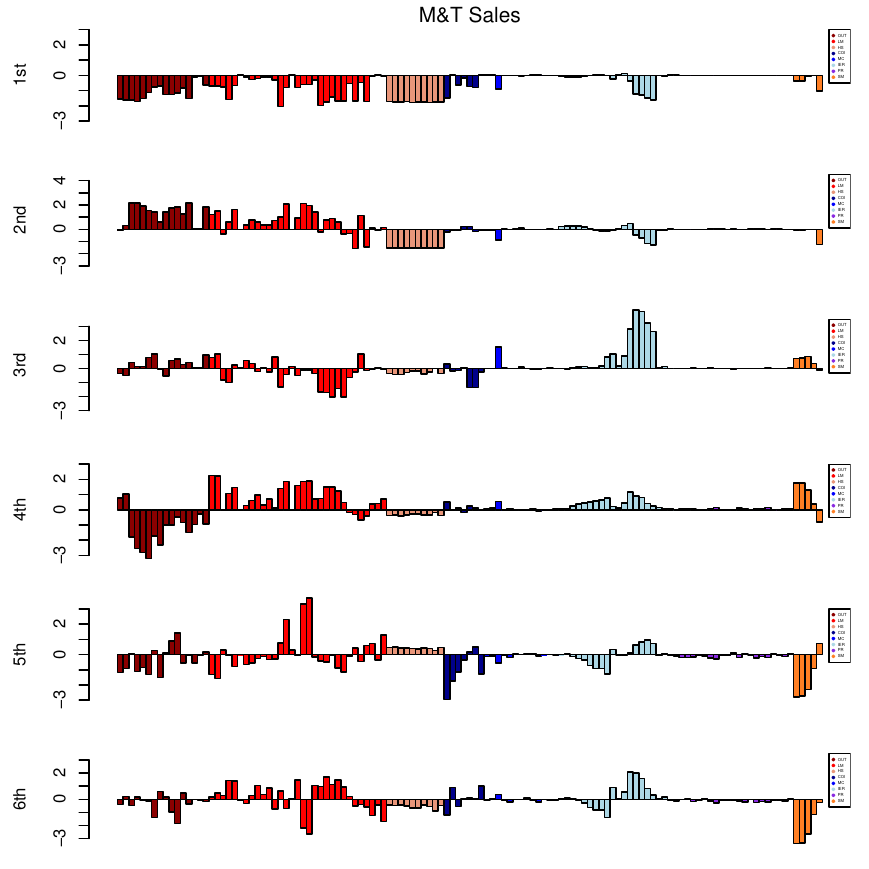}}
\end{center}
\end{figure}

\begin{figure}
\begin{center}
\caption{Bar charts of the loadings on the 123 macro variables of the 
first six sdPCA factors in predicting the 1-month ahead S\&P 500 index Volatility Change (Volatility Change), and the S\&P 500 index return (Return). The macro variables are collected at a monthly frequency from the FRED-MD data set of \cite{mccracken2016fred}, consisting of eight groups including the output and income (OUT), Labor market(LM), Housing (HS), Consumption, orders, and inventories (COI), Money and credit (MC), Interest and exchange rates (IER), Prices (PR), and Stock market (SM).  The sample period is 1962:07--2019:12.}\label{fig6-a}
{\includegraphics[width=8cm,height=14cm]{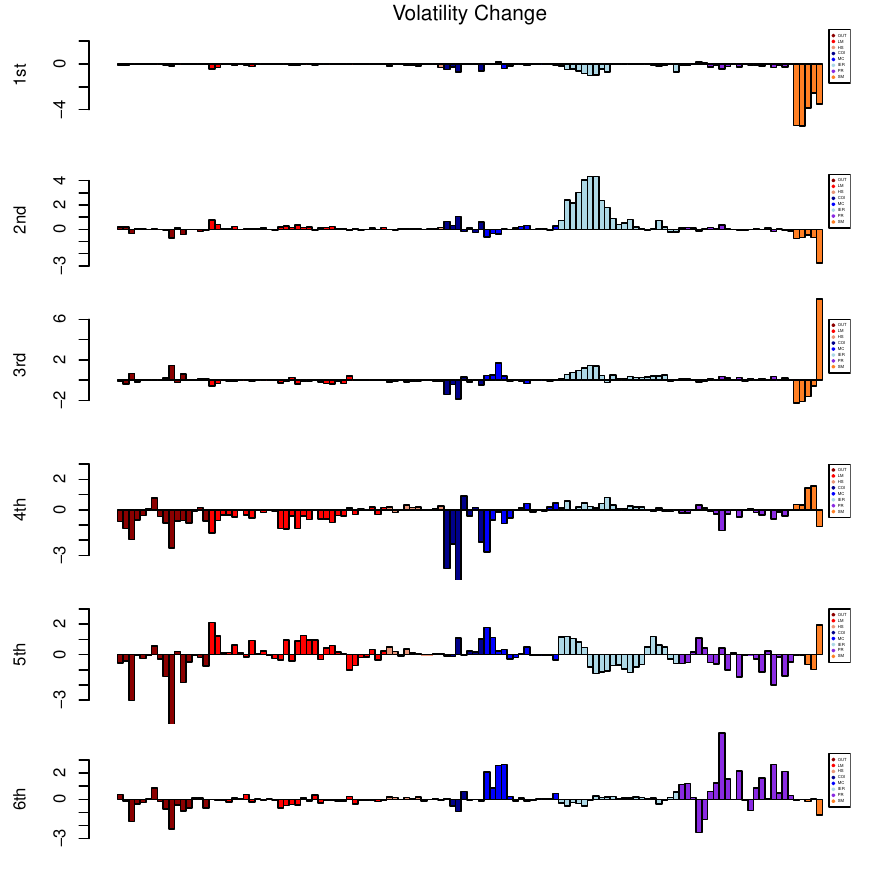}}
{\includegraphics[width=8cm,height=14cm]{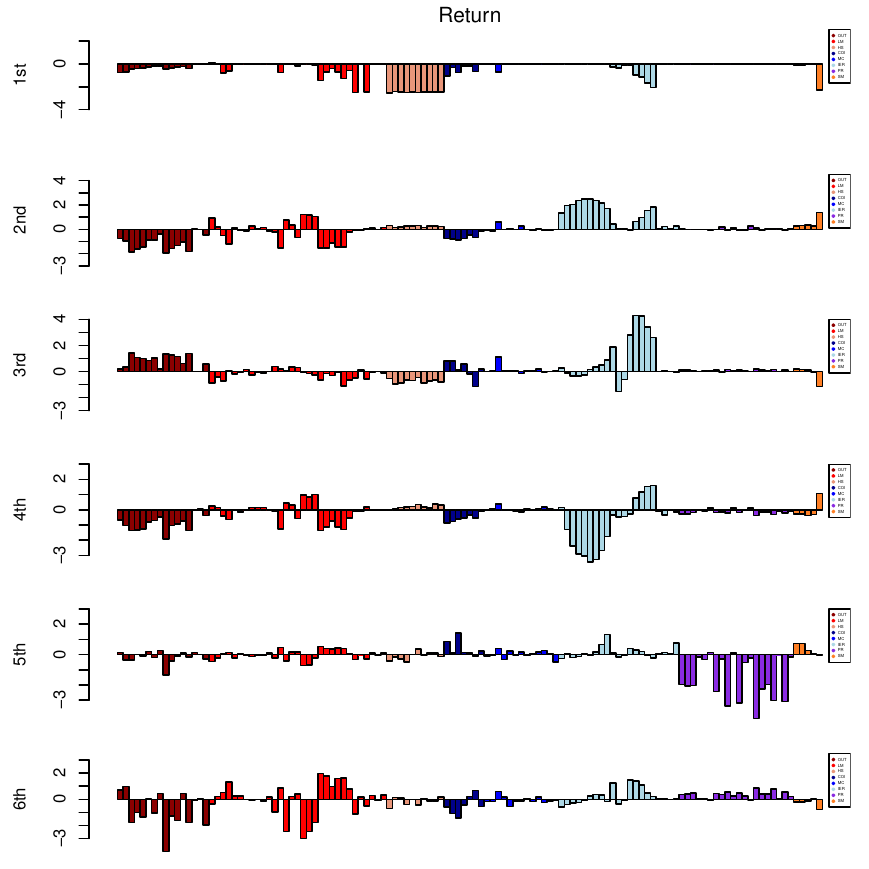}}
\end{center}
\end{figure}

For comparison purposes, we also show the first six loadings 
of the proposed sdPCA in predicting the 1-month ahead IP growth, change of UNRATE, CPI-All, M\&T Sales, Volatility Change, and the Return in Figures~\ref{fig4-a}$-$\ref{fig6-a}. From these plots, we see that the loadings are rather different from those in Figure~\ref{fig3-a}. For example, from Figure~\ref{fig4-a}, we see that the first factor in predicting IP growth has loads mainly on OUT, LM, and COI conditions while the effect of the HS seen in the first unsupervised PCA factor has decreased. Similar results are also found for UNRATE where the first sdPCA factor mainly depends on the OUT, LM, and COI. From Figure~\ref{fig5-a}, the first factor in predicting CPI-All is related mainly to the price variables resulting in 
certain adjustments to those in the first unsupervised PCA factor. The first factor in predicting M\&T sales depends relatively heavier on the OUT, LM, and HS than on the others. For the stock market data predictions, we find from Figure~\ref{fig6-a} that the first sdPCA factor in predicting the S\&P volatility change has heavier loads on the SM variables, and this is understandable as they are more directly related. On the other hand, we also find that the first sdPCA factor in predicting the Return is  more closely related to the HS conditions, implying that the stock returns and the Housing conditions are related, which, in turn, shows that investors may switch investment between the stock market and the real estate market.

\subsection{Out-of-Sample Forecasting}
  In this subsection, we assess the performance of the proposed method using out-of-sample forecasting experiments. For each target variable of interest, we split the sample into two subsamples, where the first one consists of the first 80\% of the data for modeling and the second subsample of the remaining 20\% of the data for out-of-sample prediction. We also adopt the rolling-window scheme as that in the simulation studies, that is, we train the factors and the forecasting coefficients using the first subsample to predict the next target data point. Then we repeat the above procedure after moving the next available observation of predictors and target variable from the second subsample to the first one 
  to obtain the next prediction. This rolling-window scheme is terminated when 
  there is no more observation to compute forecasting error. 

Similarly to the experiments in Section~\ref{sec3}, we compare 
the proposed method (sdPCA) with the traditional PCA (denoted by PCA), the sPCA in \cite{huang2022scaled} (denoted by sPCA), and the diffusion-index forecast in \cite{Stock2002a} (denoted by SW), where the lagged variables of the PCA factors are also included as predictors in the PCA method, and only the contemporaneous factors are used in the linear forecasting of the SW method. The forecasting performance is measured by the root MSFE defined in (\ref{rmsfe}). We use the autoregressive (AR) model with order 1 or 2 as benchmark methods in the comparison. For the forecasting of IP, UNRATE, CPI-All, and M\&T Sales, we consider $h=1,2,3,4,$ and $5$-steps ahead predictions, and we only consider $h=1$ step ahead prediction for the financial data of volatility change and the stock market return, which are of major interest in most  financial market predictions.

Tables~\ref{Table-a5}--\ref{Table-a8} report the results of $h=1,2,3,4$, and $5$-step ahead predictions of the IP, UNRATE, CPI-All, and M\&T Sales, respectively. In the comparison, please note that Model (\ref{GT-DI}) implies that the number of factors used in sdPCA is $qs$ if the number of contemporaneous factors used in PCA, sPCA, and the SW is $s$ for an integer $s>0$, where $q$ is the number of lagged variables used in Steps 1 to 3 of the proposed procedure. For the proposed sdPCA, $q=2$ and $q=3$ are employed in the empirical studies, indicating the number of lagged variables used in the forecasting. The number of factors used ranges from 1 to 3 for the methods of PCA, sPCA, and the SW, and therefore, the number of factors used in sdPCA ranges from 1 to 6 when $q=2$ and from 1 to 9 if $q=3$ according to the above discussion.   For each $h$, the smallest value is marked in boldface. The Lasso procedure is considered with the corresponding errors given in the parentheses if sdPCA does not beat other methods. We only report the results of the PCA method when $q=2$ because those with $q=3$ do not show any clear improvement in most cases due to the possibility of overfitting. On the other hand, our analysis suggests that the forecasts with more than 3 factors in PCA, sPCA, and the SW do not necessarily improve the prediction accuracy, and therefore, we only compare the results when the number of factors used ranges from 1 to 3 for the methods of PCA, sPCA, and the SW, and the corresponding number of factors used in sdPCA ranges from 1 to 6 when $q=2$ and from 1 to 9 if $q=3$.

From Table~\ref{Table-a5}, we see that the smallest prediction error is achieved by the proposed sdPCA for each $h$, and the prediction using more lagged variables ($q=3$) can improve the forecasting ability ($q=2$) in most cases. Furthermore, the PCA method tends to produce the second smallest errors since it also includes the lagged variables as predictors, which implies that the lagged variables can be used to improve the forecasting performance of IP. In addition, all the methods outperform the benchmark AR methods.

For the predictions of the UNRATE in Table~\ref{Table-a6}, we note that the sdPCA outperforms other competing methods for short-term predictions, such as the cases when $h=1,2$ and $3$, without using the Lasso approach. For $h=4$ and $5$, the errors accumulated by sdPCA are increasing, but the Lasso approach can significantly reduce the forecasting errors and a simple Lasso procedure can produce even smaller errors than all the other methods, implying that the penalized method is an effective way in selecting the factors that have more predictive power. Similar results are also found for the predictions of CPI in Table~\ref{Table-a7}, where the proposed sdPCA as well as the PCA with lagged factors can outperform other methods in short-term forecasting ($h=1$ and $2$), and the forecasting performance for long-term ahead predictions are not more accurate than those by simple AR approaches. In addition, the Lasso approach can also improve the forecasting performance as shown in the case of using sdPCA with $q=3$. Similar findings are also obtained in predicting the M\&T sales in Table~\ref{Table-a8}, and we omit the details. 

From Tables~\ref{Table-a5}--\ref{Table-a8}, we see that the proposed sdPCA might produce increased errors for $h$ = 4 and 5, especially when the number of factors used increases. There are two possible explanations. First, for 
stationary time series, such as those considered in our example, the serial dependence decays exponentially so that as $h$ increases the information 
of the target variable embedded in the predictors decreases. The forecast errors 
may increase as $h$ increases with uncertainty approaching the unconditional variance of the target variable. Second, increasing the number of factors used also increases 
the possibility of overfitting, which may lead to inferior prediction. 
 On the other hand, the Lasso procedure can significantly improve the forecasts, which confirms the overfitting issue when more factors are used without  variable selection. This highlights the importance of 
 using the Lasso procedure to select the relevant factors if we do not know how many factors to include in linear forecasting.

Finally, the one-step-ahead predictions of the stock return and the volatility change are shown in Table~\ref{Table-a9}. From Table~\ref{Table-a9}, we see that the performance of the sdPCA is comparable with that of SW in forecasting the Stock return, and the sPCA method cannot beat the benchmark methods overall. For the predictions of the volatility change, we find that the proposed sdPCA outperforms all the other competing methods. The results in Table~\ref{Table-a9} suggest that the proposed sdPCA could be helpful in forecasting financial data.

\begin{table}[htbp]
\caption{Out-of-sample  root
mean squared forecast errors in predicting IP. Four comparison methods are used: the  proposed method (sdPCA), the one using the factors and their lagged variables  extracted by traditional PCA (denoted by PCA), the scaled PCA by \cite{huang2022scaled} (denoted sPCA), and the original diffusion-index method in \cite{Stock2002b} using the factors extracted by PCA without lagged variables (denoted by SW).  We also use AR(1) and AR(2) models as benchmark methods in out-of-sample forecasting. For the proposed sdPCA, $q=2$ and $q=3$ are considered. For each $h$, the smallest MSFE is marked in boldface. 
} 
          \label{Table-a5}
{\begin{center}
\begin{tabular}{ccccccccccc}
\toprule
&\multicolumn{6}{c}{sdPCA ($q=2$)}&&\multicolumn{3}{c}{PCA ($q=2$)}\\
\cline{2-7}\cline{9-11}
$h$&1 ft&2 fts &3 fts&4 fts&5 fts&6 fts&&1 ft&2 fts&3 fts\\
1&0.080&0.078&0.079&0.080&0.079&0.076&&0.085&0.078&0.078\\
2&0.081&0.079&0.080&\boxed{\bf 0.075}&0.076&0.079&&0.086&0.082&0.081\\
3&0.085&0.082&0.084&0.082&0.083&0.083&&0.086&0.084&0.083\\
4&0.088&\boxed{\bf 0.085}&0.088&0.086&0.093&0.127&&0.086&0.086&0.088\\
5&0.090&0.087&0.089&0.087&0.089&0.228&&0.088&0.088&0.089\\
\toprule
&\multicolumn{3}{c}{sPCA}&&\multicolumn{3}{c}{SW}&&\multicolumn{2}{c}{AR}\\
\cline{2-4}\cline{6-8}\cline{10-11}
&1 ft&2 fts& 3 fts&&1 ft& 2fts& 3fts&&AR(1)&AR(2)\\
1&0.083&0.080&0.081&&0.085&0.079&0.080&&0.085&0.084\\
2&0.086&0.082&0.084&&0.086&0.082&0.082&&0.089&0.085\\
3&0.088&0.084&0.086&&0.086&0.084&0.083&&0.090&0.085\\
4&0.089&0.086&0.088&&0.086&0.086&0.086&&0.091&0.088\\
5&0.091&0.094&0.091&&0.087&0.089&0.089&&0.091&0.090\\
\toprule
&\multicolumn{9}{c}{sdPCA ($q=3$)}&\\
\cline{2-10}
&1 ft&2 fts &3 fts&4 fts&5 fts&6 fts&7 fts& 8fts&9fts&\\
1&0.078&0.077&0.077&0.074&0.075&\boxed{\bf 0.073}&0.074&0.074&0.074\\
2&0.080&0.079&0.079&0.076&0.077&0.077&0.079&0.078&0.078\\
3&0.084&0.082&0.084&\boxed{\bf 0.080}&0.082&0.083&0.083&0.081&0.081\\
4&0.086&\boxed{\bf 0.085}&0.087&\boxed{\bf 0.085}&0.089&0.232&0.181&0.182&0.134\\
5&0.089&0.088&0.090&\boxed{\bf 0.086}&0.089&0.096&0.110&0.109&0.120\\
\bottomrule
\end{tabular}
  \end{center}}
\end{table}

\begin{table}[htbp]
\caption{Out-of-sample  root
mean squared forecast errors in predicting UNRATE. Four comparison methods are used: our proposed method (sdPCA), the one using the factors and their lagged variables extracted by PCA (denoted by PCA), the scaled PCA by \cite{huang2022scaled} (denoted sPCA), and the original diffusion-index model in \cite{Stock2002b} using the factors extracted by PCA without lagged variables (denoted by SW).  We also use AR(1) and AR(2) models as benchmark methods in out-of-sample forecasting. For the proposed sdPCA, $q=2$ and $q=3$ are considered. For each $h$, the smallest MSFE is marked in boldface. The Lasso procedure is applied and the corresponding errors are placed in parentheses if  sdPCA does not outperform the  other methods.} 
          \label{Table-a6}
{\begin{center}

\begin{tabular}{ccccccccccc}
\toprule
&\multicolumn{6}{c}{sdPCA ($q=2$)}&&\multicolumn{3}{c}{PCA ($q=2$)}\\
\cline{2-7}\cline{9-11}
$h$&1 ft&2 fts &3 fts&4 fts&5 fts&6 fts&&1 ft&2 fts&3 fts\\
1&1.713&0.694&1.703&1.714&1.682&1.627&&1.976&1.773&1.767\\
2&1.772&1.764&1.795&1.781&1.757&\boxed{\bf 1.680}&&2.004&1.831&1.832\\
3&1.821&1.897&1.900&1.834&1.867&1.852&&2.009&1.850&1.893\\
4&8.769&3.082&3.657&4.013&4.887&4.783&&2.024&1.890&1.939\\
&&&&&&(1.914)\\
5&2.115&4.370&4.730&4.850&4.903&4.774&&2.029&1.931&1.951\\
&&&&&&(\boxed{\bf 1.924})\\
\toprule
&\multicolumn{3}{c}{sPCA}&&\multicolumn{3}{c}{SW}&&\multicolumn{2}{c}{AR}\\
\cline{2-4}\cline{6-8}\cline{10-11}
&1 ft&2 fts& 3 fts&&1 ft& 2fts& 3fts&&AR(1)&AR(2)\\
1&1.736&1.734&1.707&&1.973&1.784&1.791&&2.044&1.941\\
2&1.815&1.817&1.833&&1.996&1.859&1.852&&2.080&1.960\\
3&1.859&1.919&1.912&&2.006&1.877&1.897&&2.090&2.027\\
4&1.882&1.951&1.956&&2.014&1.906&1.937&&2.092&2.035\\
5&2.272&4.782&4.752&&2.028&1.939&1.955&&2.093&2.069\\
\toprule
&\multicolumn{9}{c}{sdPCA ($q=3$)}&\\
\cline{2-10}
&1 ft&2 fts &3 fts&4 fts&5 fts&6 fts&7 fts&8 fts&9 fts&\\
1&1.696&1.683&1.698&1.674&{1.600}&\boxed{\bf 1.594}&1.594&1.595&1.594\\
2&1.755&1.804&1.808&1.756&1.733&1.696&1.708&1.711&1.704\\
3&\boxed{\bf 1.803}&2.380&2.108&2.515&2.579&2.590&2.574&2.572&2.629\\
4&10.011&2.604&3.936&4.207&4.942&4.986&4.798&4.849&4.868\\
&&&&&&(\boxed{\bf 1.881})\\
5&8.353&4.835&5.299&5.286&5.082&4.976&4.991&5.032&5.129\\
\bottomrule
\end{tabular}
  \end{center}}
\end{table}

\begin{table}[htbp]
\caption{The out-sample  root
mean squared forecast errors for CPI-ALL. Four comparison methods are used: our proposed method (sdPCA), the one using the factors and their lagged variables extracted by PCA (denoted by PCA), the scales PCA by \cite{huang2022scaled} (denoted sPCA), and the original diffusion-index method in \cite{Stock2002b} using the factors extracted by PCA without lagged variables (denoted by SW).  We also use AR(1) and AR(2) models as benchmark methods to do out-of-sample forecasting. For the proposed sdPCA, $q=2$ and $q=3$ are considered. For each $h$, the smallest value is marked in boldface. The Lasso procedure is considered with the corresponding errors given in parentheses if sdPCA does not beat other methods.} 
          \label{Table-a7}
{\begin{center}

\begin{tabular}{ccccccccccc}
\toprule
&\multicolumn{6}{c}{sdPCA ($q=2$)}&&\multicolumn{3}{c}{PCA ($q=2$)}\\
\cline{2-7}\cline{9-11}
$h$&1 ft&2 fts &3 fts&4 fts&5 fts&6 fts&&1 ft&2 fts&3 fts\\
1&0.034&0.032&0.032&0.032&\boxed{\bf 0.031}&0.032&&0.034&0.034&0.034\\
2&\boxed{\bf 0.032}&0.033&0.033&0.033&0.033&0.034&&0.034&0.034&\boxed{\bf 0.032}\\
3&\boxed{\bf 0.034}&0.036&0.036&0.036&0.036&0.036&&\boxed{\bf 0.034}&0.034&0.034\\
4&0.035&0.035&0.035&0.035&0.035&0.035&&0.034&0.034&0.034\\
&&&&&&(0.034)\\
5&0.052&0.062&0.064&0.055&0.045&0.044&&\boxed{\bf 0.034}&0.034&0.035\\
&&&&&&(\boxed{\bf 0.034})\\
\toprule
&\multicolumn{3}{c}{sPCA}&&\multicolumn{3}{c}{SW}&&\multicolumn{2}{c}{AR}\\
\cline{2-4}\cline{6-8}\cline{10-11}
&1 ft&2 fts& 3 fts&&1 ft& 2fts& 3fts&&AR(1)&AR(2)\\
1&0.035&0.033&0.033&&0.034&0.034&0.035&&0.035&0.034\\
2&0.033&0.033&0.033&&0.034&0.034&0.033&&0.035&0.034\\
3&\boxed{\bf 0.034}&0.035&0.036&&\boxed{\bf 0.034}&0.034&0.034&&\boxed{\bf 0.034}&0.035\\
4&0.034&0.034&0.034&&0.034&0.034&0.034&&0.034&0.034\\
5&0.052&0.052&0.049&&\boxed{\bf 0.034}&0.034&0.034&&\boxed{\bf 0.034}&0.034\\
\toprule
&\multicolumn{9}{c}{sdPCA ($q=3$)}&\\
\cline{2-10}
&1 ft&2 fts &3 fts&4 fts&5 fts&6 fts&7 fts&8fts&9 fts\\
1&0.033&0.032&0.031&0.032&0.032&0.032&0.033&0.032&0.032\\
2&0.032&0.033&0.033&0.033&0.034&0.034&0.035&0.035&0.035\\
3&0.035&0.034&0.036&0.036&0.036&0.034&0.038&0.039&0.039\\
4&0.035&0.034&0.036&0.036&0.036&0.036&0.037&0.037&0.037\\
&&&&&&(\boxed{\bf 0.032})\\
5&0.060&0.072&0.074&0.058&0.056&0.055&0.052&0.051&0.051\\
\bottomrule
\end{tabular}
  \end{center}}
\end{table}

\begin{table}[htbp]
\caption{The out-sample  root
mean squared forecast errors for Real Manu. and Trade Industries Sales. Four comparison methods are used: our proposed method (sdPCA), the one using the factors and their lagged variables extracted by PCA (denoted by PCA), the scales PCA by \cite{huang2022scaled} (denoted sPCA), and the original diffusion-index method in \cite{Stock2002b} using the factors extracted by PCA without lagged variables  (denoted by SW).  We also use AR(1) and AR(2) models as benchmark methods to do out-of-sample forecasting. For the proposed sdPCA, $q=2$ and $q=3$ are considered. For each $h$, the smallest value is marked in boldface. The Lasso procedure is considered with the corresponding errors given in parentheses if sdPCA does not beat  other methods.} 
          \label{Table-a8}
{\begin{center}

\begin{tabular}{ccccccccccc}
\toprule
&\multicolumn{6}{c}{sdPCA ($q=2$)}&&\multicolumn{3}{c}{PCA ($q=2$)}\\
\cline{2-7}\cline{9-11}
$h$&1 ft&2 fts &3 fts&4 fts&5 fts&6 fts&&1 ft&2 fts&3 fts\\
1&0.081&\boxed{\bf 0.078}&0.081&0.082&0.081&0.082&&0.081&\boxed{\bf 0.078}&0.080\\
2&0.081&\boxed{\bf 0.079}&0.082&0.080&0.080&0.080&&0.081&0.081&0.083\\
3&0.084&\boxed{\bf 0.081}&0.085&0.083&0.083&0.084&&0.082&0.082&0.083\\
4&0.086&0.277&0.263&0.291&0.300&0.280&&\boxed{\bf 0.084}&0.084&0.084\\
&&&&&&(\boxed{\bf 0.084})\\
5&0.087&0.286&0.255&0.289&0.284&0.295&&\boxed{\bf 0.084}&0.085&0.085\\
\toprule
&\multicolumn{3}{c}{sPCA}&&\multicolumn{3}{c}{SW}&&\multicolumn{2}{c}{AR}\\
\cline{2-4}\cline{6-8}\cline{10-11}
&1 ft&2 fts& 3 fts&&1 ft& 2fts& 3fts&&AR(1)&AR(2)\\
1&0.084&0.081&0.084&&0.082&0.081&0.081&&0.090&0.090\\
2&0.084&0.080&0.083&&0.081&0.081&0.082&&0.086&0.086\\
3&0.085&0.083&0.084&&0.082&0.083&0.084&&0.087&0.087\\
4&0.087&0.090&0.088&&0.084&0.085&0.085&&0.087&0.087\\
5&0.087&0.212&0.235&&\boxed{\bf 0.084}&0.085&0.086&&0.087&0.087\\
\toprule
&\multicolumn{9}{c}{sdPCA ($q=3$)}&\\
\cline{2-10}
&1 ft&2 fts &3 fts&4 fts&5 fts&6 fts&7 fts&8 fts&9 fts\\
1&0.080&0.078&0.080&0.080&0.080&0.081&0.085&0.088&0.086\\
2&0.081&0.080&0.083&0.080&0.080&0.080&0.083&0.083&0.083\\
3&0.083&0.250&0.372&0.281&0.268&0.265&0.277&0.257&0.258\\
4&0.085&0.285&0.286&0.334&0.337&0.316&0.328&0.332&0.334\\
5&0.086&0.356&0.304&0.322&0.313&0.317&0.309&0.314&0.311\\
&&&&&&(\boxed{\bf 0.084})\\
\bottomrule
\end{tabular}
  \end{center}}
\end{table}

\begin{table}[htbp]
\caption{The out-sample  root
mean squared forecast errors for S\&P 500 index returns and change of S\&P 500 index volatility. Four comparison methods are used: our proposed method (sdPCA), the one using the factors and their lagged variables extracted by PCA (denoted by PCA), the scales PCA by \cite{huang2022scaled} (denoted sPCA), and the original diffusion-index method in \cite{Stock2002b} using the factors extracted by PCA without lagged variables (denoted by SW).  We also use AR(1) and AR(2) models as benchmark methods to do out-of-sample forecasting. For the proposed sdPCA, $q=2$ and $q=3$ are considered. For each $h$, the smallest value is marked in boldface. The Lasso procedure is considered with the corresponding errors are placed in parentheses if sdPCA does not beat  other methods.} 
          \label{Table-a9}
{\begin{center}

\begin{tabular}{ccccccccccc}
\toprule
\multicolumn{11}{c}{Stock return with dividends}\\
\hline
&\multicolumn{6}{c}{sdPCA ($q=2$)}&&\multicolumn{3}{c}{PCA ($q=2$)}\\
\cline{2-7}\cline{9-11}
$h$&1 ft&2 fts &3 fts&4 fts&5 fts&6 fts&&1 ft&2 fts&3 fts\\
1&0.495&0.505&0.514&0.506&0.534&0.517&&0.495&0.503&0.510\\
\toprule
&\multicolumn{3}{c}{sPCA}&&\multicolumn{3}{c}{SW}&&\multicolumn{2}{c}{AR}\\
\cline{2-4}\cline{6-8}\cline{10-11}
&1 ft&2 fts& 3 fts&&1 ft& 2fts& 3fts&&AR(1)&AR(2)\\
1&0.498&0.506&0.511&&\boxed{\bf 0.494}&0.504&0.506&&0.498&0.499\\
\toprule
&\multicolumn{9}{c}{sdPCA ($q=3$)}&\\
\cline{2-10}
&1 ft&2 fts &3 fts&4 fts&5 fts&6 fts&7 fts&8 fts&9 fts\\
1&\boxed{\bf 0.494}&0.506&0.508&0.510&0.518&0.525&0.510&0.515&0.522\\
\bottomrule
\toprule
\multicolumn{11}{c}{Change of Volatility}\\
\hline
&\multicolumn{6}{c}{sdPCA ($q=2$)}&&\multicolumn{3}{c}{PCA ($q=2$)}\\
\cline{2-7}\cline{9-11}
$h$&1 ft&2 fts &3 fts&4 fts&5 fts&6 fts&&1 ft&2 fts&3 fts\\
1&\boxed{\bf 0.270}&0.272&0.270&0.279&0.278&0.271&&0.279&0.276&0.278\\
\toprule
&\multicolumn{3}{c}{sPCA}&&\multicolumn{3}{c}{SW}&&\multicolumn{2}{c}{AR}\\
\cline{2-4}\cline{6-8}\cline{10-11}
&1 ft&2 fts& 3 fts&&1 ft& 2fts& 3fts&&AR(1)&AR(2)\\
1&0.273&0.271&0.278&&0.280&0.280&0.280&&0.284&0.286\\
\toprule
&\multicolumn{9}{c}{sdPCA ($q=3$)}&\\
\cline{2-10}
&1 ft&2 fts &3 fts&4 fts&5 fts&6 fts&7 fts&8 fts&9 fts\\
1&0.270&0.280&0.283&0.277&0.275&0.274&0.274&0.277&0.280\\
\bottomrule
\end{tabular}
  \end{center}}
\end{table}

\section{Conclusion} \label{sec5}
This paper introduced a new supervised dynamic PCA method for linear forecasting with many predictors, which is commonly seen in the big-data environment. The new supervised PCA provides an effective way to bridge the gap between predictors and the targeted variables of interest by scaling and combining information 
of the predictors and their lagged variables, which is in line with dynamic forecasting. Furthermore, we also proposed to use penalized methods, such as the LASSO approach, to select the significant factors that have more predictive power than the others in the linear forecasting equation.  

To highlight the prediction power achieved by the proposed method,  we showed that our estimators are consistent and outperform theoretically the 
traditional methods 
in prediction under some commonly used conditions. We conducted extensive simulations to verify that the proposed method produces satisfactory forecasting results and outperforms most of the existing methods using the traditional PCA. A real data example on predicting U.S. monthly macroeconomic variables using a large number of predictors shows that our method performs better than most of the existing ones in forecasting the U.S. industrial production (IP) growth, change in the unemployment rate (UNRATE), the consumer price index: all (CPI-All), the S\&P 500 index volatility change (Volatility Change), and the S\&P 500 index return with 123 macro variables from FRED-MD. Finally, the proposed sdPCA together with the Lasso procedure produces even more satisfactory results in many cases. Overall, the proposed procedure provides a comprehensive and effective method for dynamic forecasting when the data dimension is large.

		
	\end{onehalfspacing}

%
	
		\onehalfspacing
\bibliographystyle{econometrica}
\let\oldbibliography\thebibliography
\renewcommand{\thebibliography}[1]{%
  \oldbibliography{#1}%
  \setlength{\itemsep}{3pt}%
}
	{\footnotesize
		\bibliography{reference}
	}
	\onehalfspacing
%

\newpage

	
	\begin{titlepage}
	
	\begin{center}
	{\Huge Internet Appendix for\\Supervised Dynamic PCA: Linear Dynamic Forecasting with Many Predictors}
	\end{center}

	 \thispagestyle{empty}
		\vspace{0.5cm}
		
		\begin{abstract}
			The first part of the Internet Appendix provides detailed proofs for all the theoretical statements in the main text, and the second part presents descriptions of the real data used in the main article.
			\vspace{1cm}
			
			\noindent\textbf{Keywords:} Dynamic Forecasting, Factor Analysis, Supervised Principal Components,  Large-Dimension, LASSO

			\noindent\textbf{JEL classification:} C22, C23, C38, C53
		\end{abstract}
	\end{titlepage}

	\setcounter{page}{1}
	
	\setcounter{section}{0}
	\setcounter{subsection}{0}
	
	\renewcommand{\thesection}{IA.\Alph{section}}
	\renewcommand{\thesubsection}{\thesection.\arabic{subsection}}
	
	\setcounter{equation}{0}
	\renewcommand{\theequation}{\thesection.\arabic{equation}}
	

	
	\renewcommand{\theequation}{IA.\arabic{equation}}%
	\renewcommand{\thefigure}{IA.\arabic{figure}} \setcounter{figure}{0}
	\renewcommand{\thetable}{IA.\Roman{table}} \setcounter{table}{0}


	
	

\section{Proofs of the Theorems}

We need some lemmas first and use $C$ or $c$ to denote a generic constant the value of which may change at different places. The same notations as those of 
the main article are used throughout. 

Let $\bu_t=(u_{1,t},...,u_{N,t})'$, $\bu_i=(u_{i,1},...,u_{i,T})'$, and $\bU$ and $\bF$ be the two matrices consisting $\bu_t'$ and $\bff_t'$ as their corresponding rows. The following lemma will be used either explicitly or implicitly in the proofs of the theorems of the main article.

\begin{lemma}\label{lm0}
    Let Assumptions \ref{asm01}$-$\ref{asm5} hold. For each $t$, (i) $E\|N^{-\nu/2}\sum_i \bb_i u_{i,t}\|^2\le C$, (ii) for each $i$, $\frac{1}{NT} \bu_t'\bU'\bF=O_p(\delta_{NT}^2)$, where $\delta_{NT}=\min\{\sqrt{N},\sqrt{T}\}$, (iii) $E\|T^{-1/2} \sum_t \bff_t u_{i,t}||^2\le C$, (iv)
$\frac{1}{N^\nu T}\bu_i'\bU\bB=O_p(\frac 1 {N^\nu}) +O_p(\frac 1 {\sqrt{T N^\nu}})$;
(v) $\bB' \bU' \bF  = \sum_{i=1}^N\sum_{t=1}^T \bb_i \bff_t' u_{i,t}  =O_p( \sqrt{N^\nu T})$.
\end{lemma}

{\bf Proof.} The results are the same as those in Assumption A3 in \cite{bai2021approximate}. They can be shown via elementary argument under Assumptions 1-6. We omit the details. $\Box$
\vskip 0.2cm
Without loss of generality, we assume the data are centered so that the intercepts will be removed from the model. Then the model can be written as
\begin{equation}
    \left\{\begin{array}{c}
         \bx_t=\bB\bff_t+\bu_t  \\
         y_{t+h}=\bgg_t'\bbeta+\bve_{t+h}, 
    \end{array}\right.
\end{equation}
where $\bgg_t=(\bff_t',\bff_{t-1}',...,\bff_{t-q+1}')'$ and $\bbeta=(\bbeta_0,\bbeta_1,...,\bbeta_{q-1}')'$ for   $q\geq 1$. Let $\bx_{i,t}=(x_{i,t},x_{i,t-1},...,x_{i,t-q+1})'$, without loss of generality, we assume that
\begin{equation}\label{ident}
    \frac{1}{T}\sum_{t=q}^{T-h}\bx_{i,t}\bx_{i,t}'=\bI_{q}\,\,\text{and}\,\, \frac{1}{T}\sum_{t=q}^{T-h}\bgg_t\bgg_t'=\bI_{rq}.
\end{equation}
Letting $\wh\bgamma_i=(\wh \gamma_{i,0},...,\wh\gamma_{i,q-1})'$ and $\bxi_{i,t}=(u_{i,t},...,u_{i,t-q+1})'$, by least-squares estimation and the identification condition in (\ref{ident}), we have

\begin{align}
    \wh\bgamma_i=&\{(\bI_q\otimes\bb_i')\bbeta\}+\left\{(\bI_q\otimes\bb_i')\frac{1}{T}\sum_{t=q}^{T-h}\bgg_t\bve_{t+h}+\frac{1}{T}\sum_{t=q}^{T-h}\bxi_{i,t}\bgg_t'\bbeta+\frac{1}{T}\sum_{t=q}^{T-h}\bxi_{i,t}\bve_{t+h}\right\}\notag\\
    =&\bgamma_i+\bdelta_i.
\end{align}
Roughly speaking, each term in $\bdelta_i$ is of order $O_p(1/\sqrt{T})$. 
Letting  
\[z_{i,t}=\wh\bgamma_i'(\bI_q\otimes\bb_i')\bgg_t+\wh\bgamma_i'\bxi_{i,t},\]
we have the following lemma.
\begin{lemma}\label{lm1}
Let $\bZ_i=(z_{i,q},...,z_{i,T-h})'$ and $\wt \bV$ be the diagonal matrix consisting of the top $qr$ eigenvalues of $\sum_{i=1}^N\bZ_i\bZ_i'$ as its diagonal elements. Under Assumptions 1$-$6, if $N^{1-\nu}/T^2\rightarrow 0$, with probability tending to one, we have
\[\wt\bV\asymp N^{\nu}T.\]
\end{lemma}

{\bf Proof.} Let $\bG=(\bgg_q,...,\bgg_{T-h})'$ and $\bxi_i=(\bxi_{i,q}',...,\bxi_{i,T-h}')'$. It follows that
\[\bZ_i=\bG(\wh\bgamma_i\otimes\bb_i)+(\bI_{m}\otimes\wh\bgamma_i')\bxi_i,\]
where $m=T-q-h+1$.
Then,
\begin{align}
    \bZ_i\bZ_i'=&\bG(\wh\bgamma_i\wh\bgamma_i'\otimes\bb_i\bb_i')\bG'+\bG(\wh\bgamma_i\otimes\bb_i)\bxi_i'(\bI_m\otimes\wh\bgamma_i)\notag\\
    &+(\bI_m\otimes\wh\bgamma_i')\bxi_i(\wh\bgamma_i'\otimes\bb_i')\bG'+(\bI_m\otimes\wh\bgamma_i')\bxi_i\bxi_i'(\bI_m\otimes\wh\bgamma_i).
\end{align}
By definition, if $\wt\bG$ consists of the eigenvectors of $\sum_{i=1}^N\bZ_i\bZ_i'$ as its columns, we have
\begin{align}\label{vt}
    \wt\bV=&\wt\bG'\sum_{i=1}^N\bZ_i\bZ_i'\wt\bG\notag\\
=&\wt\bG'[\bG\sum_{i=1}^N(\wh\bgamma_i\wh\bgamma_i'\otimes\bb_i\bb_i')\bG']\wt\bG+\wt\bG'[\bG\sum_{i=1}^N(\wh\bgamma_i\otimes\bb_i)\bxi_i'(\bI_m\otimes\wh\bgamma_i)]\wt\bG\notag\\
&+\wt\bG'[\sum_{i=1}^N(\bI_m\otimes\wh\bgamma_i')\bxi_i(\wh\bgamma_i'\otimes\bb_i')\bG']\wt\bG+\wt\bG'[\sum_{i=1}^N(\bI_m\otimes\wh\bgamma_i')\bxi_i\bxi_i'(\bI_m\otimes\wh\bgamma_i)]\wt\bG\notag\\
=&I_1+I_2+I_3+I_4.
\end{align}
Let $\mathcal{I}_b$ be the set of indices whose $\bb_i$ is not zero and $\mathcal{I}_b^c$ be its complement set.  Then $|\mathcal{I}_b|=N^\nu$, $\wh\bgamma_i=\bgamma_i+\bdelta_i$ in set $\mathcal{I}_b$, and $\wh\bgamma_i=\bdelta_i$ in set $\mathcal{I}_b^c$. We first consider $I_1$ in (\ref{vt}). Note that $\wh\bgamma_i=\bgamma_i+\bdelta_i$, therefore,
\begin{align}\label{I1}
    I_1=&\wt\bG'[\bG\sum_{i\in\mathcal{I}_b}((\bgamma_i+\bdelta_i)(\bgamma_i+\bdelta_i)'\otimes\bb_i\bb_i')\bG']\wt\bG+\wt\bG'\bG\sum_{i\in\mathcal{I}_b^c}(\bdelta_i\bdelta_i'\otimes\bb_i\bb_i')\bG'\wt\bG\notag\\
=&\wt\bG'\bG(\sum_{i\in\mathcal{I}_b}\bgamma_i\bgamma_i'\otimes\bb_i\bb_i')\bG'\wt\bG+\wt\bG'\bG\sum_{i\in\mathcal{I}_b}(\bgamma_i\bdelta_i'\otimes\bb_i\bb_i')\bG'\wt\bG+\wt\bG'\bG(\sum_{i\in\mathcal{I}_b}\bdelta_i\bgamma_i'\otimes\bb_i\bb_i')\bG'\wt\bG\notag\\
    &+\wt\bG'\bG\sum_{i=1}^N(\bdelta_i\bdelta_i'\otimes\bb_i\bb_i')\bG'\wt\bG\notag\\
    =&I_{1,1}+I_{1,2}+I_{1,3}+I_{1,4}.
\end{align}
Note that $\|\wt\bG\|=O_p(1)$ and $\|\bG\|=O_p(\sqrt{T})$, by Assumptions 1$-$3, it is not hard to show that 
\begin{equation}\label{I11}
    \|I_{1,1}\|=O_P(N^\nu T).
\end{equation}
Since $\|\bdelta_i\|=O_p(1/\sqrt{T})$, we can similarly show that
\begin{equation}\label{I12}
    \|I_{1,2}\|=O_p(N^\nu T^{1/2}),\|I_{1,3}\|=O_p(N^\nu T^{1/2}),\,\,\text{and}\,\,\|I_{1,4}\|=O_p(N^\nu).
\end{equation}
It follows from (\ref{I11})--(\ref{I12}) that
\begin{equation}\label{I1:R}
    \|I_1\|=O_p(N^\nu T+N^\nu T^{1/2}+N^\nu)=O_p(N^\nu T).
\end{equation}
Consider $I_2$, which is 
\begin{align}\label{I2}
I_2=&\wt\bG'[\bG\sum_{i=1}^N(\wt\bgamma_i\otimes\bb_i)\bxi_i'(\bI_m\otimes\wt\bgamma_i)]\wt\bG\notag\\
=&\wt\bG'[\bG\sum_{i\in\mathcal{I}_b}((\wt\bgamma_i+\bdelta_i)\otimes\bb_i)\bxi_i'(\bI_m\otimes(\bgamma_i+\bdelta_i))]\wt\bG+\wt\bG'\bG\sum_{i\in\mathcal{I}_b^c}(\bdelta_i\otimes\bb_i)\bxi_i'(\bI_m\otimes\bdelta_i)\wt\bG\notag\\
=&\wt\bG'\bG\sum_{i\in\mathcal{I}_b}(\bgamma_i\otimes\bb_i)\bxi_i'(\bI_m\otimes\bgamma_i)\wt\bG+\wt\bG'\bG\sum_{i\in\mathcal{I}_b}(\bgamma_i\otimes\bb_i)\bxi_i'(\bI_m\otimes\bdelta_i)\wt\bG\notag\\
&+\wt\bG'\bG\sum_{i\in\mathcal{I}_b}(\bdelta_i\otimes\bb_i)\bxi_i'(\bI_m\otimes\bgamma_i)\wt\bG+\wt\bG'\bG\sum_{i=1}^N(\bdelta_i\otimes\bb_i)\bxi_i'(\bI_m\otimes\bdelta_i)\wt\bG\notag\\
=&I_{2,1}+I_{2,2}+I_{2,3}+I_{2,4}.
\end{align}
Note that
\begin{align}
 \sum_{i\in\mathcal{I}_b}(\bgamma_i\otimes\bb_i)\bxi_i'(\bI_m\otimes\bgamma_i)\wt\bG=&\sum_{i\in\mathcal{I}_b}(\bgamma_i\otimes\bb_i)\sum_{t=q}^{T-h}\bxi_{i,t}'\bgamma_i\wt\bgg_t'\notag\\
 =&\sum_{t=q}^{T-h}\sum_{i\in\mathcal{I}_b}(\bgamma_i\otimes\bb_i)\bgamma_i'\bxi_{i,t}\wt\bgg_t'\notag\\
 \leq&\left(\sum_{t=q}^{T-h}|\sum_{i\in\mathcal{I}_b}(\bgamma_i\otimes\bb_i)\bgamma_i'\bxi_{i,t}|^2\right)^{1/2}\left(\sum_{t=q}^{T-h}\|\wt\bgg_t\|^2\right)^{1/2}\notag\\
 =&O_p\{(T(N^{\nu/2})^2)^{1/2}\}O_p(1)\notag\\
 =&O_p(N^{\nu/2}T^{1/2}),
\end{align}
which implies that
\begin{equation}\label{I21}
    \|I_{2,1}\|=\|\wt\bG'\bG\|O_p(N^{\nu/2}T^{1/2})=O_p(N^{\nu/2}T).
\end{equation}
Recall that
\begin{align}\label{delta:d}
    \bdelta_i=&\left\{\frac{1}{T}\sum_{t=q}^{T-h}\bxi_{i,t}\bgg_t'\bbeta+\frac{1}{T}\sum_{t=q}^{T-h}\bxi_{i,t}\bve_{t+h}\right\}+\left\{(\bI_q\otimes\bb_i')\frac{1}{T}\sum_{t=q}^{T-h}\bgg_t\bve_{t+h}\right\}\notag\\
    =&\bdelta_{i,1}+\bdelta_{i,2}.
\end{align}
We now consider $I_{2,2}$. First,
\[\sum_{i\in\mathcal{I}_b}(\bgamma_i\otimes\bb_i)\bxi_{i}'(\bI_m\otimes\bdelta_{i,2})\wt\bG=O_p(N^{\nu/2}),\]
and
\begin{align}\label{I22:d}
    \sum_{i\in\mathcal{I}_b}(\bgamma_i\otimes\bb_i)\bxi_i'(\bI_m\otimes\bdelta_{i,1})\wt\bG=&\sum_{i\in\mathcal{I}_b}(\bgamma_i\otimes\bb_i)\bdelta_{i,1}'\sum_{t=q}^{T-h}\bxi_{i,t}\wt\bgg_t'\notag\\
    =&\frac{1}{T}\sum_{t=q}^{T-h}\sum_{s=q}^{T-h}w_t[\sum_{i\in\mathcal{I}_b}(\bgamma_i\otimes\bb_i)\bxi_{i,s}'\bxi_{it}]\wt\bgg_s'\notag\\
    =&\frac{1}{T}\sum_{t=q}^{T-h}\sum_{s=q}^{T-h}w_t[\sum_{i\in\mathcal{I}_b}(\bgamma_i\otimes\bb_i)(\bxi_{i,s}'\bxi_{it}-E\bxi_{i,s}'\bxi_{it})]\wt\bgg_s'\notag\\
    &+\frac{1}{T}\sum_{t=q}^{T-h}\sum_{s=q}^{T-h}w_t[\sum_{i\in\mathcal{I}_b}(\bgamma_i\otimes\bb_i)E\bxi_{i,s}'\bxi_{it}]\wt\bgg_s'\notag\\
    =&I_{2,2,1}+I_{2,2,2}.
\end{align}
By the Schwarz inequality, we have
\begin{align}\label{I221}
    I_{2,2,1}\leq&\frac{1}{T}\left\{\sum_{s=q}^{T-h}[\sum_{t=q}^{T-h}w_t\sum_{i\in\mathcal{I}_b}(\bgamma_i\otimes\bb_i)(\bxi_{i,s}'\bxi_{it}-E\bxi_{i,s}'\bxi_{it}))]^2\right\}^{1/2}\left[\sum_{s=q}^{T-h}\|\wt\bgg_s\|^2\right]^{1/2}\notag\\
    \leq&\frac{1}{T}O_p(T(N^{\nu/2} T^{1/2})^2)^{1/2}O_p(1)=O_p(N^{\nu/2}),
\end{align}
and
\begin{align}\label{I222}
    I_{2,2,2}=&\frac{1}{T}\left(\sum_{s=q}^{T-h}\|\wt\bgg_s\|^2\right)^{1/2}\left(\sum_{s=q}^{T-h}|\sum_{t=q}^{T-h}w_t[\sum_{i\in\mathcal{I}_b}E\bxi_{i,s}'\bxi_{i,t}]|^2\right)^{1/2}\notag\\
    \leq&O_p(1/T)O_p((T(N^\nu)^2)^{1/2})=O_p(N^{\nu}T^{-1/2}).
\end{align}
where $w_t=\bgg_t'\bbeta+\bve_{t+h}$. Since $\|\bG\|=O_p(T^{1/2})$, it follows that
\[I_{2,2}=O_p(N^{\nu/2}T^{1/2}+N^{\nu}).\]
Similarly, we can show that
\[I_{2,3}=O_p(N^{\nu/2}T^{1/2}+N^{\nu}).\]
Consider the last term in $I_{2,2}$,
\begin{align}\label{I24}
I_{2,4}=&\wt\bG[\bG\sum_{i=1}^N(\bdelta_{i,1}\otimes\bb_i)\bxi_i'(\bI_m\otimes\bdelta_{i,1})]\wt\bG+\wt\bG[\bG\sum_{i=1}^N(\bdelta_{i,1}\otimes\bb_i)\bxi_i'(\bI_m\otimes\bdelta_{i,2})]\wt\bG\notag\\
&+\wt\bG[\bG\sum_{i=1}^N(\bdelta_{i,2}\otimes\bb_i)\bxi_i'(\bI_m\otimes\bdelta_{i,1})]\wt\bG+\wt\bG[\bG\sum_{i=1}^N(\bdelta_{i,2}\otimes\bb_i)\bxi_i'(\bI_m\otimes\bdelta_{i,2})]\wt\bG\notag\\
=&I_{2,4,1}+I_{2,4,2}+I_{2,4,3}+I_{2,4,4}.
\end{align}
We only need to bound the first term as $\bdelta_{i,1}$ is correlated with $\bxi_i$ but $\bdelta_{i,2}$ is not. Since $\|\bG\|=O_p(\sqrt{T})$, $\|\wt\bG\|=O_p(1)$, $\|\bdelta_{i,1}\|=O_p(1/\sqrt{T})$, and $\|\bxi_{i}\|=O_p(\sqrt{T})$, we have
\begin{align}\label{I241}
    I_{2,4,1}=&O_p(T^{1/2}T^{-1}T^{1/2}N^{\nu})=O_p(N^{\nu}).
\end{align}
It follows that
\[\|I_{2,4}\|=O_p(N^\nu),\]
and, therefore,
\begin{equation}\label{I2:r}
\|I_{2}\|=O_p(N^{\nu/2}T+N^{\nu/2}T^{1/2}+N^\nu+N^\nu)=O_p(N^{\nu/2}T+N^{\nu}).    
\end{equation}
Similarly, we can show that
\begin{equation}\label{I3:R}
\|I_{3}\|=O_p(N^{\nu/2}T+N^{\nu}). 
\end{equation}

Now turn to $I_4$. Noting that $\wh\bgamma_i=\bgamma_i+\bdelta_i$ for $i\in\mathcal{I}_b$ and $\wh\bgamma_i=\bdelta_i$ for $i\in\mathcal{I}_b^c$, we have the following decomposition,
\begin{align}\label{I4}
    I_4=&\wt\bG'[\sum_{i\in\mathcal{I}_b}(\bI_{m}\otimes\bgamma_i')\bxi_i\bxi_i'(\bI_m\otimes\bgamma_i)]\wt\bG+\wt\bG'[\sum_{i\in\mathcal{I}_b}(\bI_{m}\otimes\bgamma_i')\bxi_i\bxi_i'(\bI_m\otimes\bdelta_i)]\wt\bG\notag\\
    &+\wt\bG'[\sum_{i\in\mathcal{I}_b}(\bI_{m}\otimes\bdelta_i')\bxi_i\bxi_i'(\bI_m\otimes\bgamma_i)]\wt\bG+\wt\bG'[\sum_{i\in\mathcal{I}_b}(\bI_{m}\otimes\bdelta_i')\bxi_i\bxi_i'(\bI_m\otimes\bdelta_i)]\wt\bG\notag\\
    =&I_{4,1}+I_{4,2}+I_{4,3}+I_{4,4}.
\end{align}
We consider the four terms in (\ref{I4}) one by one. First,
\begin{align}
    I_{4,1}=&\wt\bG'[\sum_{i\in\mathcal{I}_b}(\bI_m\otimes\bgamma_i')(\bxi_i\bxi_i'-E\bxi_i\bxi_i')(\bI_m\otimes\bgamma_i)]\wt\bG+\wt\bG'[\sum_{i\in\mathcal{I}_b}(\bI_m\otimes\bgamma_i')(E\bxi_i\bxi_i')(\bI_m\otimes\bgamma_i)]\wt\bG\notag\\
    =&I_{4,1,1}+I_{4,1,2}.
\end{align}
By a similar argument as that in (\ref{I221})-(\ref{I222}), we have
\begin{align}\label{I411}
    I_{4,1,1}=&\sum_{t=q}^{T-h}\sum_{s=q}^{T-h}\wt\bgg_s\sum_{i\in\mathcal{I}_b}\sum_{i\in\mathcal{I}_b}\bgamma_i'(\bxi_{i,s}\bxi_{it}'-E\bxi_{i,s}\bxi_{i,t})\bgamma_i\wt\bgg_t'\notag\\
    \leq&\left(\sum_{t=q}^{T-h}\|\wt\bgg_t\|\right)^{1/2}\left(\sum_{t=q}^{T-h}\|\sum_{s=q}^{T-h}\wt\bgg_s\sum_{i\in\mathcal{I}_b}\bgamma_i'(\bxi_{i,s}\bxi_{i,t}-E\bxi_{i,s}\bxi_{i,t})\bgamma_i\|^2\right)^{1/2}\notag\\
    \leq&\left(\sum_{t=q}^{T-h}\|\wt\bgg_t\|\right)\left(\sum_{t=q}^{T-h}\sum_{s=q}^{T-h}(\sum_{i\in\mathcal{I}_b}\bgamma_i'(\bxi_{i,s}\bxi_{i,t}-E\bxi_{i,s}\bxi_{i,t})\bgamma_i)^2\right)^{1/2}\notag\\
    =&O_p(1)O_p(T^2N^{\nu})^{1/2}=O_P(N^{\nu/2}T),
\end{align}
and
\[I_{4,1,2}=O_p(N^\nu).\]
Therefore,
\begin{equation}\label{I41:R}
    I_{4,1}=O_p(N^\nu+N^{\nu/2}T).
\end{equation}
Next,
\begin{align}\label{I42}
I_{4,2}=&\wt\bG[\sum_{i\in\mathcal{I}_b}(\bI_m\otimes\bgamma_i')\bxi_i\bxi_i'(\bI_m\otimes\bdelta_{i,1})]\wt\bG+\wt\bG[\sum_{i\in\mathcal{I}_b}(\bI_m\otimes\bgamma_i')\bxi_i\bxi_i'(\bI_m\otimes\bdelta_{i,2})]\wt\bG\notag\\
=&I_{4,2,1}+O_p(N^\nu T^{-1/2}+N^{\nu/2}T^{1/2}),
\end{align}
where the second rate follows from a similar argument for (\ref{I41:R}) and the fact that $\bdelta_{i,2}$ is uncorrelated with $\bxi_i$. For the first term, we have 
\begin{align}\label{I421}
    I_{4,2,1}=&\frac{1}{T}\sum_{t=q}^{T-h}\sum_{i\in\mathcal{I}_b}\sum_{s=q}^{T-h}\wt\bgg_t\bgamma_i'\bxi_{i,t}\bxi_{i,s}'\bdelta_{i,1}\wt\bgg_s'\notag\\
    =&\frac{1}{T}\sum_{t=q}^{T-h}\sum_{s=q}^{T-h}\sum_{l=q}^{T-h}\sum_{i\in\mathcal{I}_b}\wt\bgg_t\bgamma_i'\bxi_{i,t}\bxi_{i,s}'\bxi_{i,l}w_l\wt\bgg_s'\notag\\
    \leq&\frac{1}{T}\left(\sum_{t=q}^{T-h}\|\wt\bgg_t\|^2\right)\left(\sum_{t=q}^{T-h}\sum_{s=q}^{T-h}(\sum_{l=q}^{T-h}\sum_{i\in\mathcal{I}_b}\bgamma_i'\bxi_{i,t}\bxi_{i,s}'\bxi_{i,l}w_l)^2\right)^{1/2}\notag\\
    \leq&O_p(T^{-1})\left[\sum_{t=q}^{T-h}\sum_{s=q}^{T-h}(\sum_{l=q}^{T-h}w_l\sum_{i\in\mathcal{I}_b}\bgamma_i'(\bxi_{i,t}\bxi_{i,s}'\bxi_{i,l}-E\bxi_{i,t}\bxi_{i,s}\bxi_{i,l}))^2\right]^{1/2}\notag\\
    &+O_p(T^{-1})\left[\sum_{t=q}^{T-h}\sum_{s=q}^{T-h}(\sum_{l=q}^{T-h}w_l\sum_{i\in\mathcal{I}_b}\bgamma_i'(E\bxi_{i,t}\bxi_{i,s}\bxi_{i,l}))^2\right]^{1/2}\notag\\
    \leq& O_p(T^{-1})O_p(T^2(T^{1/2}N^{\nu/2})^2)^{1/2}+N^{\nu}T^{-1/2}=O_p(N^{\nu/2}T^{1/2}+N^\nu T^{-1/2}),
\end{align}
and 
\[I_{4,3}=O_p(N^{\nu/2}T^{1/2}+N^\nu T^{-1/2}).\]
For $I_{4,4}$, we first note that
\begin{align}
    \|I_{4,4}\|\leq& 2\|\wt\bG'[\sum_{i\in\mathcal{I}_b}(\bI_{m}\otimes\bdelta_{i,1}')\bxi_i\bxi_i'(\bI_m\otimes\bdelta_{i,1})]\wt\bG\|+2\|\wt\bG'[\sum_{i\in\mathcal{I}_b}(\bI_{m}\otimes\bdelta_{i,2}')\bxi_i\bxi_i'(\bI_m\otimes\bdelta_{i,2})]\wt\bG\|\notag\\
    =&2\|I_{4,4,1}\|+2\|I_{4,4,2}\|.
\end{align}
Recall that
\[\bdelta_{i,1}=\frac{1}{T}\sum_{l=q}^{T-h}\bxi_{i,l}w_l\,\,\text{and}\,\,\bdelta_{i,2}=(\bI_q\otimes\bb_i)'\frac{1}{T}\sum_{t=q}^{T-h}\bgg_t\bve_{t+h},\]
then, 
\begin{align}
   I_{4,4,1}=&\frac{1}{T^2} \sum_{t=q}^{T-h}\sum_{s=q}^{t-h}\sum_{l=q}^{T-h}\sum_{k=q}^{T-h}\sum_{i=1}^N\wt\bgg_t w_t w_k \bxi_{i,l}'\bxi_{i,t}\bxi_{i,s}'\bxi_{i,k}\wt\bgg_s'\notag\\
   \leq&\frac{1}{T^2}\sum_{t=q}^{T-h}\sum_{s=q}^{t-h}\sum_{l=q}^{T-h}\sum_{k=q}^{T-h}\sum_{i=1}^N\wt\bgg_t w_t w_k (\bxi_{i,l}'\bxi_{i,t}\bxi_{i,s}'\bxi_{i,k}-E\bxi_{i,l}'\bxi_{i,t}\bxi_{i,s}'\bxi_{i,k})\wt\bgg_s'\notag\\
   &+\frac{1}{T^2}\sum_{t=q}^{T-h}\sum_{s=q}^{t-h}\sum_{l=q}^{T-h}\sum_{k=q}^{T-h}\sum_{i=1}^N\wt\bgg_t w_t w_k (E\bxi_{i,l}'\bxi_{i,t}\bxi_{i,s}'\bxi_{i,k})\wt\bgg_s'\notag\\
   =&J_{4,1}+J_{4,2}.
\end{align}

\begin{align}
    J_{4,1}\leq & \frac{1}{T^2}(\sum_{t=q}^{T-h}\|\wt\bgg_t\|)\left[\sum_{t=q}^{T-h}\sum_{s=q}^{T-h}(\sum_{l=q}^{T-h}\sum_{k=q}^{T-h}\sum_{i=1}^N w_t w_k (\bxi_{i,l}'\bxi_{i,t}\bxi_{i,s}'\bxi_{i,k}-E\bxi_{i,l}'\bxi_{i,t}\bxi_{i,s}'\bxi_{i,k}))^2\right]^{1/2}\notag\\
    \leq&\frac{1}{T^2}O_p(T^2(T^2N))^{1/2}=O_p(N^{1/2}).
\end{align}
\[J_{4,2}\leq \frac{1}{T^2}(\sum_{t=q}^{T-h}\sum_{s=q}^{T-h}(\sum_{i=1}^N w_t^2E\bxi_{i,t}'\bxi_{i,t}\bxi_{i,s}'\bxi_{i,s})^2)^{1/2}=O_p(\frac{1}{T^2}(T^2N^2)^{1/2})=O_p(N/T).\]
It follows that
\[I_{4,4,1}=O_p(N^{1/2}+\frac{N}{T}).\]
\begin{align}
    I_{4,4,2}=&\sum_{t=q}^{T-h}\sum_{s=q}^{T-h}\wt\bgg_t\sum_{i=1}^N\bdelta_{i,2}'\bxi_{i,t}\bxi_{i,s}'\bdelta_{i,2}\wt\bgg_s'\notag\\
    \leq&(\sum_{t=q}^{T-h}\|\wt\bgg_t\|^2)\left(\sum_{t=q}^{T-h}(\sum_{s=q}^{T-h}\sum_{i=1}^N\bdelta_{i,2}'\bxi_{i,t}\bxi_{i,s}'\bdelta_{i,2})^2\right)^{1/2}\notag\\
    \leq&C\left(\sum_{t=q}^{T-h}\sum_{s=q}^{T-h}(\sum_{i=1}^N\bdelta_{i,2}'(\bxi_{i,t}\bxi_{i,s}'-E\bxi_{i,t}\bxi_{i,s}')\bdelta_{i,2})^2\right)^{1/2}\notag\\
    \leq&O_p(T^2(N^{\nu/2}T^{-1}))^{1/2}+O_p(N^{\nu}T^{-1})=O_p(N^{\nu/2}+N^\nu T^{-1}).
\end{align}
It follows that
\[I_{4,4}=O_p(N^{1/2}+N/T+N^{\nu}T^{-1}),\]
and therefore,
\begin{equation}\label{I4:R}
    I_4=O_p(N^{\nu/2}T+N^{\nu}+N^{1/2}+NT^{-1}).
\end{equation}
Then, Lemma \ref{lm1} follows from (\ref{I1:R}), (\ref{I2:r}), (\ref{I3:R}), and (\ref{I4:R}). This completes the proof. $\Box$


{\bf Proof of Theorem 1.}  For simplicity, we use $\wh\bG$ as the estimator $\wh\bG^{dPCA}$ in this proof. In other words, $\wh\bG'\wh\bG/T=\bI_{rq}$. By the definition of $\wt\bV$, we have $\wh\bG\wt\bV=\sum_{i=1}^N\bZ_i\bZ_i'\wh\bG$. Let $\wh\bV=\frac{1}{N^\nu T}\wt\bV$, it follows that
\begin{align}
    \wh\bG\wh\bV=&\frac{1}{N^\nu T}\left[\bG\sum_{i=1}^N(\wh\bgamma_i\wh\bgamma_i\otimes\bb_i\bb_i')\bG'+\bG\sum_{i=1}^N(\wh\bgamma_i\otimes\bb_i)\bxi_i'(\bI_m\otimes\bgamma_i)\right.\notag\\  &+\left.\sum_{i=1}^N(\bI_m\otimes\wh\bgamma_i')\bxi_i(\wh\bgamma_i'\otimes\bb_i')\bG'+\sum_{i=1}^N(\bI_m\otimes\wh\bgamma_i')\bxi_i\bxi_i'(\bI_m\otimes\wh\bgamma_i)\right]\wh\bG.
\end{align}

Letting 
\[\bH=\left(\frac{1}{N^\nu T}\sum_{i=1}^N(\wh\bgamma_i\wh\bgamma_i'\otimes\bb_i\bb_i')\bG'\wh\bG\wh\bV^{-1}\right)',\]
we have
\begin{align}\label{Ghat:d}
 \wh\bG-\bG\bH'=&\frac{1}{N^\nu T}\left[\bG\sum_{i=1}^N(\wh\bgamma_i\otimes\bb_i)\bxi_i'(\bI_m\otimes\bgamma_i)+\sum_{i=1}^N(\bI_m\otimes\wh\bgamma_i')\bxi_i(\wh\bgamma_i'\otimes\bb_i')\bG'\right.\notag\\
&\left.+\sum_{i=1}^N(\bI_m\otimes\wh\bgamma_i')\bxi_i\bxi_i'(\bI_m\otimes\wh\bgamma_i)\right]\wh\bG\wh\bV^{-1}.
\end{align}
Letting 
\[\frac{1}{\sqrt{T}}\|\wh\bG-\bG\bH'\|=\Pi_1+\Pi_2+\Pi_3,\]
we will bound all three terms in the sequel. First,
\begin{align}
    \Pi_1=&\frac{1}{N^\nu T^{3/2}}\bG\sum_{i=1}^N(\wh\bgamma_i\otimes\bb_i)\bxi_i'(\bI_m\otimes\wh\bgamma_i)\wh\bG\wh\bV^{-1}\\
    =&\frac{1}{N^\nu T^{3/2}}\bG\sum_{i=1}^N(\wh\bgamma_i\otimes\bb_i)\bxi_i'(\bI_m\otimes\wh\bgamma_i)(\wh\bG-\bG \bH')\wh\bV^{-1}+\frac{1}{N^\nu T^{3/2}}\bG\sum_{i=1}^N(\wh\bgamma_i\otimes\bb_i)\bxi_i'(\bI_m\otimes\wh\bgamma_i)\bG\bH'\wh\bV^{-1},\notag
\end{align}
it follows that
\begin{align}
    \|\Pi_1\|\leq& \frac{\|\bG\|}{\sqrt{T}}\frac{\sum_{i=1}^N (\wh\bgamma_i\otimes)\bxi_i'(\bI_m\otimes\wh\bgamma_i)\|}{N^\nu \sqrt{T}}\frac{\|\wh\bG-\bG\bH'\|}{\sqrt{T}}\|\wh\bV^{-1}\|\notag\\
    &+\frac{\|\bG\|}{\sqrt{T}}\frac{\|\sum_{i=1}^N(\wh\bgamma_i\otimes\bb_i)\bxi_i'(\bI_m\otimes\wh\bgamma_i)\bG\|}{N^\nu T}\|\bH'\wh\bV^{-1}\|\notag\\
    =&\Pi_{1,1}+\Pi_{1,2}.
\end{align}
Note that
\begin{align}
    \sum_{i=1}^N(\wh\bgamma_i\otimes\bb_i)\bxi_i'(\bI_m\otimes\wh\bgamma_i)=&\sum_{i=1}^N((\bgamma_i+\bdelta_i)\otimes\bb_i)\bxi_i'(\bI_m\otimes(\bgamma_i+\bdelta_i))\notag\\
    =&\sum_{i\in\mathcal{I}_b}(\gamma_i\otimes\bb_i)\bxi_i'(\bI_m\otimes\bgamma_i)+\sum_{i\in\mathcal{I}_b}(\bdelta_i\otimes\bb_i)\bxi_i'(\bI_m\otimes\bgamma_i)\notag\\
    &+\sum_{i\in\mathcal{I}_b}(\bgamma_i\otimes\bb_i)\bxi_i'(\bI_m\otimes\bdelta_i)+\sum_{i=1}^N(\bdelta_i\otimes\bb_i)\bxi_i'(\bI_m\otimes\bdelta_i)\notag\\
    =&K_1+K_2+K_3+K_4.
\end{align}
By Assumptions 1--3 or the results in Lemma \ref{lm0}, it is not hard to show that
\[K_1=O_p(N^{\nu/2}T^{1/2}),\]

\begin{align}
    K_2\leq&(\sum_{i\in\mathcal{I}_b}\|\bdelta_i\otimes\bb_i\|^2)^{1/2}(\sum_{i\in\mathcal{I}_b}\|\bxi_i'(\bI_m\otimes\bgamma_i)\|^2)^{1/2}
    \leq O_p(N^\nu T^{-1})^{1/2}O_p(N^\nu T)^{1/2}=O_p(N^\nu),\notag
\end{align}
and
\[K_3=O_p(N^\nu).\]
\begin{align}
    \|K_4\|\leq &\|\sum_{i=1}^N(\bdelta_i\otimes\bb_i)\bxi_i'(\bI_m\otimes\bdelta_i)\|\notag\\
    \leq&\|\sum_{i\in\mathcal{I}_b}(\bdelta_{i,1}\otimes\bb_i)\bxi_i'(\bI_m\otimes\bdelta_{i,1})\|\notag\\
    \leq&\sqrt{T}\|\sum_{i\in\mathcal{I}_b}(\bdelta_{i,1}\otimes\bb_i)\bdelta_{i,1}'\bxi_{i,t}\|\notag\\
    \leq&\sqrt{T}(\sum_{i\in\mathcal{I}_b}\|(\bdelta_{i,1}\otimes\bb_i)\bdelta_{i,1}'\|^2)^{1/2}(\sum_{i\in\mathcal{I}_b}\|\bxi_{i,t}\|^2)^{1/2}\notag\\
    \leq &O_p(\sqrt{T}(N^\nu T^{-2})^{1/2}(N^{\nu})^{1/2})=O_p(N^\nu T^{-1/2}).
\end{align}
Then,
\[\Pi_1\leq O_p(\frac{N^{\nu/2}T^{1/2}+N^\nu}{N^\nu\sqrt{T}})\frac{\|\wh\bG-\bG\bH'\|}{\sqrt{T}}=o_p(1)\frac{\|\wh\bG-\bG\bH'\|}{\sqrt{T}},\]
which is of a smaller order than $\frac{\|\wh\bG-\bG\bH'\|}{\sqrt{T}}$.
Consider $\Pi_{1,2}$, note that
\begin{align}
\sum_{i=1}^N (\wh\bgamma_i\otimes\bb_i)\bxi_i'(\bI_m\otimes\wh\bgamma_i)\bG=&\sum_{i=1}^N(\bgamma_i\otimes\bb_i)\bxi_i'(\bI_m\otimes\bgamma_i)\bG+\sum_{i=1}^N (\bgamma_i\otimes\bb_i)\bxi_i'(\bI_m\otimes\bdelta_i)\bG\notag\\
&+\sum_{i=1}^N(\bdelta_i\otimes\bb_i)\bxi_i'(\bI_m\otimes\bgamma_i)\bG+\sum_{i=1}^N (\bdelta_i\otimes\bb_i)\bxi_i'(\bI_m\otimes\bdelta_i)\bG\notag\\
=&R_1+R_2+R_3+R_4.
\end{align}
Since $\bxi_{i,t}$ is independent across $i$ and $q$-dependent across $t$, and it is uncorrelated with $\bgg_t$, we have
\[R_1=\sum_{i=1}^N(\bgamma_i\otimes\bb_i)\sum_{t=q}^{T-h}\bxi_{i,t}'\bgamma_i\bgg_t=O_p(N^{\nu/2}T^{1/2}),\]
and
\[R_2=\sum_{i=1}^N(\bgamma_i\otimes\bb_i)\bdelta_i'\sum_{t=q}^{T-h}\bxi_{i,t}\bgg_t'\leq \sum_{i=1}^N\|\bgamma_i\otimes\bb_i\bdelta_i'\|\|\sum_{t=q}^{T-h}\bxi_{i,t}\bgg_t'\|=O_p(N^{\nu}T^{-1/2}\sqrt{T})=O_p(N^\nu).\]
Similarly,
\[R_3=O_p(N^\nu).\]
\begin{align}
    R_4=&\sum_{i\in\mathcal{I}_b}(\bdelta_i\otimes\bb_i)\bdelta_i'\sum_{t=q}^{T-h}\bxi_{i,t}\bgg_t'\notag\\
    \leq&(\sum_{i\in\mathcal{I}_b}\|(\bdelta_i\otimes\bb_i)\bdelta_i'\|^2)^{1/2}(\sum_{i\in\mathcal{I}_b}\|\sum_{t=q}^{T-h}\bxi_{i,t}\bgg_t'\|^2)^{1/2}\notag\\
    \leq&(N^\nu T^{-2})^{1/2}(N^\nu T)^{1/2}=O_p(N^\nu T^{-1/2}).
\end{align}
Note that $\frac{\|\bG\|}{\sqrt{T}}=O_p(1)$ and $\|\bH'\wh\bV^{-1}\|=O_p(1)$, therefore,
\[\Pi_1\leq C\frac{N^{\nu/2}T^{1/2}+N^\nu+N^\nu+N^\nu T^{-1/2}}{N^\nu T}=O_p(N^{-\nu/2}T^{-1/2}+T^{-1}).\]
On the other hand,
\begin{align}
    \Pi_2=&\frac{1}{N^\nu T^{3/2}}\sum_{i=1}^N(\bI_m\otimes\wh\bgamma_i')\bxi_i(\wh\bgamma_i'\otimes\bb_i')\bG'\wh\bG\wh\bV^{-1}\notag\\
    \leq&\frac{1}{N^\nu T^{1/2}}\|\sum_{i=1}^N (\bI_m\otimes\wh\bgamma_i')\bxi_i(\wh\bgamma_i'\otimes\bb_i')\|\frac{\|\bG'\wh\bG\|}{T}\|\wh\bV\|,
\end{align}
where
\begin{align}
    \sum_{i=1}^N(\bI_m\otimes\wh\bgamma_i')\bxi_i(\wh\bgamma_i'\otimes\bb_i)=&\sum_{i=1}^N(\bI_m\otimes\bgamma_i')\bxi_i(\bgamma_i'\otimes\bb_i')+\sum_{i=1}^N(\bI_m\otimes\bgamma_i')\bxi_i(\bdelta_i'\otimes\bb_i')\notag\\
&+\sum_{i=1}^N(\bI_m\otimes\bdelta_i')\bxi_i(\bgamma_i'\otimes\bb_i')+\sum_{i=1}^N(\bI_m\otimes\bdelta_i')\bxi_i(\bgamma_i'\otimes\bdelta_i')\notag\\
=&\Pi_{2,1}+\Pi_{2,2}+\Pi_{2,3}+\Pi_{2,4}.
\end{align}
By a similar argument as above, we can show that
\[\Pi_{2,1}=O_p(N^{\nu/2}T^{1/2}),\]
\begin{align}
    \Pi_{2,2}\leq&C\frac{1}{T}\sum_{t=q}^{T-h}\sum_{i\in\mathcal{I}_b}(\bI_m\otimes\bgamma_i')[\bxi_i((w_t\bxi_{i,t}')\otimes\bb_i')-E\bxi_i((w_t\bxi_{i,t}')\otimes\bb_i')]\notag\\
    +&C\frac{1}{T}\sum_{t=q}^{T-h}\sum_{i\in\mathcal{I}_b}(\bI_m\otimes\bgamma_i')E\bxi_i((w_t\bxi_{i,t}')\otimes\bb_i')\notag\\
    \leq&C\frac{1}{T}(TN^\nu)^{1/2}\sqrt{T}+C\sqrt{T}\frac{1}{T}N^{\nu}=O_p(N^{\nu/2}+N^\nu T^{-1/2}).
\end{align}
Similarly, we can show that
\[\Pi_{2,3}=O_p(N^{\nu/2}+N^\nu T^{-1/2}),\,\,\text{and}\,\,\Pi_{2,4}=O_p(N^\nu T^{-1/2}).\]
Therefore,
\begin{align}\label{pi:2}
    \Pi_2\leq C\frac{N^{\nu/2}T^{1/2}+N^{\nu/2}+N^\nu T^{-1/2}+N^{\nu}T^{-1/2}}{N^\nu T^{1/2}}=O_p(N^{-\nu/2}+T^{-1}).
\end{align}

Note that
\begin{align}
    \Pi_3=&\frac{q}{N^\nu T^{-3/2}}\sum_{i=1}^N(\bI_m\otimes\wh\bgamma_i')\bxi_i\bxi_i'(\bI_m\otimes\wh\bgamma_i)(\wh\bG-\bG\bH')\wh\bV^{-1}\notag\\
    +&\frac{q}{N^\nu T^{-3/2}}\sum_{i=1}^N(\bI_m\otimes\wh\bgamma_i')\bxi_i\bxi_i'(\bI_m\otimes\wh\bgamma_i)(\bG\bH')\wh\bV^{-1}\notag\\
    \leq& \Pi_{3,1}+\Pi_{3,2}.
\end{align}
For $\Pi_{3,1}$, we first have the following decomposition,
\begin{align}
   \sum_{i=1}^N(\bI_m\otimes\wh\bgamma_i)\bxi_i\bxi_i'(\bI_m\otimes\wh\bgamma_i)=&\sum_{i\in\mathcal{I}_b}(\bI_m\otimes\bgamma_i')\bxi_i\bxi_i'(\bI_m\otimes\bgamma_i)+ \sum_{i\in\mathcal{I}_b}(\bI_m\otimes\bgamma_i')\bxi_i\bxi_i'(\bI_m\otimes\bdelta_i)\notag\\
   &+\sum_{i\in\mathcal{I}_b}(\bI_m\otimes\bdelta_i')\bxi_i\bxi_i'(\bI_m\otimes\bgamma_i)+\sum_{i\in\mathcal{I}_b}(\bI_m\otimes\bdelta_i')\bxi_i\bxi_i'(\bI_m\otimes\bdelta_i)\notag\\
   =&L_1+L_2+L_3+L_4,
\end{align}
where 
\begin{align}
    L_1=&\sum_{i\in\mathcal{I}_b}(\bI_m\otimes\bgamma_i')(\bxi_i\bxi_i'-E\bxi_i\bxi_i')(\bI_m\otimes\bgamma_i)+\sum_{i\in\mathcal{I}_b}(\bI_m\otimes\bgamma_i')(E\bxi_i\bxi_i')(\bI_m\otimes\bgamma_i)\notag\\
    \leq&C\sqrt{N^\nu T^2} +CN^\nu=O_p(N^{\nu/2}T+N^\nu).
\end{align}
The $(t,s)$-block of $L_2$ can be written as
\begin{align}
L_2(t,s)=&\sum_{i\in\mathcal{I}_b}\bgamma_i'\bxi_{i,t}\bxi_{i,s}'\bdelta_{i,1}+\sum_{i\in\mathcal{I}_b}\bgamma_i'\bxi_{i,t}\bxi_{i,s}'\bdelta_{i,2}\notag\\
\leq&C\|\sum_{i\in\mathcal{I}_b}\bgamma_i'\bxi_{i,t}\bxi_{i,s}'\bdelta_{i,1}\|\notag\\
    \leq&C\frac{1}{T}\|\sum_{l=q}^{T-h}w_l\sum_{i\in\mathcal{I}_b}\bgamma_i(\bxi_{i,t}\bxi_{i,s}'\bxi_{i,l}-E\bxi_{i,t}\bxi_{i,s}\bxi_{i,l})\|+C\frac{1}{T}\|\sum_{l=q}^{T-h}w_l\sum_{i\in\mathcal{I}_b}\bgamma_i(E\bxi_{i,t}\bxi_{i,s}\bxi_{i,l})\|\notag\\
    \leq&CT^{-1}T^{1/2}N^{\nu/2}+N^\nu T^{-1}=O_p(N^{\nu/2}T^{-1/2}+N^{\nu}T^{-1}),
\end{align}
implying that
\[L_2=O_p(N^{\nu/2}T^{1/2}+N^\nu).\]
Similarly, we can show that
\[L_3=O_p(N^{\nu/2}T^{1/2}+N^\nu).\]
By a similar argument, the $(t,s)$-block of $L_4$ satisfies
\[L_4(t,s)=O_p(N^{1/2}T^{-1}+NT^{-2}),\] 
and hence,
\[L_4=O_p(N^{1/2}+NT^{-1}).\]
Therefore,
\begin{align}
    \Pi_{3,1}\leq& C\frac{N^{\nu/2}T+N^{\nu}+N^{\nu/2}T^{1/2}+N^{1/2}+NT^{-1}}{N^\nu T}\frac{\|\wh\bG-\bG\bH'\|}{\sqrt{T}}\notag\\
    \leq&C(N^{1/2-\nu}T^{-1}+N^{1-\nu}T^{-2})\frac{\|\wh\bG-\bG\bH'\|}{\sqrt{T}}\notag\\
    \leq&o_p(1)\frac{\|\wh\bG-\bG\bH'\|}{\sqrt{T}},
\end{align}
which is of a smaller order than $\frac{\|\wh\bG-\bG\bH'\|}{\sqrt{T}}$. By a similar argument as above, we can also show that
\[\Pi_{3,2}\leq C\frac{1}{N^\nu T}(N^{\nu/2}T+N^\nu+N^{1/2}+\frac{N}{T})\frac{\|\bG\|}{\sqrt{T}}=O_p(N^{-\nu/2}+T^{-1}+N^{1/2-\nu}T^{-1}+N^{1-\nu}T^{-2}),\]
and the same result holds for $\Pi_3$. Summarizing from the rates of $\Pi_1$, $\Pi_2$, and $\Pi_3$, we have
\[\frac{1}{\sqrt{T}}\|\wh\bG-\bG\bH'\|=O_p(N^{-\nu/2}+T^{-1}+N^{1-\nu}T^{-2}).\]
This completes the proof. $\Box$

{\bf Proof of Proposition 1.} (i).
Letting $\bx_i=(x_{i,1},...,x_{i,T})'$,
by an abuse of notation, we  denote $\wt\bV$ and $\wt\bF$ as the eigenvalue and eigenvector matrices, respectively. We have
\[\wt\bV=\wt\bF'[\bF\sum_{i=1}^N\bb_i\bb_i'\bF'+\bF\sum_{i=1}^N\bb_i\be_i'+\sum_{-=1}^N\be_i\bb_i'\bF'+\sum_{i=1}^N\be_i\be_i']\wt\bF.\]
By a similar argument as the proof in Lemma \ref{lm1}, we can show that
\[\wt\bV=O_p(N^\nu T+N^{1/2}T^{1/2}+N).\]
If $N^{1-\nu}/T>c$ for some $c>0$, we know that the third term above is no 
longer a smaller term compared to the first one anymore. Then, by assumption 3, with a large probability that
\[\frac{1}{N^\nu T}|\wt\bV-\wt\bF\bF\sum_{i=1}^N\bb_i\bb_i'\bF'\wt\bF|_2>C^*.\]
Let $\wh\bF=\sqrt{T}\wt\bF$ and the corresponding rotation matrix be $\bR'$. 
We can easily show that
\[\frac{1}{\sqrt{T}}\|\wh\bF-\bF\bR'\|>C,\]
with a strictly positive probability. See also the argument in the proof of Lemma 3 in \cite{huang2022scaled}. This completes the proof of Proposition 1(i). \\
(ii) The proof of Proposition 1(ii) can be carried out in a similar way as the proof of Theorem 1 in \cite{bai2021approximate} or the proof of Lemma 4 of \cite{huang2022scaled}. We omit the details. $\Box$
\begin{lemma}\label{lm2}
 Under Assumptions 1$-$6, if $N^{1-\nu}/T^2\rightarrow 0$ and the identification condition $\bG'\bG/T=\bI_{rq}$ hold, then we have
 \[\bH\bH'-\bI_{rq}=O_p(N^{-\nu/2}+T^{-1}+N^{1-\nu}/T^2)\,\,\text{and}\,\,\bH'\bH-\bI_{rq}=O_p(N^{-\nu/2}+T^{-1}+N^{1-\nu}/T^2).\]
\end{lemma}

{\bf Proof.} Note that $\bG'\bG/T=\bI_{rq}$ and $\wh\bG'\wh\bG/T=\bI_{rq}$. Then,
\begin{align}
    \bH\bH'-\bI_{rq}=&\bH\frac{\bG'\bG}{T}\bH'-\bI_{rq}\notag\\
    =&(\frac{\bH\bG'-\wh\bG'+\wh\bG'}{\sqrt{T}})(\frac{\bG\bH'-\wh\bG+\wh\bG}{\sqrt{T}})-\bI_{rq}\notag\\
    =&O_p(\frac{\|\wh\bG-\bG\bH'\|}{\sqrt{T}})=O_p(N^{-\nu/2}+T^{-1}+N^{1-\nu}/T^2).
\end{align}
It follows that
\[\bH'\bH\bH'-\bH'=O_p(N^{-\nu/2}+T^{-1}+N^{1-\nu}/T^2),\]
and the second result follows from the fact that $\|(\bH')^{-1}\|=O_p(1)$. This completes the proof. $\Box$

{\bf Proof of Theorem 2.} 
We first consider the mean-squared forecast error (MSFE) using the proposed method. By least-squares estimation, 
\[\wh\bbeta=\frac{1}{T}\wh\bG'\by,\]
where $\by=(y_{q+h},...,y_{T})'$. By an elementary argument, if $N^{1-\nu}/T^2=o(1)$,
\begin{align}\label{pro:er}
    \by-\wh\by=&\bG\bbeta+\bve-\frac{1}{T}\wh\bG\wh\bG'\by\notag\\
    =&\bG\bbeta+\bve-\frac{1}{T}(\wh\bG-\bG\bH'+\bG\bH')(\wh\bG-\bG\bH'+\bG\bH')'\by\notag\\
    =&(\bI_{m}-\frac{1}{T}\bG\bG')\bve+O_p(w_{1,NT})+O_p(\sqrt{T}w_{1,N}),
\end{align}
where $w_{1,{NT}}=N^{-\nu/2}+T^{-1}+N^{1-\nu}/T^2$. Therefore,
\begin{equation}\label{msfe:pr}
    \text{MSFE}_{sdPCA}=\frac{1}{T}\|\by-\wh\by\|^2= \frac{1}{T}\|(\bI_m-\frac{1}{T}\bG\bG')\bve\|^2+O_p(w_{1,NT}^2).
\end{equation}

By a similar argument, if we use the traditional PCA method, we can show that
\begin{equation}\label{msfe:tr}
     \text{MSFE}_{PCA}=\frac{1}{T}\|\by-\wh\by\|^2= \frac{1}{T}\|(\bI_m-\frac{1}{T}\bG\bG')\bve\|^2+O_p(w_{2,NT}^2),
\end{equation}
under the assumption that $N^{1-\nu}/T=o(1)$, where $w_{2,NT}=O_p(N^{-\nu/2}+N^{1-\nu}/T)$.
On the other hand, if $N^{1-\nu}/T\geq c>0$ and $N^{1-\nu}/T^2=o(1)$, by the argument in the proof of Proposition 1 above, we can show that,
\[  \text{MSFE}_{PCA}=\frac{1}{T}\|\by-\wh\by\|^2\geq \frac{1}{T}\|(\bI_m-\frac{1}{T}\bG\bG')\bve\|^2+C,\]
and hence,
\[\text{MSFE}_{PCA}-\text{MSFE}_{sdPCA}\geq C>0,\]
implying that our method outperforms the traditional one in theory.

Next, we consider the method that only uses the static factors in forecasting as that in the traditional diffusion model.  The prediction model can be written as
\[\by=\bG\bbeta+\bve=\bF\bbeta_0+\bR\bbeta_*+\bve,\]
where $\bR$ contains the lagged factors, and $\bbeta_*$ consists of the associated parameters, and we only use the estimated $\wh\bF=(\bff_{q},...,\bff_{T-h})'$ as predictors. For the diffusion-index forecasts in \cite{Stock2002b}, we have
\[ \by-\wh\by=(\bI_{m}-\frac{1}{T}\bF\bF')\bve+(\bI_m-\frac{1}{T}\bF\bF')\bR\bbeta_*+O_p(\sqrt{T}w_{2,NT}).\]
Note that $\bF$ is contained in $\bG$, then 
\[\frac{1}{T}\|(\bI_{m}-\frac{1}{T}\bF\bF')\bve\|^2-\frac{1}{T}\|(\bI_{m}-\frac{1}{T}\bG\bG')\bve\|^2\geq C>0.\]
In addition, since 
$\frac{1}{T}\|\bbeta_*'\bR'\bve\|=o_p(1)$,
by a similar argument, we can show that
\[\text{MSFE}_{SW}-\text{MSFE}_{sdPCA}\geq C\|\bbeta_*\|>C>0,\]
with a strictly positive probability, where $\text{MSFE}_{SW}$ denotes the forecast errors of the diffusion-index method in \cite{Stock2002b}. This implies that our method also outperforms the diffusion-index forecasting.

Next, we consider the asymptotic forecasting performance of the sPCA method in \cite{huang2022scaled} when the correct model is our proposed one. Note that the prediction model can be written as
\[\by=\bF\bbeta_0+\bR\bbeta_*+\bve,\]
where $\bR$ contains the lagged factors, and $\bbeta_*$ consists of the associated parameters. It can be shown by a similar argument as above that
\[\text{MSFE}_{sPCA}=(\bI_{m}-\frac{1}{T}\bF\bF')\bve+(\bI_m-\frac{1}{T}\bF\bF')\bR\bbeta_*+o_p(1).\]
By (\ref{pro:er}) and the above one, we can easily show that
\[\text{MSFE}_{sPCA}-\text{MSFE}_{sdPCA}\geq\|(\bI_m-\frac{1}{T}\bF\bF')\|_{\min}\|\bR\bbeta_*\|>C>0.\]
This completes the proof. $\Box$

Denoting $\bB_\gamma'=(\bgamma_1\otimes\bb_1,...,\bgamma_N\otimes\bb_N)$, we introduce the following lemma.
\begin{lemma}\label{lm3}
    Let Assumptions 1$-$6 hold. If $N^{1-\nu}/T^2\rightarrow 0$, we have
    \[\frac{1}{T}(\wh\bF-\bF\bH')'\bF=O_p(\frac{1}{N^\nu}+\frac{1}{T}),\]
    and
    \[\frac{1}{T}(\wh\bF-\bF\bH')'\wh\bF=O_p(\frac{1}{N^\nu}+\frac{1}{T}).\]
    As a result, Under the identification conditions that $\bG'\bG/T=\bI_{rq}$ and $\bB_\gamma'\bB_\gamma$ is a diagonal matrix with distinct diagonal elements, we also have
    \[\bH=\bI_{rq}+O_p(\frac{1}{N^\nu}+\frac{1}{T}).\]
\end{lemma}
{\bf Proof.} The proof is similar to that of Lemma B.2 and Lemma B. 3 in \cite{bai2003inferential}, and the argument in the proof of (2) in \cite{bai2013principal}. We omit the details. $\Box$

{\bf Proof of Theorem 3.} By a similar argument as that in Proposition 1 of \cite{bai2021approximate}, we also have
\[\frac{1}{T}\sum_{i=1}^N\|\wh\bgg_t-\bH\bgg_t\|^2=O_p(N^{-\nu}+T^{-2}+N^{2(1-\nu)}T^{-4}),\]
which is the square of the rate of $\frac{1}{\sqrt{T}}\|\wh\bG-\bG\bH'\|$ in Theorem 1. By Lemma \ref{lm3}, we obtain
\[\frac{1}{T}\sum_{t=q}^{T-h}\|\wh\bgg_t-\bgg_t\|^2=O_p(N^{-\nu}+T^{-2}+N^{2(1-\nu)}T^{-4}).\]

Let $\wh\bbeta_{lasso}$, or simply $\wh\bbeta_{lso}$ denote the Lasso solution in the proof. We have the following basic inequality,
\[\frac{1}{T}\sum_{t=q}^{T-h}(y_{t+h}-\wh\bbeta_{lso}'\wh\bgg_t)^2+\lambda\|\wh\bbeta_{lso}\|_1\leq\frac{1}{T}\sum_{t=q}^{T-h}(y_{t+h}-\bbeta'\wh\bgg_t)^2+\lambda\|\bbeta\|_1. \]

By an elementary argument, we have
\[\frac{1}{T}\sum_{t=q}^{T-h}(\bbeta-\wh\bbeta_{lso})'\bgg_t\bgg_t(\bbeta-\wh\bbeta_{lso})\leq\frac{2}{T}\sum_{t=q}^{T-h}\ve_{t+h}\bgg_t'(\wh\bbeta_{lso}-\bbeta)+\lambda\{\|\bbeta\|_1-\|\wh\bbeta_{lso}\|_1\}+w_{1N}\|(\wh\bbeta_{lso}-\bbeta)\|_1. \]
Letting $\wh\bDelta=\wh\bbeta_{lso}-\bbeta$,  by the results in Lemma 6.3 of \cite{buhlmann2011statistics}, we have $\wh\bDelta\in C_3(S)=\{\bDelta\in R^{m}:\|\bDelta_{S^c}\|_1\leq 3\|\bDelta_{S}\|_1\}$, where  $S$ is a subset of $\{1,2,...,qr\}$ with cardinality $s^*$ consisting of the indexes of the non-zero components in $\bbeta$, and $S^c$ be its complement.  By Assumption \ref{asm01}-\ref{asm1},  H\"{o}lder's inequality, and the triangle inequality, there exists a constant $\kappa>0$ such that
\begin{equation}\label{kappa}
\kappa\|\wh\bDelta\|^2\leq2\|\frac{1}{T}\sum_{t=1}^T\bgg_{t}\bve_{t+h}'\|_{\infty}\|\wh\bDelta_i\|_1+\lambda\|\wh\bDelta\|_1+w_{1,N}\|\wh\bDelta\|_1.
\end{equation}

By Assumption \ref{asm8}, Lemma 3 in \cite{fan2013large}, and Theorem 1 in \cite{merlevede2011bernstein},   we can show that 
\[\|\frac{1}{T}\sum_{t=1}^T\bgg_{t}\ve_{t+h}\|_{\infty}=O_p(\sqrt{\frac{\log(rq)}{T}}).\]
Therefore, for any $\lambda=\lambda_T\geq \max\{M\sqrt{\frac{\log(rq)}{T}},Mw_{1,NT}\}$ with a large enough constant $M>0$,  it follows from (\ref{kappa})  that
\begin{align}\label{delta:in}
\kappa\|\wh\bDelta\|^2\leq &4\lambda_T\|\wh\bDelta\|_1\leq 4\lambda_T(\|\wh\bDelta_{S}\|_1+\|\wh\bDelta_{S^c}\|_1)\notag\\
\leq & 4\lambda_T(\|\wh\bDelta_{S}\|_1+3\|\wh\bDelta_{S}\|_1)
\leq  16\lambda_T\|\wh\bDelta_{S}\|_1\leq 16\sqrt{s^*}\lambda_T\|\wh\bDelta\|,
\end{align}
which implies that
\[\|\wh\bDelta\|\leq 16\frac{\sqrt{s^*}}{\kappa}\lambda_T.\]
This completes the proof. $\Box$

\section{Data Description}
This appendix first lists the 127 macroeconomic time series  obtained from
the Federal Reserve Monthly Database for Economic Research (FRED-MD). Four variables in boldface are removed due to the missing values and the rest 123 time series are considered in this paper. For each variable,
we report the column ID of each series, the transformation code
(tcode) used to ensure stationarity of the underlying data series, the FRED-MD mnemonics, a full variable description. The comparable series in Global Insight is given in the colummn GSI, which is from  the Global Insights Basic Economics Database. The last column specifies the group number of each series. The column tcode denotes the following data transformation for a series x: (1) no transformation;
(2)$\Delta x_t$; (3) $\Delta^2 x_t$; (4) $\ln(x_t)$; (5) $\Delta\ln(x_t)$; (6) $\Delta^2\ln(x_t)$; (7) $\Delta(x_t/x_{t-1}-1.0)$.

{\begin{center}\tiny
\begin{longtable}{ccccccc}
\caption{Data description} 
          \label{Table-data}\\
\toprule
id	&	tcode	&	fred	&	description	&	gsi	&	gsi:description	&	group	\\
1	&	5	&	RPI	&	Real Personal Income	&	M\_14386177&	PI	&	1	\\
2	&	5	&	W875RX1	&	Real personal income ex transfer receipts	&	M\_145256755	&	PI less transfers	&	1	\\
3	&	5	&	DPCERA3M086SBEA	&	Real personal consumption expenditures	&	M\_123008274	&	Real Consumption	&	4	\\
4	&	5	&	CMRMTSPLx	&	Real Manu.  and Trade Industries Sales	&	M\_110156998	&	M\&T sales	&	4	\\
5	&	5	&	RETAILx	&	Retail and Food Services Sales	&	M\_130439509	&	Retail sales	&	4	\\
6	&	5	&	INDPRO	&	IP Index	&	M\_116460980	&	IP: total	&	1	\\
7	&	5	&	IPFPNSS	&	IP: Final Products and Nonindustrial Supplies	&	M\_116460981	&	IP: products	&	1	\\
8	&	5	&	IPFINAL	&	IP: Final Products (Market Group)	&	M\_116461268	&	IP: final prod	&	1	\\
9	&	5	&	IPCONGD	&	IP: Consumer Goods	&	M\_116460982	&	IP: cons gds	&	1	\\
10	&	5	&	IPDCONGD	&	IP: Durable Consumer Goods	&	M\_116460983	&	IP: cons dble	&	1	\\
11	&	5	&	IPNCONGD	&	IP: Nondurable Consumer Goods	&	M\_116460988	&	IP: cons nondble	&	1	\\
12	&	5	&	IPBUSEQ	&	IP: Business Equipment	&	M\_116460995	&	IP: bus eqpt	&	1	\\
13	&	5	&	IPMAT	&	IP: Materials	&	M\_116461002	&	IP: matls	&	1	\\
14	&	5	&	IPDMAT	&	IP: Durable Materials	&	M\_116461004	&	IP: dble matls	&	1	\\
15	&	5	&	IPNMAT	&	IP: Nondurable Materials	&	M\_116461008	&	IP: nondble matls	&	1	\\
16	&	5	&	IPMANSICS	&	IP: Manufacturing (SIC)	&	M\_116461013	&	IP: mfg	&	1	\\
17	&	5	&	IPB51222s	&	IP: Residential Utilities	&	M\_116461276	&	IP: res util	&	1	\\
18	&	5	&	IPFUELS	&	IP: Fuels	&	M\_116461275	&	IP: fuels	&	1	\\
19	&	2	&	CUMFNS	&	Capacity Utilization:  Manufacturing	&	M\_116461602	&	Cap util	&	1	\\
20	&	2	&	HWI	&	Help-Wanted Index for United States	&		&	Help wanted indx	&	2	\\
21	&	2	&	HWIURATIO	&	Ratio of Help Wanted/No.  Unemployed	&	M\_110156531	&	Help wanted/une	&	2	\\
22	&	5	&	CLF16OV	&	Civilian Labor Force	&	M\_110156467	&	Emp CPS total	&	2	\\
23	&	5	&	CE16OV	&	Civilian Employment	&	M\_110156498	&	Emp CPS nonag	&	2	\\
24	&	2	&	UNRATE	&	Civilian Unemployment Rate	&	M\_110156541	&	U: all	&	2	\\
25	&	2	&	UEMPMEAN	&	Average Duration of Unemployment (Weeks)	&	M\_110156528	&	U: mean duration	&	2	\\
26	&	5	&	UEMPLT5	&	Civilians Unemployed - Less Than 5 Weeks	&	M\_110156527	&	U < 5 wks	&	2	\\
27	&	5	&	UEMP5TO14	&	Civilians Unemployed for 5-14 Weeks	&	M\_110156523	&	U 5-14 wks	&	2	\\
28	&	5	&	UEMP15OV	&	Civilians Unemployed - 15 Weeks \& Over	&	M\_110156524	&	U 15+ wks	&	2	\\
29	&	5	&	UEMP15T26	&	Civilians Unemployed for 15-26 Weeks	&	M\_110156525	&	U 15-26 wks	&	2	\\
30	&	5	&	UEMP27OV	&	Civilians Unemployed for 27 Weeks and Over	&	M\_110156526	&	U 27+ wks	&	2	\\
31	&	5	&	CLAIMSx	&	Initial Claims	&	M\_15186204	&	UI claims	&	2	\\
32	&	5	&	PAYEMS	&	All Employees:  Total nonfarm	&	M\_123109146	&	Emp:  total	&	2	\\
33	&	5	&	USGOOD	&	All Employees:  Goods-Producing Industries	&	M\_123109172	&	Emp:  gds prod	&	2	\\
34	&	5	&	CES1021000001	&	All Employees:  Mining and Logging:  Mining	&	M\_123109244	&	Emp:  mining	&	2	\\
35	&	5	&	USCONS	&	All Employees:  Construction	&	M\_123109331	&	Emp:  const	&	2	\\
36	&	5	&	MANEMP	&	All Employees:  Manufacturing	&	M\_123109542	&	Emp:  mfg	&	2	\\
37	&	5	&	DMANEMP	&	All Employees:  Durable goods	&	M\_123109573	&	Emp:  dble gds	&	2	\\
38	&	5	&	NDMANEMP	&	All Employees:  Nondurable goods	&	M\_123110741	&	Emp:  nondbles	&	2	\\
39	&	5	&	SRVPRD	&	All Employees:  Service-Providing Industries	&	M\_123109193	&	Emp:  services	&	2	\\
40	&	5	&	USTPU	&	All Employees:  Trade, Transportation \& Utilities	&	M\_123111543	&	Emp:  TTU	&	2	\\
41	&	5	&	USWTRADE	&	All Employees:  Wholesale Trade	&	M\_123111563	&	Emp:  wholesale	&	2	\\
42	&	5	&	USTRADE	&	All Employees:  Retail Trade	&	M\_123111867	&	Emp:  retail	&	2	\\
43	&	5	&	USFIRE	&	All Employees:  Financial Activities	&	M\_123112777	&	Emp:  FIRE	&	2	\\
44	&	5	&	USGOVT	&	All Employees:  Government	&	M\_123114411	&	Emp:  Govt	&	2	\\
45	&	1	&	CES0600000007	&	Avg Weekly Hours :  Goods-Producing	&	M\_140687274	&	Avg hrs	&	2	\\
46	&	2	&	AWOTMAN	&	Avg Weekly Overtime Hours :  Manufacturing	&	M\_123109554	&	Overtime:  mfg	&	2	\\
47	&	1	&	AWHMAN	&	Avg Weekly Hours :  Manufacturing	&	M\_14386098	&	Avg hrs:  mfg	&	2	\\
48	&	4	&	HOUST	&	Housing Starts:  Total New Privately Owned	&	M\_110155536	&	Starts:  nonfarm	&	3	\\
49	&	4	&	HOUSTNE	&	Housing Starts, Northeast	&	M\_110155538	&	Starts:  NE	&	3	\\
50	&	4	&	HOUSTMW	&	Housing Starts, Midwest	&	M\_110155537	&	Starts:  MW	&	3	\\
51	&	4	&	HOUSTS	&	Housing Starts, South	&	M\_110155543	&	Starts:  South	&	3	\\
52	&	4	&	HOUSTW	&	Housing Starts, West	&	M\_110155544	&	Starts:  West	&	3	\\
53	&	4	&	PERMIT	&	New Private Housing Permits (SAAR)	&	M\_110155532	&	BP: total	&	3	\\
54	&	4	&	PERMITNE	&	New Private Housing Permits, Northeast (SAAR)	&	M\_110155531	&	BP: NE	&	3	\\
55	&	4	&	PERMITMW	&	New Private Housing Permits, Midwest (SAAR)	&	M\_110155530	&	BP: MW	&	3	\\
56	&	4	&	PERMITS	&	New Private Housing Permits, South (SAAR)	&	M\_110155533	&	BP: South	&	3	\\
57	&	4	&	PERMITW	&	New Private Housing Permits, West (SAAR)	&	M\_110155534	&	BP: West	&	3	\\
\bf 58	&	\bf 5	&\bf 	ACOGNO	&	\bf New Orders for Consumer Goods	&\bf 	M\_14385863	&	\bf Orders:  cons gds	&	\bf 4	\\
59	&	5	&	AMDMNOx	&	New Orders for Durable Goods	&	M\_14386110	&	Orders:  dble gds	&	4	\\
\bf 60	&\bf 	5	&	\bf ANDENOx	&\bf 	New Orders for Nondefense Capital Goods	&\bf 	M\_178554409	&\bf 	Orders:  cap gds	&\bf 	4	\\
61	&	5	&	AMDMUOx	&	Unfilled Orders for Durable Goods	&	M\_14385946	&	Unf orders:  dble	&	4	\\
62	&	5	&	BUSINVx	&	Total Business Inventories	&	M\_15192014	&	M\&T invent	&	4	\\
63	&	2	&	ISRATIOx	&	Total Business:  Inventories to Sales Ratio	&	M\_15191529	&	M\&T invent/sales	&	4	\\
64	&	6	&	M1SL	&	M1 Money Stock	&	M\_110154984	&	M1	&	5	\\
65	&	6	&	M2SL	&	M2 Money Stock	&	M\_110154985	&	M2	&	5	\\
66	&	5	&	M2REAL	&	Real M2 Money Stock	&	M\_110154985	&	M2 (real)	&	5	\\
67	&	6	&	BOGMBASE	&	Monetary Base	&	M\_110154995	&	MB	&	5	\\
68	&	6	&	TOTRESNS	&	Total Reserves of Depository Institutions	&	M\_110155011	&	Reserves tot	&	5	\\
69	&	7	&	NONBORRES	&	Reserves Of Depository Institutions	&	M\_110155009	&	Reserves nonbor	&	5	\\
70	&	6	&	BUSLOANS	&	Commercial and Industrial Loans	&	BUSLOANS	&	C\&I loan plus	&	5	\\
71	&	6	&	REALLN	&	Real Estate Loans at All Commercial Banks	&	BUSLOANS	&	DC\&I loans	&	5	\\
72	&	6	&	NONREVSL	&	Total Nonrevolving Credit	&	M\_110154564	&	Cons credit	&	5	\\
73	&	2	&	CONSPI	&	Nonrevolving consumer credit to Personal Income	&	M\_110154569	&	Inst cred/PI	&	5	\\
74	&	5	&	S\&P 500	&	S\&P ís Common Stock Price Index: Composite	&	M\_110155044	&	S\&P 500	&	8	\\
75	&	5	&	S\&P: indust	&	S\&P ís Common Stock Price Index: Industrials	&	M\_110155047	&	S\&P: indust	&	8	\\
76	&	2	&	S\&P div yield	&	S\&P ís Composite Common Stock: Dividend Yield	&		&	S\&P div yield	&	8	\\
77	&	5	&	S\&P PE ratio	&	S\&P ís Composite Common Stock: Price-Earnings Ratio	&		&	S\&P PE ratio	&	8	\\
78	&	2	&	FEDFUNDS	&	Effective Federal Funds Rate	&	M\_110155157	&	Fed Funds	&	6	\\
79	&	2	&	CP3Mx	&	3-Month AA Financial Commercial Paper Rate	&	CPF3M	&	Comm paper	&	6	\\
80	&	2	&	TB3MS	&	3-Month Treasury Bill:	&	M\_110155165	&	3 mo T-bill	&	6	\\
81	&	2	&	TB6MS	&	6-Month Treasury Bill:	&	M\_110155166	&	6 mo T-bill	&	6	\\
82	&	2	&	GS1	&	1-Year Treasury Rate	&	M\_110155168	&	1 yr T-bond	&	6	\\
83	&	2	&	GS5	&	5-Year Treasury Rate	&	M\_110155174	&	5 yr T-bond	&	6	\\
84	&	2	&	GS10	&	10-Year Treasury Rate	&	M\_110155169	&	10 yr T-bond	&	6	\\
85	&	2	&	AAA	&	Moodyís Seasoned Aaa Corporate Bond Yield	&		&	Aaa bond	&	6	\\
86	&	2	&	BAA	&	Moodyís Seasoned Baa Corporate Bond Yield	&		&	Baa bond	&	6	\\
87	&	1	&	COMPAPFFx	&	3-Month Commercial Paper Minus FEDFUNDS	&		&	CP-FF spread	&	6	\\
88	&	1	&	TB3SMFFM	&	3-Month Treasury C Minus FEDFUNDS	&		&	3 mo-FF spread	&	6	\\
89	&	1	&	TB6SMFFM	&	6-Month Treasury C Minus FEDFUNDS	&		&	6 mo-FF spread	&	6	\\
90	&	1	&	T1YFFM	&	1-Year Treasury C Minus FEDFUNDS	&		&	1 yr-FF spread	&	6	\\
91	&	1	&	T5YFFM	&	5-Year Treasury C Minus FEDFUNDS	&		&	5 yr-FF spread	&	6	\\
92	&	1	&	T10YFFM	&	10-Year Treasury C Minus FEDFUNDS	&		&	10 yr-FF spread	&	6	\\
93	&	1	&	AAAFFM	&	Moodyís Aaa Corporate Bond Minus FEDFUNDS	&		&	Aaa-FF spread	&	6	\\
94	&	1	&	BAAFFM	&	Moody's Baa Corporate Bond Minus FEDFUNDS	&		&	Baa-FF spread	&	6	\\
\bf 95	&\bf 	5	&	\bf TWEXAFEGSMTHx	&	\bf Trade Weighted U.S. Dollar Index	&		&	\bf Ex rate: avg	&	\bf 6	\\
96	&	5	&	EXSZUSx	&	Switzerland / U.S. Foreign Exchange Rate	&	M\_110154768	&	Ex rate: Switz	&	6	\\
97	&	5	&	EXJPUSx	&	Japan / U.S. Foreign Exchange Rate	&	M\_110154755	&	Ex rate:  Japan	&	6	\\
98	&	5	&	EXUSUKx	&	U.S. / U.K. Foreign Exchange Rate	&	M\_110154772	&	Ex rate:  UK	&	6	\\
99	&	5	&	EXCAUSx	&	Canada / U.S. Foreign Exchange Rate	&	M\_110154744	&	EX rate:  Canada	&	6	\\
100	&	6	&	WPSFD49207	&	PPI: Finished Goods	&	M\_110157517	&	PPI: fin gds	&	7	\\
101	&	6	&	WPSFD49502	&	PPI: Finished Consumer Goods	&	M\_110157508	&	PPI: cons gds	&	7	\\
102	&	6	&	WPSID61	&	PPI: Intermediate Materials	&	M\_110157527	&	PPI: int matls	&	7	\\
103	&	6	&	WPSID62	&	PPI: Crude Materials	&	M\_110157500	&	PPI: crude matls	&	7	\\
104	&	6	&	OILPRICEx	&	Crude Oil, spliced WTI and Cushing	&	M\_110157273	&	Spot market price	&	7	\\
105	&	6	&	PPICMM	&	PPI: Metals and metal products:	&	M\_110157335	&	PPI: nonferrous	&	7	\\
106	&	6	&	CPIAUCSL	&	CPI : All Items	&	M\_110157323	&	CPI-U: all	&	7	\\
107	&	6	&	CPIAPPSL	&	CPI : Apparel	&	M\_110157299	&	CPI-U: apparel	&	7	\\
108	&	6	&	CPITRNSL	&	CPI : Transportation	&	M\_110157302	&	CPI-U: transp	&	7	\\
109	&	6	&	CPIMEDSL	&	CPI : Medical Care	&	M\_110157304	&	CPI-U: medical	&	7	\\
110	&	6	&	CUSR0000SAC	&	CPI : Commodities	&	M\_110157314	&	CPI-U: comm.	&	7	\\
111	&	6	&	CUSR0000SAD	&	CPI : Durables	&	M\_110157315	&	CPI-U: dbles	&	7	\\
112	&	6	&	CUSR0000SAS	&	CPI : Services	&	M\_110157325	&	CPI-U: services	&	7	\\
113	&	6	&	CPIULFSL	&	CPI : All Items Less Food	&	M\_110157328	&	CPI-U: ex food	&	7	\\
114	&	6	&	CUSR0000SA0L2	&	CPI : All items less shelter	&	M\_110157329	&	CPI-U: ex shelter	&	7	\\
115	&	6	&	CUSR0000SA0L5	&	CPI : All items less medical care	&	M\_110157330	&	CPI-U: ex med	&	7	\\
116	&	6	&	PCEPI	&	Personal Cons.  Expend.:  Chain Index	&	gmdc	&	PCE defl	&	7	\\
117	&	6	&	DDURRG3M086SBEA	&	Personal Cons.  Exp:  Durable goods	&	gmdcd	&	PCE defl:  dlbes	&	7	\\
118	&	6	&	DNDGRG3M086SBEA	&	Personal Cons.  Exp:  Nondurable goods	&	gmdcn	&	PCE defl:  nondble	&	7	\\
119	&	6	&	DSERRG3M086SBEA	&	Personal Cons.  Exp:  Services	&	gmdcs	&	PCE defl:  service	&	7	\\
120	&	6	&	CES0600000008	&	Avg Hourly Earnings :  Goods-Producing	&	M\_123109182	&	AHE: goods	&	2	\\
121	&	6	&	CES2000000008	&	Avg Hourly Earnings :  Construction	&	M\_123109341	&	AHE: const	&	2	\\
122	&	6	&	CES3000000008	&	Avg Hourly Earnings :  Manufacturing	&	M\_123109552	&	AHE: mfg	&	2	\\
\bf 123	&\bf 	2	&	\bf UMCSENTx	&	\bf Consumer Sentiment Index	&\bf 	hhsntn	&	\bf Consumer expect	&\bf 	4	\\
124	&	6	&	DTCOLNVHFNM	&	Consumer Motor Vehicle Loans Outstanding	&	N.A.	&	N.A.	&	5	\\
125	&	6	&	DTCTHFNM	&	Total Consumer Loans and Leases Outstanding	&	N.A.	&	N.A.	&	5	\\
126	&	6	&	INVEST	&	Securities in Bank Credit at All Commercial Banks	&	N.A.	&	N.A.	&	5	\\
127	&	1	&	VIXCLSx	&	VIX	&		&		&	8	\\
\bottomrule
 
\end{longtable}
 \end{center}}

\end{document}